\documentclass[submission,copyright,creativecommons]{eptcs}
\providecommand{\event}{QPL 2022}
\usepackage{graphicx}

\usepackage{amsmath,amsthm,amssymb}
\usepackage{mathtools}
\usepackage{xcolor}
\usepackage[font=small,labelfont=bf,skip=0pt,justification=justified]{caption}
\usepackage{stackengine}
\usepackage{stmaryrd}
\usepackage[most]{tcolorbox}
\usepackage[export]{adjustbox}
\usepackage{tikzit}
\input{zx.tikzdefs}

\tikzstyle{box}=[shape=rectangle, fill=white, draw=black, minimum height=4mm, minimum width=4mm, font={\footnotesize}]
\tikzstyle{Z dot}=[inner sep=0mm, minimum size=2mm, shape=circle, draw=black, fill={rgb,255: red,221; green,255; blue,221}]
\tikzstyle{Z phase dot}=[minimum size=5mm, font={\footnotesize\boldmath}, shape=rectangle, rounded corners=2mm, inner sep=1mm, outer sep=-2mm, scale=0.8, tikzit shape=circle, draw=black, fill={rgb,255: red,221; green,255; blue,221}, tikzit draw=blue]
\tikzstyle{X dot}=[Z dot, shape=circle, draw=black, fill={rgb,255: red,255; green,136; blue,136}]
\tikzstyle{X phase dot}=[Z phase dot, tikzit shape=circle, tikzit draw=blue, fill={rgb,255: red,255; green,136; blue,136}, font={\footnotesize\boldmath}]
\tikzstyle{hadamard}=[fill=yellow, draw=black, shape=rectangle, inner sep=0.6mm, minimum height=1.5mm, minimum width=1.5mm, rounded corners=0.3mm]
\tikzstyle{qutrit Z phase}=[minimum size=5mm, shape=circle split, draw=black, fill={rgb,255: red,221; green,255; blue,221}, font={\tiny}, inner sep=0.75pt, tikzit draw=blue]
\tikzstyle{qutrit Z phase rectangle}=[minimum size=5mm, shape=rectangle split, rectangle split parts=2, rounded corners=1mm, draw=black, fill={rgb,255: red,221; green,255; blue,221}, font={\tiny}, inner sep=1pt, outer sep=-0.5mm, tikzit draw=blue, tikzit shape=rectangle]
\tikzstyle{qutrit X phase}=[minimum size=5mm, shape=circle split, draw=black, fill={rgb,255: red,255; green,136; blue,136}, font={\tiny}, inner sep=0.75pt, tikzit draw=blue]
\tikzstyle{qutrit hadamard}=[fill=yellow, draw=black, shape=rectangle, inner sep=0.5mm, minimum height=2.5mm, minimum width=2.5mm, font={\tiny{1}}, tikzit draw=blue, rounded corners=0.5mm]
\tikzstyle{qutrit hadamard adjoint}=[fill=yellow, draw=black, shape=rectangle, inner sep=0.5mm, minimum height=2.5mm, minimum width=2.5mm, font={\tiny{2}}, tikzit draw=red, rounded corners=0.5mm]
\tikzstyle{qutrit hadamard unknown}=[fill=yellow, draw=black, shape=rectangle, inner sep=0.5mm, minimum height=2.5mm, minimum width=2.5mm, font={\tiny}, tikzit draw=green, rounded corners=0.5mm]
\tikzstyle{diamond}=[draw=black, fill={rgb,255: red,250; green,255; blue,190}, shape=diamond, minimum size=2.5mm, inner sep=0mm, tikzit shape=diamond]
\tikzstyle{diamond non-zero}=[draw=black, fill={rgb,255: red,250; green,255; blue,190}, draw=black, shape=diamond, minimum size=2.5mm, inner sep=0.2mm, tikzit draw=blue]
\tikzstyle{red font}=[text=red, tikzit draw=red]
\tikzstyle{vertex}=[inner sep=0mm, minimum size=1mm, shape=circle, draw=black, fill=black]
\tikzstyle{vertex set}=[inner sep=0mm, minimum size=1.5mm, shape=circle, draw=black, fill=white, font={\footnotesize\boldmath}]

\tikzstyle{hadamard edge}=[-, dashed, dash pattern=on 2.5pt off 0.75pt, thick, draw={rgb,255: red,68; green,136; blue,255}]
\tikzstyle{hadamard adjoint adge}=[-, dashed, dash pattern=on 1pt off 0.75pt, thick, draw={rgb,255: red,185; green,25; blue,255}]
\tikzstyle{brace edge}=[-, tikzit draw=blue, decorate, decoration={brace,amplitude=1mm,raise=-1mm}]
\tikzstyle{diredge}=[->]
\tikzstyle{empty}=[-, dashed, draw={rgb,255: red,128; green,128; blue,128}]
\tikzstyle{bang}=[-, draw={rgb,255: red,255; green,128; blue,0}, tikzit draw={rgb,255: red,255; green,128; blue,0}]

\usepackage{hyperref} 

\theoremstyle{definition}
\newtheorem{theorem}{Theorem}[section]
\newtheorem{corollary}[theorem]{Corollary}
\newtheorem{lemma}[theorem]{Lemma}
\newtheorem*{lemma*}{Lemma}
\newtheorem*{proposition*}{Proposition}
\newtheorem{proposition}[theorem]{Proposition}

\newtheorem{definition}[theorem]{Definition}

\newcommand{\defeq}{\vcentcolon=}
\newcommand{\eqdef}{=\vcentcolon}
\newcommand{\xeq}[1]{\mathrel{\stackon[5pt]{$=$}{$\scriptstyle{#1}$}}}
\newcommand{\xeqq}[2]{\mathrel{\stackon[5pt]{$=$}{\stackon[5pt]{$\scriptstyle{#1}$}{$\scriptstyle{#2}$}}}}

\newcommand{\qutritZphase}[2]{\,\tikz{\node[style=qutrit Z phase] (x) {$#1$ \nodepart{lower} $#2$};}\,}

\newcommand{\qutritZspider}[2]{\,\begin{tikzpicture}
	\begin{pgfonlayer}{nodelayer}
		\node [style=none] (1) at (0, 0.75) {...};
		\node [style=none] (5) at (-0.75, 1) {};
		\node [style=none] (7) at (0.75, 1) {};
		\node [style=none] (8) at (0, -0.75) {...};
		\node [style=none] (9) at (-0.75, -1) {};
		\node [style=qutrit Z phase] (10) at (0, 0) {$#1$ \nodepart{lower} $#2$};
		\node [style=none] (11) at (0.75, -1) {};
	\end{pgfonlayer}
	\begin{pgfonlayer}{edgelayer}
		\draw [bend left=15] (10) to (11.center);
		\draw [bend right=15] (10) to (9.center);
		\draw [bend right=15] (10) to (7.center);
		\draw [bend left=15] (10) to (5.center);
	\end{pgfonlayer}
\end{tikzpicture}}
\newcommand{\qutritState}[3]{\, \begin{tikzpicture}
	\begin{pgfonlayer}{nodelayer}
		\node [style=none] (1) at (0, 0.75) {};
		\node [style=qutrit #1 phase] (2) at (0, -0.5) {$#2$ \nodepart{lower} $#3$};
	\end{pgfonlayer}
	\begin{pgfonlayer}{edgelayer}
		\draw (1.center) to (2);
	\end{pgfonlayer}
\end{tikzpicture}}
\newcommand{\qutritZstate}[2]{\qutritState{Z}{#1}{#2}}
\newcommand{\qutritXstate}[2]{\qutritState{X}{#1}{#2}}

\newcommand{\qubitPhaseGate}[2]{\,\begin{tikzpicture}
	\begin{pgfonlayer}{nodelayer}
		\node [style=#1 phase dot] (0) at (0, 0) {$#2$};
		\node [style=none] (1) at (0, 1) {};
		\node [style=none] (2) at (0, -1) {};
	\end{pgfonlayer}
	\begin{pgfonlayer}{edgelayer}
		\draw (1.center) to (2.center);
	\end{pgfonlayer}
\end{tikzpicture}}
\newcommand{\qubitZphase}[1]{\qubitPhaseGate{Z}{#1}}
\newcommand{\qubitXphase}[1]{\qubitPhaseGate{X}{#1}}

\newcommand{\someSpider}[1]{$\mathcal{#1}$-spider}
\newcommand{\someSpiders}[1]{\someSpider{#1}s}
\newcommand{\Mspider}[0]{\someSpider{M}}
\newcommand{\Nspider}[0]{\someSpider{N}}
\newcommand{\Pspider}[0]{\someSpider{P}}
\newcommand{\Mspiders}[0]{\someSpiders{M}}
\newcommand{\Nspiders}[0]{\someSpiders{N}}
\newcommand{\Pspiders}[0]{\someSpiders{P}}

\newcommand{\qutritRuleFusion}{\hyperlink{qutrit_rule_fusion}{\mathbf{(f)}}}
\newcommand{\qutritRuleId}{\hyperlink{qutrit_rule_id}{\mathbf{(id)}}}

\newcommand{\qutritRuleZeroCopy}{\hyperlink{qutrit_rule_0_copy}{\mathbf{(cp_0)}}}

\newcommand{\qutritRuleMCopy}{\hyperlink{qutrit_rule_m_copy}{\mathbf{(cp_{\mathcal{M}})}}}
\newcommand{\qutritRuleCommute}{\hyperlink{qutrit_rule_commute}{\mathbf{(cm)}}}
\newcommand{\qutritRuleHadamard}{\hyperlink{qutrit_rule_hadamard}{\mathbf{(H)}}}
\newcommand{\qutritRuleEuler}{\hyperlink{qutrit_rule_euler}{\mathbf{(E)}}}
\newcommand{\qutritRuleColourChange}{\hyperlink{qutrit_rule_colour_change}{\mathbf{(cc)}}}
\newcommand{\qutritRuleSnake}{\hyperlink{qutrit_rule_snake}{\mathbf{(s)}}}

\newcommand{\eliminatePSpidersStatement}{
	Given any graph-like ZX-diagram containing an interior \Pspider\ $x$ with phase \qutritZphase{p}{p} for $p \in \{1,2\}$, suppose we perform a $p$-local complementation at $x$. Then the new ZX-diagram is related to the old one by the following equality:
	\begin{equation}
		\begin{tikzpicture}
	\begin{pgfonlayer}{nodelayer}
		\node [style=qutrit Z phase] (0) at (-1.75, 0.25) {$\alpha_k$ \nodepart{lower} $\beta_k$};
		\node [style=none] (1) at (-1.75, 1.25) {...};
		\node [style=none] (2) at (-2.5, 1.5) {};
		\node [style=none] (3) at (-1, 1.5) {};
		\node [style=qutrit Z phase] (4) at (1.75, 0.25) {$\gamma_k$ \nodepart{lower} $\delta_k$};
		\node [style=none] (5) at (1.75, 1.25) {...};
		\node [style=none] (6) at (1, 1.5) {};
		\node [style=none] (7) at (2.5, 1.5) {};
		\node [style=none] (13) at (-1.75, 2.075) {\textcolor{bang_orange}{$\scriptstyle{k \in [K_1]}$}};
		\node [style=none] (14) at (-2.5, -0.5) {};
		\node [style=none] (15) at (-2.75, -0.25) {};
		\node [style=none] (16) at (-1, -0.5) {};
		\node [style=none] (17) at (-0.75, -0.25) {};
		\node [style=none] (18) at (-1, 1.75) {};
		\node [style=none] (19) at (-0.75, 1.5) {};
		\node [style=none] (20) at (-2.5, 1.75) {};
		\node [style=none] (21) at (-2.75, 1.5) {};
		\node [style=none] (45) at (1.75, 2.075) {\textcolor{bang_orange}{$\scriptstyle{k \in [K_2]}$}};
		\node [style=none] (46) at (1, -0.5) {};
		\node [style=none] (47) at (0.75, -0.25) {};
		\node [style=none] (48) at (2.5, -0.5) {};
		\node [style=none] (49) at (2.75, -0.25) {};
		\node [style=none] (50) at (2.5, 1.75) {};
		\node [style=none] (51) at (2.75, 1.5) {};
		\node [style=none] (52) at (1, 1.75) {};
		\node [style=none] (53) at (0.75, 1.5) {};
		\node [style=none] (64) at (1, 1.75) {};
		\node [style=qutrit Z phase] (65) at (0, -1.75) {$p$ \nodepart{lower} $p$};
	\end{pgfonlayer}
	\begin{pgfonlayer}{edgelayer}
		\draw [bend right=15, looseness=0.75] (2.center) to (0);
		\draw [bend right=15, looseness=0.75] (0) to (3.center);
		\draw [bend right=15, looseness=0.75] (6.center) to (4);
		\draw [bend right=15, looseness=0.75] (4) to (7.center);
		\draw [style=bang] (17.center)
			 to (19.center)
			 to [bend right=45, looseness=0.75] (18.center)
			 to (20.center)
			 to [bend right=45, looseness=0.75] (21.center)
			 to (15.center)
			 to [bend right=45, looseness=0.75] (14.center)
			 to (16.center)
			 to [bend right=45, looseness=0.75] cycle;
		\draw [style=bang] (49.center)
			 to (51.center)
			 to [bend right=45, looseness=0.75] (50.center)
			 to (52.center)
			 to [bend right=45, looseness=0.75] (53.center)
			 to (47.center)
			 to [bend right=45, looseness=0.75] (46.center)
			 to (48.center)
			 to [bend right=45, looseness=0.75] cycle;
		\draw [style=hadamard edge] (0) to (65);
		\draw [style=hadamard adjoint adge] (65) to (4);
	\end{pgfonlayer}
\end{tikzpicture}
 \quad = \quad \begin{tikzpicture}
	\begin{pgfonlayer}{nodelayer}
		\node [style=qutrit Z phase rectangle] (0) at (-1.75, -1) {$\alpha_k - p$ \nodepart{second} $\beta_k - p$};
		\node [style=none] (1) at (-1.75, 0) {...};
		\node [style=none] (2) at (-2.5, 0.25) {};
		\node [style=none] (3) at (-1, 0.25) {};
		\node [style=qutrit Z phase rectangle] (4) at (1.75, -1) {$\gamma_k - p$ \nodepart{second} $\delta_k - p$};
		\node [style=none] (5) at (1.75, 0) {...};
		\node [style=none] (6) at (1, 0.25) {};
		\node [style=none] (7) at (2.5, 0.25) {};
		\node [style=none] (13) at (-1.75, 1.825) {\textcolor{bang_orange}{$\scriptstyle{k \in [K_1]}$}};
		\node [style=none] (14) at (-2.5, -1.75) {};
		\node [style=none] (15) at (-2.75, -1.5) {};
		\node [style=none] (16) at (-1, -1.75) {};
		\node [style=none] (17) at (-0.75, -1.5) {};
		\node [style=none] (18) at (-1, 1.5) {};
		\node [style=none] (19) at (-0.75, 1.25) {};
		\node [style=none] (20) at (-2.5, 1.5) {};
		\node [style=none] (21) at (-2.75, 1.25) {};
		\node [style=none] (45) at (1.75, 1.825) {\textcolor{bang_orange}{$\scriptstyle{k \in [K_2]}$}};
		\node [style=none] (46) at (1, -1.75) {};
		\node [style=none] (47) at (0.75, -1.5) {};
		\node [style=none] (48) at (2.5, -1.75) {};
		\node [style=none] (49) at (2.75, -1.5) {};
		\node [style=none] (50) at (2.5, 1.5) {};
		\node [style=none] (51) at (2.75, 1.25) {};
		\node [style=none] (52) at (1, 1.5) {};
		\node [style=none] (53) at (0.75, 1.25) {};
		\node [style=none] (60) at (-1.75, 1.5) {};
		\node [style=none] (61) at (-2.75, 0.5) {};
		\node [style=none] (62) at (-1.75, 0.5) {};
		\node [style=none] (64) at (1, 1.5) {};
		\node [style=qutrit hadamard unknown] (69) at (0, -1) {$-p$};
		\node [style=none] (70) at (-2.25, 0.5) {};
		\node [style=none] (71) at (-2.25, 1.5) {};
		\node [style=qutrit hadamard unknown] (72) at (-2.25, 1) {$p$};
		\node [style=none] (73) at (1.75, 1.5) {};
		\node [style=none] (74) at (0.75, 0.5) {};
		\node [style=none] (75) at (1.75, 0.5) {};
		\node [style=none] (76) at (1.25, 0.5) {};
		\node [style=none] (77) at (1.25, 1.5) {};
		\node [style=qutrit hadamard unknown] (78) at (1.25, 1) {$p$};
	\end{pgfonlayer}
	\begin{pgfonlayer}{edgelayer}
		\draw [bend right=15, looseness=0.75] (2.center) to (0);
		\draw [bend right=15, looseness=0.75] (0) to (3.center);
		\draw [bend right=15, looseness=0.75] (6.center) to (4);
		\draw [bend right=15, looseness=0.75] (4) to (7.center);
		\draw [style=bang] (17.center)
			 to (19.center)
			 to [bend right=45, looseness=0.75] (18.center)
			 to (20.center)
			 to [bend right=45, looseness=0.75] (21.center)
			 to (15.center)
			 to [bend right=45, looseness=0.75] (14.center)
			 to (16.center)
			 to [bend right=45, looseness=0.75] cycle;
		\draw [style=bang] (49.center)
			 to (51.center)
			 to [bend right=45, looseness=0.75] (50.center)
			 to (52.center)
			 to [bend right=45, looseness=0.75] (53.center)
			 to (47.center)
			 to [bend right=45, looseness=0.75] (46.center)
			 to (48.center)
			 to [bend right=45, looseness=0.75] cycle;
		\draw [style=bang] (60.center) to (62.center);
		\draw [style=bang] (61.center) to (62.center);
		\draw (0) to (4);
		\draw (71.center) to (70.center);
		\draw [style=bang] (73.center) to (75.center);
		\draw [style=bang] (74.center) to (75.center);
		\draw (77.center) to (76.center);
	\end{pgfonlayer}
\end{tikzpicture}

	\end{equation}
}

\newcommand{\eliminateNSpidersStatement}{
	Given any graph-like ZX-diagram containing an interior \Nspider\ $x$ with phase \qutritZphase{0}{n} or \qutritZphase{n}{0} for $n \in \{1,2\}$, suppose we perform a $(-n)$-local complementation at $x$. Then, treating the two cases separately, the new ZX-diagrams are related to the old ones by the equalities:
	\begin{equation}
		\scalebox{0.8}{\begin{tikzpicture}
	\begin{pgfonlayer}{nodelayer}
		\node [style=qutrit Z phase] (0) at (-1.25, 0) {$\alpha_k$ \nodepart{lower} $\beta_k$};
		\node [style=none] (1) at (-1.25, 1) {...};
		\node [style=none] (2) at (-2, 1.25) {};
		\node [style=none] (3) at (-0.5, 1.25) {};
		\node [style=qutrit Z phase] (4) at (1.25, 0) {$\gamma_k$ \nodepart{lower} $\delta_k$};
		\node [style=none] (5) at (1.25, 1) {...};
		\node [style=none] (6) at (0.5, 1.25) {};
		\node [style=none] (7) at (2, 1.25) {};
		\node [style=qutrit Z phase] (8) at (0, -1.75) {$0$ \nodepart{lower} $n$};
		\node [style=none] (13) at (-1.25, 1.825) {\textcolor{bang_orange}{$\scriptstyle{k \in [K_1]}$}};
		\node [style=none] (14) at (-2, -0.75) {};
		\node [style=none] (15) at (-2.25, -0.5) {};
		\node [style=none] (16) at (-0.5, -0.75) {};
		\node [style=none] (17) at (-0.25, -0.5) {};
		\node [style=none] (18) at (-0.5, 1.5) {};
		\node [style=none] (19) at (-0.25, 1.25) {};
		\node [style=none] (20) at (-2, 1.5) {};
		\node [style=none] (21) at (-2.25, 1.25) {};
		\node [style=none] (45) at (1.25, 1.825) {\textcolor{bang_orange}{$\scriptstyle{k \in [K_2]}$}};
		\node [style=none] (46) at (0.5, -0.75) {};
		\node [style=none] (47) at (0.25, -0.5) {};
		\node [style=none] (48) at (2, -0.75) {};
		\node [style=none] (49) at (2.25, -0.5) {};
		\node [style=none] (50) at (2, 1.5) {};
		\node [style=none] (51) at (2.25, 1.25) {};
		\node [style=none] (52) at (0.5, 1.5) {};
		\node [style=none] (53) at (0.25, 1.25) {};
	\end{pgfonlayer}
	\begin{pgfonlayer}{edgelayer}
		\draw [bend right=15, looseness=0.75] (2.center) to (0);
		\draw [bend right=15, looseness=0.75] (0) to (3.center);
		\draw [bend right=15, looseness=0.75] (6.center) to (4);
		\draw [bend right=15, looseness=0.75] (4) to (7.center);
		\draw [style=hadamard edge] (0) to (8);
		\draw [style=hadamard adjoint adge] (8) to (4);
		\draw [style=bang] (17.center)
			 to (19.center)
			 to [bend right=45, looseness=0.75] (18.center)
			 to (20.center)
			 to [bend right=45, looseness=0.75] (21.center)
			 to (15.center)
			 to [bend right=45, looseness=0.75] (14.center)
			 to (16.center)
			 to [bend right=45, looseness=0.75] cycle;
		\draw [style=bang] (49.center)
			 to (51.center)
			 to [bend right=45, looseness=0.75] (50.center)
			 to (52.center)
			 to [bend right=45, looseness=0.75] (53.center)
			 to (47.center)
			 to [bend right=45, looseness=0.75] (46.center)
			 to (48.center)
			 to [bend right=45, looseness=0.75] cycle;
	\end{pgfonlayer}
\end{tikzpicture}
} = 
		\scalebox{0.8}{\begin{tikzpicture}
	\begin{pgfonlayer}{nodelayer}
		\node [style=qutrit Z phase rectangle] (0) at (-2.25, -1) {$\alpha_k + n  + 2$ \nodepart{second} $\beta_k + n + 1$};
		\node [style=none] (1) at (-2.25, 0) {...};
		\node [style=none] (2) at (-3, 0.25) {};
		\node [style=none] (3) at (-1.5, 0.25) {};
		\node [style=qutrit Z phase rectangle] (4) at (2.25, -1) {$\gamma_k + n + 1$ \nodepart{second} $\delta_k + n + 2$};
		\node [style=none] (5) at (2.25, 0) {...};
		\node [style=none] (6) at (1.5, 0.25) {};
		\node [style=none] (7) at (3, 0.25) {};
		\node [style=none] (13) at (-2.25, 1.825) {\textcolor{bang_orange}{$\scriptstyle{k \in [K_1]}$}};
		\node [style=none] (14) at (-3.5, -1.75) {};
		\node [style=none] (15) at (-3.75, -1.5) {};
		\node [style=none] (16) at (-1, -1.75) {};
		\node [style=none] (17) at (-0.75, -1.5) {};
		\node [style=none] (18) at (-1, 1.5) {};
		\node [style=none] (19) at (-0.75, 1.25) {};
		\node [style=none] (20) at (-3.5, 1.5) {};
		\node [style=none] (21) at (-3.75, 1.25) {};
		\node [style=none] (45) at (2.25, 1.825) {\textcolor{bang_orange}{$\scriptstyle{k \in [K_{2}]}$}};
		\node [style=none] (46) at (1, -1.75) {};
		\node [style=none] (47) at (0.75, -1.5) {};
		\node [style=none] (48) at (3.5, -1.75) {};
		\node [style=none] (49) at (3.75, -1.5) {};
		\node [style=none] (50) at (3.5, 1.5) {};
		\node [style=none] (51) at (3.75, 1.25) {};
		\node [style=none] (52) at (1, 1.5) {};
		\node [style=none] (53) at (0.75, 1.25) {};
		\node [style=none] (64) at (1, 1.5) {};
		\node [style=qutrit hadamard unknown] (69) at (0, -1) {$n$};
		\node [style=none] (70) at (2.25, 1.5) {};
		\node [style=none] (71) at (0.75, 0.5) {};
		\node [style=none] (72) at (2.25, 0.5) {};
		\node [style=none] (73) at (1.5, 0.5) {};
		\node [style=none] (74) at (1.5, 1.5) {};
		\node [style=qutrit hadamard unknown] (75) at (1.5, 1) {$-n$};
		\node [style=none] (76) at (-2.25, 1.5) {};
		\node [style=none] (77) at (-3.75, 0.5) {};
		\node [style=none] (78) at (-2.25, 0.5) {};
		\node [style=none] (79) at (-3, 0.5) {};
		\node [style=none] (80) at (-3, 1.5) {};
		\node [style=qutrit hadamard unknown] (81) at (-3, 1) {$-n$};
	\end{pgfonlayer}
	\begin{pgfonlayer}{edgelayer}
		\draw [bend right=15, looseness=0.75] (2.center) to (0);
		\draw [bend right=15, looseness=0.75] (0) to (3.center);
		\draw [bend right=15, looseness=0.75] (6.center) to (4);
		\draw [bend right=15, looseness=0.75] (4) to (7.center);
		\draw [style=bang] (17.center)
			 to (19.center)
			 to [bend right=45, looseness=0.75] (18.center)
			 to (20.center)
			 to [bend right=45, looseness=0.75] (21.center)
			 to (15.center)
			 to [bend right=45, looseness=0.75] (14.center)
			 to (16.center)
			 to [bend right=45, looseness=0.75] cycle;
		\draw [style=bang] (49.center)
			 to (51.center)
			 to [bend right=45, looseness=0.75] (50.center)
			 to (52.center)
			 to [bend right=45, looseness=0.75] (53.center)
			 to (47.center)
			 to [bend right=45, looseness=0.75] (46.center)
			 to (48.center)
			 to [bend right=45, looseness=0.75] cycle;
		\draw (0) to (4);
		\draw [style=bang] (70.center) to (72.center);
		\draw [style=bang] (71.center) to (72.center);
		\draw (74.center) to (73.center);
		\draw [style=bang] (76.center) to (78.center);
		\draw [style=bang] (77.center) to (78.center);
		\draw (80.center) to (79.center);
	\end{pgfonlayer}
\end{tikzpicture}
} ~~, \qquad
		\scalebox{0.8}{\begin{tikzpicture}
	\begin{pgfonlayer}{nodelayer}
		\node [style=qutrit Z phase] (0) at (-1.25, 0) {$\alpha_k$ \nodepart{lower} $\beta_k$};
		\node [style=none] (1) at (-1.25, 1) {...};
		\node [style=none] (2) at (-2, 1.25) {};
		\node [style=none] (3) at (-0.5, 1.25) {};
		\node [style=qutrit Z phase] (4) at (1.25, 0) {$\gamma_k$ \nodepart{lower} $\delta_k$};
		\node [style=none] (5) at (1.25, 1) {...};
		\node [style=none] (6) at (0.5, 1.25) {};
		\node [style=none] (7) at (2, 1.25) {};
		\node [style=qutrit Z phase] (8) at (0, -1.75) {$n$ \nodepart{lower} $0$};
		\node [style=none] (13) at (-1.25, 1.825) {\textcolor{bang_orange}{$\scriptstyle{k \in [K_1]}$}};
		\node [style=none] (14) at (-2, -0.75) {};
		\node [style=none] (15) at (-2.25, -0.5) {};
		\node [style=none] (16) at (-0.5, -0.75) {};
		\node [style=none] (17) at (-0.25, -0.5) {};
		\node [style=none] (18) at (-0.5, 1.5) {};
		\node [style=none] (19) at (-0.25, 1.25) {};
		\node [style=none] (20) at (-2, 1.5) {};
		\node [style=none] (21) at (-2.25, 1.25) {};
		\node [style=none] (45) at (1.25, 1.825) {\textcolor{bang_orange}{$\scriptstyle{k \in [K_2]}$}};
		\node [style=none] (46) at (0.5, -0.75) {};
		\node [style=none] (47) at (0.25, -0.5) {};
		\node [style=none] (48) at (2, -0.75) {};
		\node [style=none] (49) at (2.25, -0.5) {};
		\node [style=none] (50) at (2, 1.5) {};
		\node [style=none] (51) at (2.25, 1.25) {};
		\node [style=none] (52) at (0.5, 1.5) {};
		\node [style=none] (53) at (0.25, 1.25) {};
	\end{pgfonlayer}
	\begin{pgfonlayer}{edgelayer}
		\draw [bend right=15, looseness=0.75] (2.center) to (0);
		\draw [bend right=15, looseness=0.75] (0) to (3.center);
		\draw [bend right=15, looseness=0.75] (6.center) to (4);
		\draw [bend right=15, looseness=0.75] (4) to (7.center);
		\draw [style=hadamard edge] (0) to (8);
		\draw [style=hadamard adjoint adge] (8) to (4);
		\draw [style=bang] (17.center)
			 to (19.center)
			 to [bend right=45, looseness=0.75] (18.center)
			 to (20.center)
			 to [bend right=45, looseness=0.75] (21.center)
			 to (15.center)
			 to [bend right=45, looseness=0.75] (14.center)
			 to (16.center)
			 to [bend right=45, looseness=0.75] cycle;
		\draw [style=bang] (49.center)
			 to (51.center)
			 to [bend right=45, looseness=0.75] (50.center)
			 to (52.center)
			 to [bend right=45, looseness=0.75] (53.center)
			 to (47.center)
			 to [bend right=45, looseness=0.75] (46.center)
			 to (48.center)
			 to [bend right=45, looseness=0.75] cycle;
	\end{pgfonlayer}
\end{tikzpicture}
} =
		\scalebox{0.8}{\begin{tikzpicture}
	\begin{pgfonlayer}{nodelayer}
		\node [style=qutrit Z phase rectangle] (0) at (-2.25, -1) {$\alpha_k + n  + 1$ \nodepart{second} $\beta_k + n + 2$};
		\node [style=none] (1) at (-2.25, 0) {...};
		\node [style=none] (2) at (-3, 0.25) {};
		\node [style=none] (3) at (-1.5, 0.25) {};
		\node [style=qutrit Z phase rectangle] (4) at (2.25, -1) {$\gamma_k + n + 2$ \nodepart{second} $\delta_k + n + 1$};
		\node [style=none] (5) at (2.25, 0) {...};
		\node [style=none] (6) at (1.5, 0.25) {};
		\node [style=none] (7) at (3, 0.25) {};
		\node [style=none] (13) at (-2.25, 1.825) {\textcolor{bang_orange}{$\scriptstyle{k \in [K_1]}$}};
		\node [style=none] (14) at (-3.5, -1.75) {};
		\node [style=none] (15) at (-3.75, -1.5) {};
		\node [style=none] (16) at (-1, -1.75) {};
		\node [style=none] (17) at (-0.75, -1.5) {};
		\node [style=none] (18) at (-1, 1.5) {};
		\node [style=none] (19) at (-0.75, 1.25) {};
		\node [style=none] (20) at (-3.5, 1.5) {};
		\node [style=none] (21) at (-3.75, 1.25) {};
		\node [style=none] (45) at (2.25, 1.825) {\textcolor{bang_orange}{$\scriptstyle{k \in [K_{2}]}$}};
		\node [style=none] (46) at (1, -1.75) {};
		\node [style=none] (47) at (0.75, -1.5) {};
		\node [style=none] (48) at (3.5, -1.75) {};
		\node [style=none] (49) at (3.75, -1.5) {};
		\node [style=none] (50) at (3.5, 1.5) {};
		\node [style=none] (51) at (3.75, 1.25) {};
		\node [style=none] (52) at (1, 1.5) {};
		\node [style=none] (53) at (0.75, 1.25) {};
		\node [style=none] (64) at (1, 1.5) {};
		\node [style=qutrit hadamard unknown] (69) at (0, -1) {$n$};
		\node [style=none] (70) at (2.25, 1.5) {};
		\node [style=none] (71) at (0.75, 0.5) {};
		\node [style=none] (72) at (2.25, 0.5) {};
		\node [style=none] (73) at (1.5, 0.5) {};
		\node [style=none] (74) at (1.5, 1.5) {};
		\node [style=qutrit hadamard unknown] (75) at (1.5, 1) {$-n$};
		\node [style=none] (76) at (-2.25, 1.5) {};
		\node [style=none] (77) at (-3.75, 0.5) {};
		\node [style=none] (78) at (-2.25, 0.5) {};
		\node [style=none] (79) at (-3, 0.5) {};
		\node [style=none] (80) at (-3, 1.5) {};
		\node [style=qutrit hadamard unknown] (81) at (-3, 1) {$-n$};
	\end{pgfonlayer}
	\begin{pgfonlayer}{edgelayer}
		\draw [bend right=15, looseness=0.75] (2.center) to (0);
		\draw [bend right=15, looseness=0.75] (0) to (3.center);
		\draw [bend right=15, looseness=0.75] (6.center) to (4);
		\draw [bend right=15, looseness=0.75] (4) to (7.center);
		\draw [style=bang] (17.center)
			 to (19.center)
			 to [bend right=45, looseness=0.75] (18.center)
			 to (20.center)
			 to [bend right=45, looseness=0.75] (21.center)
			 to (15.center)
			 to [bend right=45, looseness=0.75] (14.center)
			 to (16.center)
			 to [bend right=45, looseness=0.75] cycle;
		\draw [style=bang] (49.center)
			 to (51.center)
			 to [bend right=45, looseness=0.75] (50.center)
			 to (52.center)
			 to [bend right=45, looseness=0.75] (53.center)
			 to (47.center)
			 to [bend right=45, looseness=0.75] (46.center)
			 to (48.center)
			 to [bend right=45, looseness=0.75] cycle;
		\draw (0) to (4);
		\draw [style=bang] (70.center) to (72.center);
		\draw [style=bang] (71.center) to (72.center);
		\draw (74.center) to (73.center);
		\draw [style=bang] (76.center) to (78.center);
		\draw [style=bang] (77.center) to (78.center);
		\draw (80.center) to (79.center);
	\end{pgfonlayer}
\end{tikzpicture}
}
	\end{equation}
}

\newcommand{\eliminateMSpidersStatement}{
	Given any graph-like ZX-diagram containing two interior \Mspiders\ $i$ and $j$ connected by edge $ij$ of weight $w_{i,j} \eqdef w \in \{1,2\}$, suppose we perform a proper $\pm w$-pivot along $ij$ (both choices give the same result). For $w=1$, the new ZX-diagram is related to the old one by the following equality:
	\begin{equation}
		\begin{split}
			& \hspace{60pt} \scalebox{0.9}{\input{eliminate/M_spiders/bang/w_1/LHS.tikz}} \\[15pt]
		 	= &\quad \scalebox{0.9}{\input{eliminate/M_spiders/bang/w_1/RHS.tikz}}
		 \end{split}
	\end{equation}
	The case $w=2$ differs only as follows: in the first diagram (the left hand side of the equation), the edge $ij$ is purple (by definition), while in the lower diagram (the right hand side of the equation), a $\pm 2 = \mp 1$ will replace all occurences of $\pm 1$, and the roles of purple and blue will be swapped throughout.
}

\newcommand{\HEdgesAreModThreeStatement}{
	The following equations hold in the qutrit ZX-calculus:
	\begin{equation}
		\input{hadamard_lemmas/3_h_edges_vanish/blue.tikz} \ = \ 
		\input{hadamard_lemmas/3_h_edges_vanish/disconnected.tikz} \ = \ 
		\input{hadamard_lemmas/3_h_edges_vanish/purple.tikz} \ ,
		\hspace{25pt}
		\begin{tikzpicture}
	\begin{pgfonlayer}{nodelayer}
		\node [style=none] (11) at (-0.75, 2.25) {};
		\node [style=none] (12) at (0.75, 2.25) {};
		\node [style=none] (13) at (0, 2) {...};
		\node [style=qutrit Z phase] (14) at (0, 1) {$\alpha$ \nodepart{lower} $\beta$};
		\node [style=none] (15) at (-0.75, -2.25) {};
		\node [style=none] (16) at (0.75, -2.25) {};
		\node [style=none] (17) at (0, -2) {...};
		\node [style=qutrit Z phase] (18) at (0, -1) {$\gamma$ \nodepart{lower} $\delta$};
	\end{pgfonlayer}
	\begin{pgfonlayer}{edgelayer}
		\draw [bend right=15] (11.center) to (14);
		\draw [bend right=15] (14) to (12.center);
		\draw [bend left=15] (15.center) to (18);
		\draw [bend left=15] (18) to (16.center);
		\draw [style=hadamard edge, bend right=45] (14) to (18);
		\draw [style=hadamard edge, bend left=45] (14) to (18);
	\end{pgfonlayer}
\end{tikzpicture}
 \ = \ 
		\begin{tikzpicture}
	\begin{pgfonlayer}{nodelayer}
		\node [style=none] (11) at (-0.75, 2.25) {};
		\node [style=none] (12) at (0.75, 2.25) {};
		\node [style=none] (13) at (0, 2) {...};
		\node [style=qutrit Z phase] (14) at (0, 1) {$\alpha$ \nodepart{lower} $\beta$};
		\node [style=none] (15) at (-0.75, -2.25) {};
		\node [style=none] (16) at (0.75, -2.25) {};
		\node [style=none] (17) at (0, -2) {...};
		\node [style=qutrit Z phase] (18) at (0, -1) {$\gamma$ \nodepart{lower} $\delta$};
	\end{pgfonlayer}
	\begin{pgfonlayer}{edgelayer}
		\draw [bend right=15] (11.center) to (14);
		\draw [bend right=15] (14) to (12.center);
		\draw [bend left=15] (15.center) to (18);
		\draw [bend left=15] (18) to (16.center);
		\draw [style=hadamard adjoint adge] (14) to (18);
	\end{pgfonlayer}
\end{tikzpicture}
 \ ,
		\hspace{25pt}
		\begin{tikzpicture}
	\begin{pgfonlayer}{nodelayer}
		\node [style=none] (11) at (-0.75, 2.25) {};
		\node [style=none] (12) at (0.75, 2.25) {};
		\node [style=none] (13) at (0, 2) {...};
		\node [style=qutrit Z phase] (14) at (0, 1) {$\alpha$ \nodepart{lower} $\beta$};
		\node [style=none] (15) at (-0.75, -2.25) {};
		\node [style=none] (16) at (0.75, -2.25) {};
		\node [style=none] (17) at (0, -2) {...};
		\node [style=qutrit Z phase] (18) at (0, -1) {$\gamma$ \nodepart{lower} $\delta$};
	\end{pgfonlayer}
	\begin{pgfonlayer}{edgelayer}
		\draw [bend right=15] (11.center) to (14);
		\draw [bend right=15] (14) to (12.center);
		\draw [bend left=15] (15.center) to (18);
		\draw [bend left=15] (18) to (16.center);
		\draw [style=hadamard adjoint adge, bend right=45] (14) to (18);
		\draw [style=hadamard adjoint adge, bend left=45] (14) to (18);
	\end{pgfonlayer}
\end{tikzpicture}
 \ = \  
		\begin{tikzpicture}
	\begin{pgfonlayer}{nodelayer}
		\node [style=none] (11) at (-0.75, 2.25) {};
		\node [style=none] (12) at (0.75, 2.25) {};
		\node [style=none] (13) at (0, 2) {...};
		\node [style=qutrit Z phase] (14) at (0, 1) {$\alpha$ \nodepart{lower} $\beta$};
		\node [style=none] (15) at (-0.75, -2.25) {};
		\node [style=none] (16) at (0.75, -2.25) {};
		\node [style=none] (17) at (0, -2) {...};
		\node [style=qutrit Z phase] (18) at (0, -1) {$\gamma$ \nodepart{lower} $\delta$};
	\end{pgfonlayer}
	\begin{pgfonlayer}{edgelayer}
		\draw [bend right=15] (11.center) to (14);
		\draw [bend right=15] (14) to (12.center);
		\draw [bend left=15] (15.center) to (18);
		\draw [bend left=15] (18) to (16.center);
		\draw [style=hadamard edge] (14) to (18);
	\end{pgfonlayer}
\end{tikzpicture}

	\end{equation}
}

\newcommand{\qutritPivotEqualityStatement}{
	Given $a \in \mathbb{Z}_3$ and a graph state $(G, W)$ containing connected nodes $i$ and $j$, define $N_{=}(i, j) \defeq \left\{x \in N(i) \cap N(j) \mid w_{x,i} = w_{x,j} \right\}$ and $N_{\neq}(i, j) \defeq \left\{x \in N(i) \cap N(j) \mid w_{x,i} \neq w_{x,j} \right\}$. Then the following equation relates $G$ and its proper $a$-pivot along $ij$:
	\begin{equation}
		\begin{tikzpicture}
	\begin{pgfonlayer}{nodelayer}
		\node [style=none] (0) at (0, 0) {=};
		\node [style=none] (1) at (1, 0.25) {};
		\node [style=none] (2) at (14, 0.25) {};
		\node [style=none] (3) at (7, -2) {};
		\node [style=none] (4) at (1, -0.25) {};
		\node [style=none] (5) at (1.25, 0.25) {};
		\node [style=none] (6) at (3.75, 0.25) {};
		\node [style=none] (10) at (7, -0.75) {${G \wedge_a ij}$};
		\node [style=none] (12) at (3.75, 2) {};
		\node [style=qutrit Z phase] (32) at (7.5, 1) {$a$ \nodepart{lower} $a$};
		\node [style=none] (33) at (6.25, 0.25) {};
		\node [style=none] (34) at (6.25, 2) {};
		\node [style=qutrit Z phase] (35) at (10, 1) {$a$ \nodepart{lower} $a$};
		\node [style=none] (36) at (8.75, 1) {...};
		\node [style={scale=0.7}] (38) at (5, 2.5) {$\overbrace{\hspace{60pt}}^{N_{=}(i, j)}$};
		\node [style={scale=0.7}] (39) at (2.5, 2.5) {$\overbrace{\quad}^{j}$};
		\node [style=none] (62) at (11.25, 0.25) {};
		\node [style=none] (63) at (11.25, 2) {};
		\node [style=none] (65) at (13.75, 0.25) {};
		\node [style=none] (66) at (13.75, 2) {};
		\node [style=none] (68) at (12.5, 1) {...};
		\node [style=none] (79) at (2.5, 0.25) {};
		\node [style=none] (80) at (2.5, 2) {};
		\node [style={scale=0.7}] (82) at (1.25, 2.5) {$\overbrace{\quad}^{i}$};
		\node [style=qutrit hadamard unknown] (109) at (1.25, 1) {$-a$};
		\node [style=qutrit hadamard unknown] (110) at (2.5, 1) {$a$};
		\node [style=none] (111) at (7.5, 0.25) {};
		\node [style=none] (112) at (7.5, 2) {};
		\node [style=qutrit Z phase] (113) at (3.75, 1) {$-a$ \nodepart{lower} $-a$};
		\node [style=none] (114) at (10, 0.25) {};
		\node [style=none] (115) at (10, 2) {};
		\node [style=qutrit Z phase] (116) at (6.25, 1) {$-a$ \nodepart{lower} $-a$};
		\node [style=none] (117) at (5, 1) {...};
		\node [style={scale=0.7}] (118) at (8.75, 2.5) {$\overbrace{\hspace{60pt}}^{N_{\neq}(i, j)}$};
		\node [style=none] (119) at (-6.25, 0.25) {};
		\node [style=none] (120) at (-0.75, 0.25) {};
		\node [style=none] (121) at (-3.75, -2) {};
		\node [style=none] (122) at (-6.25, -0.25) {};
		\node [style=none] (123) at (-6, 0.25) {};
		\node [style=none] (125) at (-3.75, -0.75) {$G$};
		\node [style=none] (126) at (-6, 0.75) {};
		\node [style=none] (137) at (-1, 0.25) {};
		\node [style=none] (138) at (-1, 0.75) {};
		\node [style=none] (139) at (-3.75, 0.5) {...};
		\node [style=none] (155) at (1.25, 2) {};
	\end{pgfonlayer}
	\begin{pgfonlayer}{edgelayer}
		\draw (1.center) to (4.center);
		\draw (4.center) to (3.center);
		\draw (3.center) to (2.center);
		\draw (1.center) to (2.center);
		\draw (6.center) to (12.center);
		\draw (33.center) to (34.center);
		\draw (62.center) to (63.center);
		\draw (65.center) to (66.center);
		\draw (79.center) to (80.center);
		\draw (111.center) to (112.center);
		\draw (114.center) to (115.center);
		\draw (119.center) to (122.center);
		\draw (122.center) to (121.center);
		\draw (121.center) to (120.center);
		\draw (119.center) to (120.center);
		\draw (123.center) to (126.center);
		\draw (137.center) to (138.center);
		\draw (5.center) to (155.center);
	\end{pgfonlayer}
\end{tikzpicture}

	\end{equation}
}

\title{Simplification Strategies for the Qutrit ZX-Calculus}

\author{Alex Townsend-Teague $^{1, 3}$ and Konstantinos Meichanetzidis $^{1,2}$
	\institute{
		$^1$ Department of Computer Science, Oxford OX1 3QD, University of Oxford, UK \\ 
		$^2$ Quantinuum, 17 Beaumont St. Oxford OX1 2NA, UK\\
		$^3$ Dahlem Center for Complex Quantum Systems, Freie Universit\"{a}t Berlin, 14195 Berlin, Germany}
	}
\def\titlerunning{Simplification Strategies for the Qutrit ZX-Calculus}
\def\authorrunning{Alex Townsend-Teague and Konstantinos Meichanetzidis}

\begin{document}
\maketitle
\begin{abstract}
	The ZX-calculus is a graphical language for suitably represented tensor networks, called ZX-diagrams.
	Calculations are performed by transforming ZX-diagrams with rewrite rules.
	The ZX-calculus has found applications in reasoning about quantum circuits, condensed matter systems, quantum algorithms, quantum error correcting codes, and counting problems.
	A key notion is the stabiliser fragment of the ZX-calculus, a subfamily of ZX-diagrams for which rewriting can be done efficiently in terms of derived simplifying rewrites.
	Recently, higher dimensional qudits - in particular, qutrits - have gained prominence within quantum computing research.
	The main contribution of this work is the derivation of
	efficient rewrite strategies for the stabiliser fragment of the qutrit ZX-calculus.
	Notably, this constitutes a first non-trivial step towards the simplification of qutrit quantum circuits. We then give further unexpected areas in which these rewrite strategies provide complexity-theoretic insight; namely, we reinterpret known results about evaluating the Jones polynomial, an important link invariant in knot theory, and counting graph colourings.
\end{abstract}

\section{Introduction}

The ZX-calculus has its origins in
quantum foundations \cite{Coecke2011,vandewetering2020zxcalculus}.
It is a graphical language that allows for reasoning about ZX-diagrams,
which are interpreted as tensor networks over
a (semi)ring \cite{wang2020completeness}.
The calculus
is defined by a set of rewrite rules
that transform the diagrams.
Note that the set of rewrite rules depends on
the (semi)ring over which they are interpreted.
Importantly, ZX rewrites are sound and complete.

In the context of quantum computing,
the Gottesman-Knill theorem states that stabilizer quantum
circuits can be
simulated efficiently on classical computers \cite{Aaronson2004},
where by simulation here we mean the exact computation of quantum probability amplitudes.
An alternative
proof of the Gottesman-Knill theorem for qubits has been obtained graphically in the qubit ZX-calculus.
Stabilizer circuits can be expressed by the \emph{stabilizer
fragment},
and stabilizer ZX-diagrams can be rewritten efficiently
using derived rewrite strategies \cite{graph_theoretic_simplification}.
Moreover, these rewrites can be viewed as theorems
that are useful for simplifying or simulating universal quantum circuits, not just stabilizer circuits.
The key simplifying rewrites
for qubit stabilizer ZX-diagrams
are called \emph{local complementation} and \emph{pivot},
the latter being a composition of three of the former \cite{graph_theoretic_simplification, Duncan_2014, Van_den_Nest_2005}.
In this work, we recall the qutrit version of local complementation \cite{harny_completeness} and derive the corresponding pivot rule.
We then derive simplification strategies
which allow for the efficient rewriting of qutrit stabilizer ZX-diagrams,
the core result of this work.

Interestingly, though the motivation for studying ZX diagrams stems from quantum computation, they have broader applicability as they
can express arbitrary linear maps;
every quantum circuit is a ZX-diagram but not vice versa.
A plethora of hard problems in physics and computer science
regard interacting many-body multi-state systems - both classical and quantum -
and reduce to exactly computing a single scalar.
These range from quantum amplitudes,
partition functions in classical statistical mechanics,
counting problems, probabilistic inference, and many more.
Problem instances can be encoded as closed
tensor networks over an appropriate (semi)ring.
The scalar can be evaluated by full tensor contraction,
which in general is \#P-hard \cite{Damm2002}.

If a scalar of interest is expressed as a closed ZX-diagram,
one can rewrite the diagram by applying rewrite rules; one's goal is then to perform full diagram simplification in order to compute the desired scalar.
Again, simplifying arbitrary closed ZX-diagrams is a hard problem.
Note that
given an arbitrary tensor network, one can attempt to invent rewrite rules by inspecting the contents of the tensors and performing linear-algebra operations \cite{gray2020hyperoptimized}.


In the spirit of \cite{debeaudrap2020tensor},
where the ZH-calculus \cite{backens2018zh}, a cousin of ZX, was employed to rederive known complexity results about counting problems, we treat the ZX-calculus as a library of rewrite rules, which we import and use to reason about the complexity of problem families of interest.
Here we present two case studies: evaluating the Jones polynomial at lattice roots of unity, and graph colouring.
Both of these problem families reduce to evaluating
closed tensor networks and show a transition in complexity
at a particular dimension, below which the tensor network corresponds to a stabilizer ZX-diagram.

We underline that throughout this work, all rewrites are valid
up to a scalar, but keeping track of multiplicative scalar factors that arise under rewriting can be done efficiently.

\section{Simplifying Qubit ZX-Diagrams}



Qubit ZX-diagrams are generated by \emph{spiders}, whose \emph{legs}
or \emph{wires} carry vector spaces of dimension $d=2$.
Diagrams are read bottom-to-top; bottom open wires (not connected to anything) are \emph{input wires} and top ones are \emph{outputs}.
Diagrams with only outputs are called \emph{states} and those with only inputs are called \emph{effects}.
A \emph{closed} diagram is one with no inputs nor outputs and represents a scalar.

Spiders can be \emph{composed};
placing diagrams side by side represents parallel composition ($\otimes$),
with concrete interpretation the tensor product.
Vertically stacking diagrams corresponds to sequential composition ($\circ$) and concretely means matrix multiplication.
Specifically, it means tensor contraction along spider legs; the common tensor indices represented by these wires are summed over.
The concrete representation of these operations of course depends on the (semi)ring over which the spiders are interpreted as tensors.
For spiders $S$ and $S'$ we write:
$\left\llbracket S \otimes S' \right\rrbracket = \left\llbracket S \right\rrbracket \otimes \left\llbracket S' \right\rrbracket ~, ~~
	\left\llbracket S \circ S' \right\rrbracket = \left\llbracket S \right\rrbracket \cdot \left\llbracket S' \right\rrbracket $.


Spiders come in two species: green $Z$-spiders and red $X$-spiders, decorated by a \emph{phase} $\alpha\in[0,2\pi)$. When $\alpha=0$, we will omit it.
The \emph{standard representation} of the spider generators as tensors over $\mathbb{C}$ is:
\begin{equation}\label{eq:qubit_standard_interpretation}
	\left\llbracket~ \scalebox{0.75}{\begin{tikzpicture}
	\begin{pgfonlayer}{nodelayer}
		\node [style=none] (1) at (0, 1) {...};
		\node [style=none] (5) at (-1, 1.5) {};
		\node [style=none] (7) at (1, 1.5) {};
		\node [style=none] (8) at (0, -1) {...};
		\node [style=none] (9) at (-1, -1.5) {};
		\node [style=Z phase dot] (10) at (0, 0) {$\alpha$};
		\node [style=none] (11) at (1, -1.5) {};
		\node [style=none] (13) at (0, 2.25) {$\overbrace{\hspace{30pt}}^{\tiny{m}}$};
		\node [style=none] (14) at (0, -2.25) {$\underbrace{\hspace{30pt}}_{\tiny{n}}$};
	\end{pgfonlayer}
	\begin{pgfonlayer}{edgelayer}
		\draw [bend left, looseness=0.75] (10) to (11.center);
		\draw [bend right, looseness=0.75] (10) to (9.center);
		\draw [bend right, looseness=0.75] (10) to (7.center);
		\draw [bend left, looseness=0.75] (10) to (5.center);
	\end{pgfonlayer}
\end{tikzpicture}
} ~\right\rrbracket = 
	\ket{0}^{m}\bra{0}^{n} + 
	e^{i\alpha}\ket{1}^{m}\bra{1}^{n} ,
	~~
	\left\llbracket~ \scalebox{0.75}{\begin{tikzpicture}
	\begin{pgfonlayer}{nodelayer}
		\node [style=none] (1) at (0, 1) {...};
		\node [style=none] (5) at (-1, 1.5) {};
		\node [style=none] (7) at (1, 1.5) {};
		\node [style=none] (8) at (0, -1) {...};
		\node [style=none] (9) at (-1, -1.5) {};
		\node [style=X phase dot] (10) at (0, 0) {$\alpha$};
		\node [style=none] (11) at (1, -1.5) {};
		\node [style=none] (13) at (0, 2.25) {$\overbrace{\hspace{30pt}}^{\tiny{m}}$};
		\node [style=none] (14) at (0, -2.25) {$\underbrace{\hspace{30pt}}_{\tiny{n}}$};
	\end{pgfonlayer}
	\begin{pgfonlayer}{edgelayer}
		\draw [bend left, looseness=0.75] (10) to (11.center);
		\draw [bend right, looseness=0.75] (10) to (9.center);
		\draw [bend right, looseness=0.75] (10) to (7.center);
		\draw [bend left, looseness=0.75] (10) to (5.center);
	\end{pgfonlayer}
\end{tikzpicture}
} ~\right\rrbracket = 
	\ket{+}^{m}\bra{+}^{n} + 
	e^{i\alpha}\ket{-}^{m}\bra{-}^{n}
\end{equation}
where $\{\ket{0}$, $\ket{1}\}$ is the \emph{$Z$-basis} and
$\ket{\pm}=\ket{0}\pm\ket{1}$ the \emph{$X$-basis} in $\mathbb{C}^2$, in Dirac notation.
The Hadamard gate $H$, whose function is to switch between the $Z$ and $X$ bases, is denoted as a yellow box.
Often we will instead draw a dashed blue line to represent a \emph{Hadamard edge}:
\vspace{-5pt}
\begin{equation}
	\begin{tikzpicture}
	\begin{pgfonlayer}{nodelayer}
		\node [style=none] (0) at (0, -1) {};
		\node [style=none] (1) at (0, 1) {};
		\node [style=hadamard] (6) at (0, 0) {};
	\end{pgfonlayer}
	\begin{pgfonlayer}{edgelayer}
		\draw (0.center) to (1.center);
	\end{pgfonlayer}
\end{tikzpicture}
 \quad = \quad
	\begin{tikzpicture}
	\begin{pgfonlayer}{nodelayer}
		\node [style=none] (0) at (0, -1) {};
		\node [style=none] (1) at (0, 1) {};
	\end{pgfonlayer}
	\begin{pgfonlayer}{edgelayer}
		\draw [style=hadamard edge] (0.center) to (1.center);
	\end{pgfonlayer}
\end{tikzpicture}
 \quad = \quad
	\begin{tikzpicture}
	\begin{pgfonlayer}{nodelayer}
		\node [style=none] (0) at (0, -1.75) {};
		\node [style=none] (1) at (0, 1.75) {};
		\node [style=Z phase dot] (3) at (0, 1) {$\frac{\pi}{2}$};
		\node [style=Z phase dot] (5) at (0, -1) {$\frac{\pi}{2}$};
		\node [style=X phase dot] (6) at (0, 0) {$\frac{\pi}{2}$};
	\end{pgfonlayer}
	\begin{pgfonlayer}{edgelayer}
		\draw (0.center) to (1.center);
	\end{pgfonlayer}
\end{tikzpicture}
 ~~~,
	\qquad 
	\left\llbracket ~~~ \begin{tikzpicture}
	\begin{pgfonlayer}{nodelayer}
		\node [style=none] (0) at (0, -1) {};
		\node [style=none] (1) at (0, 1) {};
		\node [style=hadamard] (6) at (0, 0) {};
	\end{pgfonlayer}
	\begin{pgfonlayer}{edgelayer}
		\draw (0.center) to (1.center);
	\end{pgfonlayer}
\end{tikzpicture}
 ~~~ \right\rrbracket \simeq 
	\ket{0}\bra{0}+\ket{0}\bra{1}+\ket{1}\bra{0}-\ket{1}\bra{1}
\end{equation}
The ZX-calculus is \emph{universal} for multilinear maps;
any tensor with entries in $\mathbb{C}$
has a corresponding ZX-diagram.
The rewrite rules of the ZX-calculus (see Fig.\ref{fig:qubit_ZX_rules} in Appendix \ref{app:qubit_zx_calculus}) allow manipulation of the diagrams by \emph{rewrites}. Importantly, up to a scalar, the rewrites are \emph{sound}; that is, they preserve the concrete tensor representation over $\mathbb{C}$.
The ZX-calculus is also \emph{complete} in the sense that any true equation between tensors can be proven only in terms of rewrites.
An incredibly useful feature of the qubit ZX-diagrams
is that \emph{only topology matters};
the concrete tensor semantics of the diagram are invariant under
deformations of the network as long as the inter-spider connectivity is respected.

The \emph{stabilizer fragment} of the calculus consists of all diagrams in which all phases are $\alpha=\frac{\pi n}{2}$, $n\in\mathbb{Z}$.
In \cite[Theorem 5.4]{graph_theoretic_simplification} the authors give an efficient algorithm for simplifying any qubit ZX-diagram (see Appendix \ref{app:qubit_zx_calculus}).
The algorithm consists of consecutive applications of
spider-eliminating rewrites.
In particular,
this algorithm will efficiently simplify any closed stabilizer diagram until it contains at most one spider, at which point the scalar it represents can be easily read off. 
By `efficiently' we mean via a sequence of spider elimination rewrites whose cost is polynomial in the initial number of spiders and their legs. Note each such rewrite updates only a polynomial number of edges in the diagram, which prevents an overwhelming memory cost of the simplification procedure.
\section{Simplifying Qutrit ZX-Diagrams}\label{sec:qutrits}

We now turn to the qutrit ZX-calculus and examine the analogous story to that of the previous section,
but for the case where the dimension of the vector space carried by the wires is $d=3$.

\begin{figure}
	\begin{tcolorbox}[colback=white]
		\begin{equation*}
			\input{qutrit_rules/fusion/all.tikz} \quad \hypertarget{qutrit_rule_fusion}{\mathbf{(f)}}
		\end{equation*}
		\begin{equation*}
			\input{qutrit_rules/identity/lhs.tikz} \quad = \quad 
			\input{qutrit_rules/identity/rhs.tikz} \quad \hypertarget{qutrit_rule_id}{\mathbf{(id)}}
			\hspace{60pt}
			\input{qutrit_rules/twisted_cup/lhs.tikz} \quad = \quad 
			\input{qutrit_rules/twisted_cup/rhs.tikz} \quad \hypertarget{qutrit_rule_twisted_cup}{\mathbf{(t)}}
		\end{equation*}
		\vspace{5pt}
		\begin{equation*}
			\input{qutrit_rules/0_copy/lhs.tikz} \quad = \quad 
			\input{qutrit_rules/0_copy/rhs.tikz} \quad \hypertarget{qutrit_rule_0_copy}{\mathbf{(cp_0)}}
			\hspace{60pt}
			\input{qutrit_rules/bialgebra/lhs.tikz} \quad = \quad 
			\input{qutrit_rules/bialgebra/rhs.tikz} \quad \hypertarget{qutrit_rule_bialgebra}{\mathbf{(b)}}
		\end{equation*}
		\vspace{5pt}
		\begin{equation*}
			\input{qutrit_rules/m_copy/1_2_lhs.tikz} \quad = \quad 
			\input{qutrit_rules/m_copy/1_2_rhs.tikz} \quad \hypertarget{qutrit_rule_m_copy}{\mathbf{(cp_{\mathcal{M}})}}
			\hspace{60pt}
			\input{qutrit_rules/m_copy/2_1_lhs.tikz} \quad = \quad 
			\input{qutrit_rules/m_copy/2_1_rhs.tikz} \quad \mathbf{(cp_{\mathcal{M}})}
		\end{equation*}
		\vspace{5pt}
		\begin{equation*}
			\input{qutrit_rules/commute/1_2_lhs.tikz} \quad = \quad 
			\input{qutrit_rules/commute/1_2_rhs.tikz} \quad \hypertarget{qutrit_rule_commute}{\mathbf{(cm)}}
			\hspace{60pt}
			\input{qutrit_rules/commute/2_1_lhs.tikz} \quad = \quad 
			\input{qutrit_rules/commute/2_1_rhs.tikz} \quad {\mathbf{(cm)}}
		\end{equation*}
		\begin{equation*}
			\input{qutrit_rules/hadamard/h_hdagger.tikz} \quad = \quad 
			\input{qutrit_rules/hadamard/identity.tikz} \quad = \quad 
			\input{qutrit_rules/hadamard/hdagger_h.tikz} \quad \hypertarget{qutrit_rule_hadamard}{\mathbf{(H)}}
			\hspace{60pt}
			\input{qutrit_rules/hadamard/euler/h.tikz} \quad = \quad 
			\input{qutrit_rules/hadamard/euler/decomposition.tikz} \quad \hypertarget{qutrit_rule_euler}{\mathbf{(E)}}
		\end{equation*}
		\begin{equation*}
			\input{qutrit_rules/colour_change/lhs.tikz} \quad = \quad 
			\input{qutrit_rules/colour_change/rhs.tikz} \quad \hypertarget{qutrit_rule_colour_change}{\mathbf{(cc)}}
			\hspace{60pt}
			\input{qutrit_rules/colour_change/flip_lhs.tikz} \quad = \quad 
			\input{qutrit_rules/colour_change/flip_rhs.tikz} \quad \mathbf{(cc)}
		\end{equation*}
		\vspace{5pt}
		\begin{equation*}
			\input{qutrit_rules/snake/snake.tikz} \quad = \quad 
			\input{qutrit_rules/snake/hopf_ish.tikz} \quad = \quad 
			\input{qutrit_rules/snake/hadamards.tikz} \quad = \quad 
			\input{qutrit_rules/snake/hadamard_adjoints.tikz} \quad \hypertarget{qutrit_rule_snake}{\mathbf{(s)}}
		\end{equation*}
	\end{tcolorbox}
	\vspace{5pt}
	\caption{Rewrite rules for the qutrit ZX-calculus where spiders are interpreted as tensors over $\mathbb{C}$.}
	\label{fig:qutrit_ZX_rules}
\end{figure}

Qutrit spiders again come in two species,
$Z$ (green) and $X$ (red),
with the three-dimensional \emph{Z-basis} $\{\ket{0},\ket{1},\ket{2}\}$.
Let $\omega = e^{i \frac{2\pi}{3}}$ denote the third root of unity
with $\bar\omega = \omega^2$ its complex conjugate.
The qutrit \emph{$X$-basis} is:
$\{ \ket{+} = \frac{1}{\sqrt{3}} \left(\ket{0} + \ket{1} + \ket{2}\right)$,
$\ket{\omega} = \frac{1}{\sqrt{3}} \left(\ket{0} + \omega\ket{1} + \bar{\omega}\ket{2}\right)$,
$\ket{\bar{\omega}} = \frac{1}{\sqrt{3}} \left(\ket{0} + \bar{\omega}\ket{1} + \omega\ket{2}\right) \} $.
Spiders now carry two phases $\alpha$ and $\beta$,
and have the following \emph{standard representation} as linear maps:
\begingroup
	\allowdisplaybreaks
	\setlength{\jot}{5pt}
		\begin{align}
			&\left\llbracket \quad \input{qutrit_generators/spiders/Z_a_b_labelled.tikz} \quad \right\rrbracket = 
			\ket{0}^{\otimes m}\bra{0}^{\otimes n} + 
			e^{i\alpha}\ket{1}^{\otimes m}\bra{1}^{\otimes n} + 
			e^{i\beta}\ket{2}^{\otimes m}\bra{2}^{\otimes n} \\
			&\left\llbracket \quad \input{qutrit_generators/spiders/X_a_b_labelled.tikz} \quad \right\rrbracket = 
			\ket{+}^{\otimes m}\bra{+}^{\otimes n} + 
			e^{i\alpha}\ket{\omega}^{\otimes m}\bra{\omega}^{\otimes n} + 
			e^{i\beta}\ket{\bar{\omega}}^{\otimes m}\bra{\bar{\omega}}^{\otimes n}
		\end{align}
\endgroup
When $\alpha = \beta = 0$ we will again omit the angles entirely, and just draw a small green or red dot. Throughout our work, whenever we use an integer $n$ as a spider decoration, this is a shorthand for $\frac{2\pi}{3}n$. Since spider phases hold mod $2\pi$, these integer decorations hold mod $3$. Unless otherwise stated, we will use Greek letters to denote general angles, and Roman letters for these integer shorthands.

Hadamard gates are no longer self-adjoint, so we change our notation: we let a yellow box decorated with a $1$ (mod $3$) denote a Hadamard gate, while decorating with a $2$ (mod $3$) denotes its adjoint. We will shortly explain this choice. We also use a dashed blue line for the \emph{Hadamard edge} (\emph{$H$-edge}) and a purple dashed line for its adjoint (\emph{$H^\dagger$-edge}). Those familar with the ZH-calculus should note that this notation is \emph{not} analogous to the $H$-boxes therein.  
\begin{equation}\label{eq:qutrit_dashed_lines}
		\begin{tikzpicture}
	\begin{pgfonlayer}{nodelayer}
		\node [style=qutrit hadamard unknown] (1) at (0, 0) {1};
		\node [style=none] (2) at (0, 1) {};
		\node [style=none] (3) at (0, -1) {};
	\end{pgfonlayer}
	\begin{pgfonlayer}{edgelayer}
		\draw (2.center) to (3.center);
	\end{pgfonlayer}
\end{tikzpicture}
 \ = \ 
		\begin{tikzpicture}
	\begin{pgfonlayer}{nodelayer}
		\node [style=none] (1) at (0, 1) {};
		\node [style=none] (2) at (0, -1) {};
	\end{pgfonlayer}
	\begin{pgfonlayer}{edgelayer}
		\draw [style=hadamard edge] (1.center) to (2.center);
	\end{pgfonlayer}
\end{tikzpicture}
 \ = \ 
		\begin{tikzpicture}
	\begin{pgfonlayer}{nodelayer}
		\node [style=none] (3) at (0, 2) {};
		\node [style=none] (4) at (0, -2) {};
		\node [style=qutrit X phase] (5) at (0, 0) {$2$ \nodepart{lower} $2$};
		\node [style=qutrit Z phase] (6) at (0, 1.25) {$2$ \nodepart{lower} $2$};
		\node [style=qutrit Z phase] (7) at (0, -1.25) {$2$ \nodepart{lower} $2$};
	\end{pgfonlayer}
	\begin{pgfonlayer}{edgelayer}
		\draw (3.center) to (4.center);
	\end{pgfonlayer}
\end{tikzpicture}
 ~~ , 
		\hspace{50pt}
		\begin{tikzpicture}
	\begin{pgfonlayer}{nodelayer}
		\node [style=qutrit hadamard unknown] (1) at (0, 0) {2};
		\node [style=none] (2) at (0, 1) {};
		\node [style=none] (3) at (0, -1) {};
	\end{pgfonlayer}
	\begin{pgfonlayer}{edgelayer}
		\draw (2.center) to (3.center);
	\end{pgfonlayer}
\end{tikzpicture}
 \ = \ 
		\begin{tikzpicture}
	\begin{pgfonlayer}{nodelayer}
		\node [style=none] (1) at (0, 1) {};
		\node [style=none] (2) at (0, -1) {};
	\end{pgfonlayer}
	\begin{pgfonlayer}{edgelayer}
		\draw [style=hadamard adjoint adge] (1.center) to (2.center);
	\end{pgfonlayer}
\end{tikzpicture}
 \ = \ 
		\begin{tikzpicture}
	\begin{pgfonlayer}{nodelayer}
		\node [style=none] (3) at (0, 2) {};
		\node [style=none] (4) at (0, -2) {};
		\node [style=qutrit X phase] (5) at (0, 0) {$1$ \nodepart{lower} $1$};
		\node [style=qutrit Z phase] (6) at (0, 1.25) {$1$ \nodepart{lower} $1$};
		\node [style=qutrit Z phase] (7) at (0, -1.25) {$1$ \nodepart{lower} $1$};
	\end{pgfonlayer}
	\begin{pgfonlayer}{edgelayer}
		\draw (3.center) to (4.center);
	\end{pgfonlayer}
\end{tikzpicture}
 ~~ ,
		\hspace{50pt}
		\begin{tikzpicture}
	\begin{pgfonlayer}{nodelayer}
		\node [style=qutrit hadamard unknown] (1) at (0, 0) {0};
		\node [style=none] (2) at (0, 1) {};
		\node [style=none] (3) at (0, -1) {};
	\end{pgfonlayer}
	\begin{pgfonlayer}{edgelayer}
		\draw (2.center) to (3.center);
	\end{pgfonlayer}
\end{tikzpicture}
 \ = \ 
		\begin{tikzpicture}
	\begin{pgfonlayer}{nodelayer}
		\node [style=Z dot] (5) at (0, 0.5) {};
		\node [style=Z dot] (6) at (0, -0.5) {};
		\node [style=none] (7) at (0, -1) {};
		\node [style=none] (8) at (0, 1) {};
	\end{pgfonlayer}
	\begin{pgfonlayer}{edgelayer}
		\draw (8.center) to (5);
		\draw (6) to (7.center);
	\end{pgfonlayer}
\end{tikzpicture}

\end{equation}
The last equation above says that in a graph-like diagram (which we will define shortly), a Hadamard edge decorated by a $0$ is in fact not an edge at all, thanks to spider fusion. 
A very important difference from the qubit case is that in qutrit ZX-calculus there is no \emph{plain} cap or cup:
\begin{equation}
	\begin{tikzpicture}
	\begin{pgfonlayer}{nodelayer}
		\node [style=Z dot] (0) at (0, -0.5) {};
		\node [style=none] (1) at (-1, 0.5) {};
		\node [style=none] (2) at (1, 0.5) {};
	\end{pgfonlayer}
	\begin{pgfonlayer}{edgelayer}
		\draw [bend right=90, looseness=1.75] (1.center) to (2.center);
	\end{pgfonlayer}
\end{tikzpicture}
 \quad \neq \quad \begin{tikzpicture}
	\begin{pgfonlayer}{nodelayer}
		\node [style=X dot] (0) at (0, -0.5) {};
		\node [style=none] (1) at (-1, 0.5) {};
		\node [style=none] (2) at (1, 0.5) {};
	\end{pgfonlayer}
	\begin{pgfonlayer}{edgelayer}
		\draw [bend right=90, looseness=1.75] (1.center) to (2.center);
	\end{pgfonlayer}
\end{tikzpicture}
 \quad , \hspace{50pt}
	\begin{tikzpicture}
	\begin{pgfonlayer}{nodelayer}
		\node [style=Z dot] (0) at (0, 0.5) {};
		\node [style=none] (1) at (-1, -0.5) {};
		\node [style=none] (2) at (1, -0.5) {};
	\end{pgfonlayer}
	\begin{pgfonlayer}{edgelayer}
		\draw [bend left=90, looseness=1.75] (1.center) to (2.center);
	\end{pgfonlayer}
\end{tikzpicture}
 \quad \neq \quad \begin{tikzpicture}
	\begin{pgfonlayer}{nodelayer}
		\node [style=X dot] (0) at (0, 0.5) {};
		\node [style=none] (1) at (-1, -0.5) {};
		\node [style=none] (2) at (1, -0.5) {};
	\end{pgfonlayer}
	\begin{pgfonlayer}{edgelayer}
		\draw [bend left=90, looseness=1.75] (1.center) to (2.center);
	\end{pgfonlayer}
\end{tikzpicture}

\end{equation}
This has several consequences. Firstly, the maxim that `only topology matters' no longer applies. That is, it is now important to make clear the distinction between a spider's input and output wires, unlike in the qubit case where we could freely interchange the two.
This gives the qutrit calculus a slightly more rigid flavour than its qubit counterpart. 
That said, this rigidity can be loosened; in particular, this distinction is irrelevant for $H$- and $H^\dagger$-edges \cite{qutrit_euler}:
	\begin{equation}
		\scalebox{0.8}{\begin{tikzpicture}
	\begin{pgfonlayer}{nodelayer}
		\node [style=none] (0) at (-1.75, 1.75) {...};
		\node [style=none] (1) at (-2.5, 2) {};
		\node [style=none] (2) at (-1, 2) {};
		\node [style=none] (3) at (-1.75, -1.75) {...};
		\node [style=none] (4) at (-2.5, -2) {};
		\node [style=qutrit Z phase] (5) at (-1.75, -0.75) {$\alpha$ \nodepart{lower} $\beta$};
		\node [style=none] (6) at (-1, -2) {};
		\node [style=none] (7) at (1.75, 1.75) {...};
		\node [style=none] (8) at (1, 2) {};
		\node [style=none] (9) at (2.5, 2) {};
		\node [style=none] (10) at (1.75, -1.75) {...};
		\node [style=none] (11) at (1, -2) {};
		\node [style=qutrit Z phase] (12) at (1.75, 0.75) {$\gamma$ \nodepart{lower} $\delta$};
		\node [style=none] (13) at (2.5, -2) {};
	\end{pgfonlayer}
	\begin{pgfonlayer}{edgelayer}
		\draw [bend left=15] (5) to (6.center);
		\draw [bend right=15] (5) to (4.center);
		\draw [bend right=15] (5) to (2.center);
		\draw [bend left=15] (5) to (1.center);
		\draw [bend left=15] (12) to (13.center);
		\draw [bend right=15] (12) to (11.center);
		\draw [bend right=15] (12) to (9.center);
		\draw [bend left=15] (12) to (8.center);
		\draw [style=hadamard edge] (5) to (12);
	\end{pgfonlayer}
\end{tikzpicture}
} = 
		\scalebox{0.8}{\begin{tikzpicture}
	\begin{pgfonlayer}{nodelayer}
		\node [style=none] (16) at (-1.75, 1.75) {...};
		\node [style=none] (17) at (-2.5, 2) {};
		\node [style=none] (18) at (-1, 2) {};
		\node [style=none] (19) at (-1.75, -1.75) {...};
		\node [style=none] (20) at (-2.5, -2) {};
		\node [style=qutrit Z phase] (21) at (-1.75, 0.75) {$\alpha$ \nodepart{lower} $\beta$};
		\node [style=none] (22) at (-1, -2) {};
		\node [style=none] (25) at (1.75, 1.75) {...};
		\node [style=none] (26) at (1, 2) {};
		\node [style=none] (27) at (2.5, 2) {};
		\node [style=none] (28) at (1.75, -1.75) {...};
		\node [style=none] (29) at (1, -2) {};
		\node [style=qutrit Z phase] (30) at (1.75, -0.75) {$\gamma$ \nodepart{lower} $\delta$};
		\node [style=none] (31) at (2.5, -2) {};
	\end{pgfonlayer}
	\begin{pgfonlayer}{edgelayer}
		\draw [bend left=15] (21) to (22.center);
		\draw [bend right=15] (21) to (20.center);
		\draw [bend right=15] (21) to (18.center);
		\draw [bend left=15] (21) to (17.center);
		\draw [bend left=15] (30) to (31.center);
		\draw [bend right=15] (30) to (29.center);
		\draw [bend right=15] (30) to (27.center);
		\draw [bend left=15] (30) to (26.center);
		\draw [style=hadamard edge] (21) to (30);
	\end{pgfonlayer}
\end{tikzpicture}
} ~,
		\hspace{50pt}
		\scalebox{0.8}{\begin{tikzpicture}
	\begin{pgfonlayer}{nodelayer}
		\node [style=none] (0) at (-1.75, 1.75) {...};
		\node [style=none] (1) at (-2.5, 2) {};
		\node [style=none] (2) at (-1, 2) {};
		\node [style=none] (3) at (-1.75, -1.75) {...};
		\node [style=none] (4) at (-2.5, -2) {};
		\node [style=qutrit Z phase] (5) at (-1.75, -0.75) {$\alpha$ \nodepart{lower} $\beta$};
		\node [style=none] (6) at (-1, -2) {};
		\node [style=none] (7) at (1.75, 1.75) {...};
		\node [style=none] (8) at (1, 2) {};
		\node [style=none] (9) at (2.5, 2) {};
		\node [style=none] (10) at (1.75, -1.75) {...};
		\node [style=none] (11) at (1, -2) {};
		\node [style=qutrit Z phase] (12) at (1.75, 0.75) {$\gamma$ \nodepart{lower} $\delta$};
		\node [style=none] (13) at (2.5, -2) {};
	\end{pgfonlayer}
	\begin{pgfonlayer}{edgelayer}
		\draw [bend left=15] (5) to (6.center);
		\draw [bend right=15] (5) to (4.center);
		\draw [bend right=15] (5) to (2.center);
		\draw [bend left=15] (5) to (1.center);
		\draw [bend left=15] (12) to (13.center);
		\draw [bend right=15] (12) to (11.center);
		\draw [bend right=15] (12) to (9.center);
		\draw [bend left=15] (12) to (8.center);
		\draw [style=hadamard adjoint adge] (5) to (12);
	\end{pgfonlayer}
\end{tikzpicture}
} = 
		\scalebox{0.8}{\begin{tikzpicture}
	\begin{pgfonlayer}{nodelayer}
		\node [style=none] (16) at (-1.75, 1.75) {...};
		\node [style=none] (17) at (-2.5, 2) {};
		\node [style=none] (18) at (-1, 2) {};
		\node [style=none] (19) at (-1.75, -1.75) {...};
		\node [style=none] (20) at (-2.5, -2) {};
		\node [style=qutrit Z phase] (21) at (-1.75, 0.75) {$\alpha$ \nodepart{lower} $\beta$};
		\node [style=none] (22) at (-1, -2) {};
		\node [style=none] (25) at (1.75, 1.75) {...};
		\node [style=none] (26) at (1, 2) {};
		\node [style=none] (27) at (2.5, 2) {};
		\node [style=none] (28) at (1.75, -1.75) {...};
		\node [style=none] (29) at (1, -2) {};
		\node [style=qutrit Z phase] (30) at (1.75, -0.75) {$\gamma$ \nodepart{lower} $\delta$};
		\node [style=none] (31) at (2.5, -2) {};
	\end{pgfonlayer}
	\begin{pgfonlayer}{edgelayer}
		\draw [bend left=15] (21) to (22.center);
		\draw [bend right=15] (21) to (20.center);
		\draw [bend right=15] (21) to (18.center);
		\draw [bend left=15] (21) to (17.center);
		\draw [bend left=15] (30) to (31.center);
		\draw [bend right=15] (30) to (29.center);
		\draw [bend right=15] (30) to (27.center);
		\draw [bend left=15] (30) to (26.center);
		\draw [style=hadamard adjoint adge] (21) to (30);
	\end{pgfonlayer}
\end{tikzpicture}
}
	\end{equation}
The full set of rules of the qutrit ZX-calculus is shown in Figure \ref{fig:qutrit_ZX_rules}, and a more rigorous definition of the calculus as a whole is found in Appendix \ref{app:qutrit_zx_calculus}.

\subsection{Graph-Like Qutrit ZX Diagrams}


A \emph{graph-like} qutrit ZX-diagram is one where
every spider is green,
spiders are only connected by blue Hadamard edges (\emph{$H$-edges})
or their purple adjoints (\emph{$H^\dagger$-edges}),
every pair of spiders is connected by at most one $H$-edge or $H^\dagger$-edge,
every input and output is connected to a spider,
and every spider is connected to at most one input or output.
A graph-like qutrit ZX-diagram is a \emph{graph state} when every spider has zero phases and is connected to an output. 


Note the difference compared to the qubit case: we need not worry about self-loops beacuse the qutrit ZX-calculus doesn't define a `plain' cap or cup. But this comes at a cost: spiders in the qutrit case fuse more fussily. Specifically, when two spiders of the same colour are connected by at least one plain edge and at least one $H$- or $H^\dagger$-edge, fusion is not possible. Instead, should we want to ensure we have a graph-like diagram, we can replace the plain wire:
\begin{equation}\label{eq:spiders_reluctant_to_fuse}
	\begin{tikzpicture}
	\begin{pgfonlayer}{nodelayer}
		\node [style=none] (8) at (0, 1) {};
		\node [style=none] (9) at (0, -1) {};
	\end{pgfonlayer}
	\begin{pgfonlayer}{edgelayer}
		\draw (8.center) to (9.center);
	\end{pgfonlayer}
\end{tikzpicture}
 \quad \xeq{\qutritRuleHadamard} \quad
	\begin{tikzpicture}
	\begin{pgfonlayer}{nodelayer}
		\node [style=none] (3) at (0, 1.5) {};
		\node [style=none] (4) at (0, -1.5) {};
		\node [style=qutrit hadamard] (5) at (0, -0.75) {};
		\node [style=qutrit hadamard adjoint] (6) at (0, 0.75) {};
	\end{pgfonlayer}
	\begin{pgfonlayer}{edgelayer}
		\draw (3.center) to (4.center);
	\end{pgfonlayer}
\end{tikzpicture}
 \quad \xeq{\qutritRuleId} \quad
	\begin{tikzpicture}
	\begin{pgfonlayer}{nodelayer}
		\node [style=none] (3) at (0, 1.5) {};
		\node [style=none] (4) at (0, -1.5) {};
		\node [style=qutrit hadamard] (5) at (0, -0.75) {};
		\node [style=qutrit hadamard adjoint] (6) at (0, 0.75) {};
		\node [style=Z dot] (7) at (0, 0) {};
	\end{pgfonlayer}
	\begin{pgfonlayer}{edgelayer}
		\draw (3.center) to (4.center);
	\end{pgfonlayer}
\end{tikzpicture}
 \quad = \quad
	\begin{tikzpicture}
	\begin{pgfonlayer}{nodelayer}
		\node [style=none] (0) at (0, -1.5) {};
		\node [style=none] (1) at (0, 1.5) {};
		\node [style=Z dot] (2) at (0, 0) {};
	\end{pgfonlayer}
	\begin{pgfonlayer}{edgelayer}
		\draw [style=hadamard adjoint adge] (0.center) to (2);
		\draw [style=hadamard edge] (1.center) to (2);
	\end{pgfonlayer}
\end{tikzpicture}

\end{equation}
Indeed, we can show that every qutrit ZX-diagram is equivalent to a graph-like one. The following equations, derived in Appendix \ref{lem:h_edges_are_mod_3_appendix}, are vital to this:
\begin{equation}\label{eq:h_edges_are_mod_3}
	\scalebox{0.8}{\input{hadamard_lemmas/3_h_edges_vanish/blue.tikz}} \ = \ 
	\scalebox{0.8}{\input{hadamard_lemmas/3_h_edges_vanish/disconnected.tikz}} \ = \ 
	\scalebox{0.8}{\input{hadamard_lemmas/3_h_edges_vanish/purple.tikz}} \ ,
	\qquad
	\scalebox{0.8}{\begin{tikzpicture}
	\begin{pgfonlayer}{nodelayer}
		\node [style=none] (11) at (-0.75, 2.25) {};
		\node [style=none] (12) at (0.75, 2.25) {};
		\node [style=none] (13) at (0, 2) {...};
		\node [style=qutrit Z phase] (14) at (0, 1) {$\alpha$ \nodepart{lower} $\beta$};
		\node [style=none] (15) at (-0.75, -2.25) {};
		\node [style=none] (16) at (0.75, -2.25) {};
		\node [style=none] (17) at (0, -2) {...};
		\node [style=qutrit Z phase] (18) at (0, -1) {$\gamma$ \nodepart{lower} $\delta$};
	\end{pgfonlayer}
	\begin{pgfonlayer}{edgelayer}
		\draw [bend right=15] (11.center) to (14);
		\draw [bend right=15] (14) to (12.center);
		\draw [bend left=15] (15.center) to (18);
		\draw [bend left=15] (18) to (16.center);
		\draw [style=hadamard edge, bend right=45] (14) to (18);
		\draw [style=hadamard edge, bend left=45] (14) to (18);
	\end{pgfonlayer}
\end{tikzpicture}
} \ = \ 
	\scalebox{0.8}{\begin{tikzpicture}
	\begin{pgfonlayer}{nodelayer}
		\node [style=none] (11) at (-0.75, 2.25) {};
		\node [style=none] (12) at (0.75, 2.25) {};
		\node [style=none] (13) at (0, 2) {...};
		\node [style=qutrit Z phase] (14) at (0, 1) {$\alpha$ \nodepart{lower} $\beta$};
		\node [style=none] (15) at (-0.75, -2.25) {};
		\node [style=none] (16) at (0.75, -2.25) {};
		\node [style=none] (17) at (0, -2) {...};
		\node [style=qutrit Z phase] (18) at (0, -1) {$\gamma$ \nodepart{lower} $\delta$};
	\end{pgfonlayer}
	\begin{pgfonlayer}{edgelayer}
		\draw [bend right=15] (11.center) to (14);
		\draw [bend right=15] (14) to (12.center);
		\draw [bend left=15] (15.center) to (18);
		\draw [bend left=15] (18) to (16.center);
		\draw [style=hadamard adjoint adge] (14) to (18);
	\end{pgfonlayer}
\end{tikzpicture}
} \ ,
	\qquad
	\scalebox{0.8}{\begin{tikzpicture}
	\begin{pgfonlayer}{nodelayer}
		\node [style=none] (11) at (-0.75, 2.25) {};
		\node [style=none] (12) at (0.75, 2.25) {};
		\node [style=none] (13) at (0, 2) {...};
		\node [style=qutrit Z phase] (14) at (0, 1) {$\alpha$ \nodepart{lower} $\beta$};
		\node [style=none] (15) at (-0.75, -2.25) {};
		\node [style=none] (16) at (0.75, -2.25) {};
		\node [style=none] (17) at (0, -2) {...};
		\node [style=qutrit Z phase] (18) at (0, -1) {$\gamma$ \nodepart{lower} $\delta$};
	\end{pgfonlayer}
	\begin{pgfonlayer}{edgelayer}
		\draw [bend right=15] (11.center) to (14);
		\draw [bend right=15] (14) to (12.center);
		\draw [bend left=15] (15.center) to (18);
		\draw [bend left=15] (18) to (16.center);
		\draw [style=hadamard adjoint adge, bend right=45] (14) to (18);
		\draw [style=hadamard adjoint adge, bend left=45] (14) to (18);
	\end{pgfonlayer}
\end{tikzpicture}
} \ = \  
	\scalebox{0.8}{\begin{tikzpicture}
	\begin{pgfonlayer}{nodelayer}
		\node [style=none] (11) at (-0.75, 2.25) {};
		\node [style=none] (12) at (0.75, 2.25) {};
		\node [style=none] (13) at (0, 2) {...};
		\node [style=qutrit Z phase] (14) at (0, 1) {$\alpha$ \nodepart{lower} $\beta$};
		\node [style=none] (15) at (-0.75, -2.25) {};
		\node [style=none] (16) at (0.75, -2.25) {};
		\node [style=none] (17) at (0, -2) {...};
		\node [style=qutrit Z phase] (18) at (0, -1) {$\gamma$ \nodepart{lower} $\delta$};
	\end{pgfonlayer}
	\begin{pgfonlayer}{edgelayer}
		\draw [bend right=15] (11.center) to (14);
		\draw [bend right=15] (14) to (12.center);
		\draw [bend left=15] (15.center) to (18);
		\draw [bend left=15] (18) to (16.center);
		\draw [style=hadamard edge] (14) to (18);
	\end{pgfonlayer}
\end{tikzpicture}
}
\end{equation}
This justifies our notation for Hadamard gates: we can think of Hadamard edges (in blue) as $1$-weighted edges and their adjoints (purple) as $2$-weighted edges, then work modulo $3$, since every triple of parallel edges disappears. Where the previous equations relate single $H$- and $H^\dagger$-boxes across multiple edges, the next three relate multiple $H$- and $H^\dagger$- boxes on single edges. They hold for $h \in \{1, 2\}$, and are proved via simple applications of rules $\qutritRuleId$, $\qutritRuleHadamard$ and $\qutritRuleSnake$.
\begin{equation}\label{eq:h_boxes_are_mod_4}
	\begin{tikzpicture}
	\begin{pgfonlayer}{nodelayer}
		\node [style=none] (4) at (-1.5, -1.75) {};
		\node [style=none] (5) at (-1.5, 1.75) {};
		\node [style=qutrit hadamard unknown] (6) at (-1.5, 1.125) {$h$};
		\node [style=qutrit hadamard unknown] (7) at (-1.5, 0.375) {$h$};
		\node [style=qutrit hadamard unknown] (8) at (-1.5, -0.375) {$h$};
		\node [style=qutrit hadamard unknown] (9) at (-1.5, -1.15) {$h$};
		\node [style=none] (10) at (1.5, 1.75) {};
		\node [style=none] (11) at (1.5, -1.75) {};
		\node [style=none] (12) at (0, 0) {$=$};
	\end{pgfonlayer}
	\begin{pgfonlayer}{edgelayer}
		\draw (11.center) to (10.center);
		\draw (5.center) to (6);
		\draw (6) to (7);
		\draw (7) to (8);
		\draw (8) to (9);
		\draw (9) to (4.center);
	\end{pgfonlayer}
\end{tikzpicture}
 \quad ,
	\hspace{50pt}
	\begin{tikzpicture}
	\begin{pgfonlayer}{nodelayer}
		\node [style=none] (5) at (-1.5, 1.25) {};
		\node [style=qutrit hadamard unknown] (6) at (-1.5, 0.75) {$h$};
		\node [style=qutrit hadamard unknown] (7) at (-1.5, 0) {$h$};
		\node [style=qutrit hadamard unknown] (8) at (-1.5, -0.75) {$h$};
		\node [style=none] (10) at (1.5, 1.25) {};
		\node [style=none] (11) at (1.5, -1.25) {};
		\node [style=none] (12) at (0, 0) {=};
		\node [style=none] (13) at (-1.5, -1.25) {};
		\node [style=qutrit hadamard unknown] (14) at (1.5, 0) {$-h$};
	\end{pgfonlayer}
	\begin{pgfonlayer}{edgelayer}
		\draw [in=270, out=90] (11.center) to (10.center);
		\draw (5.center) to (6);
		\draw (6) to (7);
		\draw (7) to (8);
		\draw (13.center) to (8);
	\end{pgfonlayer}
\end{tikzpicture}
 \quad ,
	\hspace{50pt}
	\begin{tikzpicture}
	\begin{pgfonlayer}{nodelayer}
		\node [style=none] (5) at (-1.5, 1) {};
		\node [style=qutrit hadamard unknown] (6) at (-1.5, 0.375) {$h$};
		\node [style=qutrit hadamard unknown] (7) at (-1.5, -0.375) {$h$};
		\node [style=none] (10) at (1.5, 1.25) {};
		\node [style=none] (11) at (1.5, -1.25) {};
		\node [style=none] (12) at (0, 0) {=};
		\node [style=none] (13) at (-1.5, -1) {};
		\node [style=qutrit hadamard unknown] (14) at (1.5, 0.75) {$h$};
		\node [style=qutrit hadamard unknown] (15) at (1.5, -0.75) {$h$};
		\node [style=Z dot] (16) at (1.5, 0) {};
	\end{pgfonlayer}
	\begin{pgfonlayer}{edgelayer}
		\draw [in=270, out=90] (11.center) to (10.center);
		\draw (5.center) to (6);
		\draw (6) to (7);
		\draw (7) to (13.center);
	\end{pgfonlayer}
\end{tikzpicture}

\end{equation}
\begin{proposition}\label{prop:every_diagram_is_graph_like_qutrit}
	Every qutrit ZX-diagram is equivalent to one that is graph-like.
	\begin{proof}
		First use the colour change rule to turn all $X$-spiders into $Z$-spiders. Then use \eqref{eq:h_boxes_are_mod_4} to remove excess $H$- and $H^\dagger$-boxes, inserting a spider between any remaining consecutive pair of such boxes, so that all spiders are connected only by plain edges, $H$-edges or $H^\dagger$-edges. Fuse together as many as possible, and apply \eqref{eq:spiders_reluctant_to_fuse} where fusion is not possible, so that no plain edge connects two spiders. Apply \eqref{eq:h_edges_are_mod_3} to all connected pairs of spiders until at most one $H$- or $H^\dagger$-edge remains between them. Finally, to ensure every input and output is connected to a spider and every spider is connected to at most one input or output, we can use $\qutritRuleHadamard$ and $\qutritRuleId$ to add a few spiders, $H$- and $H^\dagger$-edges as needed: 
		\begin{equation*}
			\begin{tikzpicture}
	\begin{pgfonlayer}{nodelayer}
		\node [style=none] (0) at (0, 0) {=};
		\node [style=none] (1) at (1.5, 1.5) {};
		\node [style=none] (2) at (1.5, -1.5) {};
		\node [style=none] (3) at (-1.5, 1) {};
		\node [style=none] (4) at (-1.5, -1) {};
		\node [style=Z dot] (5) at (1.5, 0) {};
		\node [style=Z dot] (8) at (1.5, 1) {};
		\node [style=Z dot] (9) at (1.5, -1) {};
	\end{pgfonlayer}
	\begin{pgfonlayer}{edgelayer}
		\draw (3.center) to (4.center);
		\draw [style=hadamard edge] (8) to (5);
		\draw [style=hadamard adjoint adge] (5) to (9);
		\draw (9) to (2.center);
		\draw (1.center) to (8);
	\end{pgfonlayer}
\end{tikzpicture}
 \quad ,
			\hspace{50pt}
			\begin{tikzpicture}
	\begin{pgfonlayer}{nodelayer}
		\node [style=none] (0) at (0, 0) {=};
		\node [style=none] (1) at (1.5, 1) {};
		\node [style=none] (2) at (1.5, -1) {};
		\node [style=none] (3) at (-1.5, 1) {};
		\node [style=none] (4) at (-1.5, -1) {};
		\node [style=qutrit hadamard unknown] (6) at (1.5, 0.5) {$h$};
		\node [style=Z dot] (8) at (1.5, -0.5) {};
		\node [style=qutrit hadamard unknown] (10) at (-1.5, 0) {$h$};
	\end{pgfonlayer}
	\begin{pgfonlayer}{edgelayer}
		\draw (3.center) to (4.center);
		\draw (1.center) to (2.center);
	\end{pgfonlayer}
\end{tikzpicture}
 \quad ,
			\hspace{50pt}
			\begin{tikzpicture}
	\begin{pgfonlayer}{nodelayer}
		\node [style=none] (1) at (-2.75, -1.25) {};
		\node [style=none] (2) at (-1.25, -1.25) {};
		\node [style=none] (4) at (-2, -1) {...};
		\node [style=none] (6) at (1.25, -1.5) {};
		\node [style=none] (9) at (2, -1.25) {...};
		\node [style=none] (14) at (1.375, -0.75) {};
		\node [style=qutrit Z phase] (16) at (-2, 0) {$\alpha$ \nodepart{lower} $\beta$};
		\node [style=none] (17) at (-2.75, 1.25) {};
		\node [style=none] (18) at (-1.25, 1.25) {};
		\node [style=none] (19) at (-2, 1) {...};
		\node [style=none] (21) at (1.25, 1.5) {};
		\node [style=none] (22) at (2.75, 1.5) {};
		\node [style=none] (23) at (2, 1.25) {...};
		\node [style=none] (24) at (0, 0) {=};
		\node [style=Z dot] (28) at (1.375, -0.75) {};
		\node [style=none] (29) at (2.75, -1.5) {};
		\node [style=qutrit Z phase] (31) at (2, 0.25) {$\alpha$ \nodepart{lower} $\beta$};
		\node [style=Z dot] (32) at (2.625, -0.75) {};
	\end{pgfonlayer}
	\begin{pgfonlayer}{edgelayer}
		\draw [bend right=15] (17.center) to (16);
		\draw [bend right=15] (16) to (18.center);
		\draw [bend left=15] (1.center) to (16);
		\draw [bend left=15] (16) to (2.center);
		\draw [style=hadamard adjoint adge, bend right=15, looseness=0.75] (14.center) to (6.center);
		\draw [bend right=15] (21.center) to (31);
		\draw [bend right=15] (31) to (22.center);
		\draw [style=hadamard edge, bend right=15, looseness=0.50] (31) to (14.center);
		\draw [style=hadamard adjoint adge, bend left=15, looseness=0.75] (32) to (29.center);
		\draw [style=hadamard edge, bend left=15, looseness=0.50] (31) to (32);
	\end{pgfonlayer}
\end{tikzpicture}

		\end{equation*}
	\end{proof}
\end{proposition}

A graph state is described fully by its underlying multigraph, or equivalently by an adjacency matrix, where edges take weights in $\mathbb{Z}_3$\ \cite[Lemma 4.2]{harny_completeness}. Nodes correspond to phaseless green spiders, edges of weight $1$ correspond to Hadamard edges, and edges of weight $2$ correspond to $H^\dagger$-edges. As in the qubit case, graph states admit a \emph{local complementation} operation\ \cite[Definition 2.6]{harny_completeness}, though the effect is now slightly more complicated. We'll give the intuition after the formal definition:
\begin{definition}\label{def:local_complementation_qutrit}
	Given $a \in \mathbb{Z}_3$ and a graph state $G$ with adjacency matrix $W = (w_{i,j})$, the \emph{$a$-local complentation} at node $x$ is the new graph state $G *_a x$, whose adjacency matrix $W' = (w'_{i,j})$ is given by $w'_{i,j} = w_{i,j} + aw_{i,x}w_{j,x}$.
\end{definition}
So only those edges between neighbours of node $x$ are affected. Specifically, for two nodes $i$ and $j$ both connected to $x$ by the same colour edge, $a$-local complementation at $x$ increases weight $w_{i,j}$ by $a$. If instead $i$ and $j$ are connected to $x$ by edges of different colours, $a$-local complementation at $x$ decreases $w_{i,j}$ by $a$. This is shown graphically below, and holds with the roles of blue and purple interchanged:
\begin{equation}
	\begin{tikzpicture}
	\begin{pgfonlayer}{nodelayer}
		\node [style=Z dot] (18) at (-2.75, -0.75) {};
		\node [style=Z dot] (19) at (-4.5, 1) {};
		\node [style=Z dot] (20) at (-1, 1) {};
		\node [style={text=red}] (21) at (-2.55, -1.325) {$*_a$};
		\node [style=none] (22) at (-2.75, -0.25) {};
		\node [style=none] (23) at (-4.5, 1.5) {};
		\node [style=none] (24) at (-1, 1.5) {};
		\node [style=Z dot] (25) at (2.75, -0.75) {};
		\node [style=Z dot] (26) at (1, 1) {};
		\node [style=Z dot] (27) at (4.5, 1) {};
		\node [style=none] (28) at (2.75, -0.25) {};
		\node [style=none] (29) at (1, 1.5) {};
		\node [style=none] (30) at (4.5, 1.5) {};
		\node [style=none] (31) at (0, 0) {$\mapsto$};
		\node [style=qutrit hadamard unknown] (32) at (-2.75, 1) {$w$};
		\node [style=qutrit hadamard unknown] (33) at (2.75, 1) {$w+a$};
	\end{pgfonlayer}
	\begin{pgfonlayer}{edgelayer}
		\draw [style=hadamard edge] (19) to (18);
		\draw (19) to (20);
		\draw (18) to (22.center);
		\draw (19) to (23.center);
		\draw (20) to (24.center);
		\draw [style=hadamard edge] (26) to (25);
		\draw (26) to (27);
		\draw (25) to (28.center);
		\draw (26) to (29.center);
		\draw (27) to (30.center);
		\draw [style=hadamard edge] (18) to (20);
		\draw [style=hadamard edge] (25) to (27);
	\end{pgfonlayer}
\end{tikzpicture}
 \quad ,
	\hspace{25pt}
	\begin{tikzpicture}
	\begin{pgfonlayer}{nodelayer}
		\node [style=Z dot] (0) at (-2.75, -0.75) {};
		\node [style=Z dot] (1) at (-4.5, 1) {};
		\node [style=Z dot] (2) at (-1, 1) {};
		\node [style={text=red}] (3) at (-2.55, -1.325) {$*_a$};
		\node [style=none] (4) at (-2.75, -0.25) {};
		\node [style=none] (5) at (-4.5, 1.5) {};
		\node [style=none] (6) at (-1, 1.5) {};
		\node [style=Z dot] (8) at (2.75, -0.75) {};
		\node [style=Z dot] (9) at (1, 1) {};
		\node [style=Z dot] (10) at (4.5, 1) {};
		\node [style=none] (11) at (2.75, -0.25) {};
		\node [style=none] (12) at (1, 1.5) {};
		\node [style=none] (13) at (4.5, 1.5) {};
		\node [style=none] (17) at (0, 0) {$\mapsto$};
		\node [style=qutrit hadamard unknown] (18) at (-2.75, 1) {$w$};
		\node [style=qutrit hadamard unknown] (19) at (2.75, 1) {$w-a$};
	\end{pgfonlayer}
	\begin{pgfonlayer}{edgelayer}
		\draw [style=hadamard edge] (1) to (0);
		\draw (1) to (2);
		\draw (0) to (4.center);
		\draw (1) to (5.center);
		\draw (2) to (6.center);
		\draw [style=hadamard edge] (9) to (8);
		\draw (9) to (10);
		\draw (8) to (11.center);
		\draw (9) to (12.center);
		\draw (10) to (13.center);
		\draw [style=hadamard adjoint adge] (0) to (2);
		\draw [style=hadamard adjoint adge] (8) to (10);
	\end{pgfonlayer}
\end{tikzpicture}

\end{equation}
\begin{theorem}\label{thm:local_comp_equality} \cite[Theorem 4.4, Corollary 4.5]{harny_completeness}~
	Given $a \in \mathbb{Z}_3$ and a graph state $(G, W)$ containing a node $x$, let $N(x)$ denote the neighbours of $x$; that is, nodes $i$ with weight $w_{i,x} \in \{1, 2\}$. Then the following equality holds:
	\begin{equation}
		\begin{tikzpicture}
	\begin{pgfonlayer}{nodelayer}
		\node [style=none] (0) at (0, 0) {=};
		\node [style=none] (1) at (1.25, 0.25) {};
		\node [style=none] (2) at (8.25, 0.25) {};
		\node [style=none] (3) at (4.5, -1.75) {};
		\node [style=none] (4) at (1.25, -0.25) {};
		\node [style=none] (5) at (1.5, 0.25) {};
		\node [style=none] (6) at (2.75, 0.25) {};
		\node [style=none] (8) at (6.5, 0.25) {};
		\node [style=none] (9) at (8, 0.25) {};
		\node [style=none] (10) at (4.5, -0.5) {$G *_a x$};
		\node [style=none] (11) at (1.5, 1.75) {};
		\node [style=none] (12) at (2.75, 1.75) {};
		\node [style=none] (14) at (6.5, 1.75) {};
		\node [style=none] (15) at (8, 1.75) {};
		\node [style=qutrit X phase] (31) at (1.5, 1) {$a$ \nodepart{lower} $a$};
		\node [style=qutrit Z phase] (32) at (2.75, 1) {$-a$ \nodepart{lower} $-a$};
		\node [style=none] (33) at (5.25, 0.25) {};
		\node [style=none] (34) at (5.25, 1.75) {};
		\node [style=qutrit Z phase] (35) at (5.25, 1) {$-a$ \nodepart{lower} $-a$};
		\node [style=none] (36) at (4, 1) {...};
		\node [style=none] (37) at (7.25, 1) {...};
		\node [style={scale=0.7}] (38) at (4, 2.25) {$\overbrace{\hspace{60pt}}^{N(x)}$};
		\node [style={scale=0.7}] (39) at (1.5, 2.25) {$\overbrace{\quad}^{x}$};
		\node [style=none] (40) at (-8, 0.25) {};
		\node [style=none] (41) at (-1, 0.25) {};
		\node [style=none] (42) at (-4.75, -1.75) {};
		\node [style=none] (43) at (-8, -0.25) {};
		\node [style=none] (44) at (-7.75, 0.25) {};
		\node [style=none] (45) at (-6.5, 0.25) {};
		\node [style=none] (46) at (-2.75, 0.25) {};
		\node [style=none] (47) at (-1.25, 0.25) {};
		\node [style=none] (48) at (-4.75, -0.5) {$G$};
		\node [style=none] (49) at (-7.75, 0.75) {};
		\node [style=none] (50) at (-6.5, 0.75) {};
		\node [style=none] (51) at (-2.75, 0.75) {};
		\node [style=none] (52) at (-1.25, 0.75) {};
		\node [style=none] (55) at (-4, 0.25) {};
		\node [style=none] (56) at (-4, 0.75) {};
		\node [style=none] (58) at (-5.25, 0.5) {...};
		\node [style=none] (59) at (-2, 0.5) {...};
		\node [style={scale=0.7}] (60) at (-5.25, 1.25) {$\overbrace{\hspace{60pt}}^{N(x)}$};
		\node [style={scale=0.7}] (61) at (-7.75, 1.25) {$\overbrace{\quad}^{x}$};
	\end{pgfonlayer}
	\begin{pgfonlayer}{edgelayer}
		\draw (1.center) to (4.center);
		\draw (4.center) to (3.center);
		\draw (3.center) to (2.center);
		\draw (1.center) to (2.center);
		\draw (5.center) to (11.center);
		\draw (6.center) to (12.center);
		\draw (8.center) to (14.center);
		\draw (9.center) to (15.center);
		\draw (33.center) to (34.center);
		\draw (40.center) to (43.center);
		\draw (43.center) to (42.center);
		\draw (42.center) to (41.center);
		\draw (40.center) to (41.center);
		\draw (44.center) to (49.center);
		\draw (45.center) to (50.center);
		\draw (46.center) to (51.center);
		\draw (47.center) to (52.center);
		\draw (55.center) to (56.center);
	\end{pgfonlayer}
\end{tikzpicture}

	\end{equation}
\end{theorem}


\begin{definition}\label{def:local_pivot_qutrit}
	Given $a,b,c \in \mathbb{Z}_3$ and a graph state $G$ containing nodes $i$ and $j$, the \emph{$(a,b,c)$-pivot} along $ij$ is the new graph state $G \wedge_{(a,b,c)} ij \defeq ((G *_a i) *_b j) *_c i$. 
\end{definition}
This pivot operation again leads to an equality, up to introducing some extra gates on outputs, whose proof is found in Appendix \ref{thm:local_pivot_equality_appendix}. Here we shall only consider an $(a,-a,a)$-pivot along an edge $ij$ of non-zero weight, for $a \in \{1, 2\}$. We will call this a \emph{proper $a$-pivot} along $ij$, and denote it $G \wedge_a ij$.
\begin{theorem}\label{thm:local_pivot_equality}
	\qutritPivotEqualityStatement
\end{theorem}

\subsection{Qutrit Elimination Theorems}

We classify spiders into three families exactly as in \cite[Theorem 3.1]{harny_completeness}:
\begin{equation}
	\mathcal{M} = \left\{\scalebox{0.8}{\qutritZspider{0}{0}}, \scalebox{0.8}{\qutritZspider{1}{2}}, \scalebox{0.8}{\qutritZspider{2}{1}}\right\},
	~
	\mathcal{N} = \left\{\scalebox{0.8}{\qutritZspider{0}{1}}, \scalebox{0.8}{\qutritZspider{1}{0}}, \scalebox{0.8}{\qutritZspider{0}{2}}, \scalebox{0.8}{\qutritZspider{2}{0}}\right\},
	~
	\mathcal{P} = \left\{\scalebox{0.8}{\qutritZspider{1}{1}}, \scalebox{0.8}{\qutritZspider{2}{2}}\right\}
\end{equation}
We call a spider in a graph-like ZX-diagram \emph{interior} if it isn't connected to an input or output. Given any graph-like ZX-diagram, we will show that we can eliminate standalone interior $\mathcal{P}$- and \Nspiders\ by local complementation, and pairs of connected interior \Mspiders\ by pivoting. 

First, we define a modification of the \emph{!-box} notation
(pronounced \emph{`bang-box'}),
as introduced in \cite{dixon2009graphical} for general string diagrams. A !-box is a compressed notation for a family of diagrams; the contents of the !-box
are `repeated' or `unfolded' $n\geq 0$ times.
Following the style of \cite{backens2018zh}, we allow a parameter denoting the maximum number of copies.
We also extend this notation as follows: we decorate the !-box with an edge-type, which denotes that the spiders unfolded are all-to-all connected by an edge of this type. In qutrit ZX-diagrams this notation is well-defined in certain scenarios: in particular, when the denoted edge-type is a $H$- or $H^{\dagger}$-edge (since then the distinction between a spider's input and output wires disappears) and the main contents of the !-box is equivalent to a single spider (since then there is no ambiguity about which spiders are connected). For example:
\begin{equation}\label{eq:bang_box}
	\left\{ \begin{tikzpicture}
	\begin{pgfonlayer}{nodelayer}
		\node [style=Z dot] (0) at (-0.75, -1.5) {};
		\node [style=X dot] (1) at (0.75, -1.5) {};
		\node [style=qutrit Z phase] (2) at (0, 0) {$\alpha_k$ \nodepart{lower} $\beta_k$};
		\node [style=none] (7) at (0, 1.125) {};
		\node [style=none] (8) at (0, 1.575) {\textcolor{bang_orange}{$\scriptstyle{k \in {[K]}}$}};
		\node [style=none] (19) at (-1, -0.75) {};
		\node [style=none] (20) at (-1.25, -0.5) {};
		\node [style=none] (21) at (1, -0.75) {};
		\node [style=none] (22) at (1.25, -0.5) {};
		\node [style=none] (23) at (1, 1.25) {};
		\node [style=none] (24) at (1.25, 1) {};
		\node [style=none] (25) at (-1, 1.25) {};
		\node [style=none] (26) at (-1.25, 1) {};
		\node [style=none] (27) at (-1.25, 0.75) {};
		\node [style=none] (28) at (-0.75, 1.25) {};
		\node [style=none] (29) at (-0.75, 0.75) {};
		\node [style=none] (30) at (-1, 0.75) {};
	\end{pgfonlayer}
	\begin{pgfonlayer}{edgelayer}
		\draw [bend left=15] (0) to (2);
		\draw [bend left=15, looseness=0.75] (2) to (1);
		\draw (7.center) to (2);
		\draw [style=bang] (22.center)
			 to (24.center)
			 to [bend right=45, looseness=0.75] (23.center)
			 to (25.center)
			 to [bend right=45, looseness=0.75] (26.center)
			 to (20.center)
			 to [bend right=45, looseness=0.75] (19.center)
			 to (21.center)
			 to [bend right=45, looseness=0.75] cycle;
		\draw [style=bang] (27.center) to (29.center);
		\draw [style=bang] (28.center) to (29.center);
		\draw [style=hadamard edge] (25.center) to (30.center);
	\end{pgfonlayer}
\end{tikzpicture}
 : K \in \{0, 1, 2, 3\} \right\} =
	\left\{ ~
		\scalebox{0.8}{\begin{tikzpicture}
	\begin{pgfonlayer}{nodelayer}
		\node [style=Z dot] (0) at (-0.75, -1.25) {};
		\node [style=X dot] (1) at (0.75, -1.25) {};
	\end{pgfonlayer}
\end{tikzpicture}
} \ , \
		\scalebox{0.8}{\begin{tikzpicture}
	\begin{pgfonlayer}{nodelayer}
		\node [style=Z dot] (0) at (-0.75, -1.25) {};
		\node [style=X dot] (1) at (0.75, -1.25) {};
		\node [style=qutrit Z phase] (2) at (0, 0.25) {$\alpha_1$ \nodepart{lower} $\beta_1$};
		\node [style=none] (7) at (0, 1.5) {};
	\end{pgfonlayer}
	\begin{pgfonlayer}{edgelayer}
		\draw (7.center) to (2);
		\draw [bend left=15] (0) to (2);
		\draw [bend left=15] (2) to (1);
	\end{pgfonlayer}
\end{tikzpicture}
} \ , \
		\scalebox{0.8}{\begin{tikzpicture}
	\begin{pgfonlayer}{nodelayer}
		\node [style=Z dot] (0) at (-0.875, -1.25) {};
		\node [style=X dot] (1) at (0.875, -1.25) {};
		\node [style=qutrit Z phase] (2) at (-0.875, 0.25) {$\alpha_1$ \nodepart{lower} $\beta_1$};
		\node [style=none] (7) at (-0.875, 1.5) {};
		\node [style=qutrit Z phase] (10) at (0.875, 0.25) {$\alpha_2$ \nodepart{lower} $\beta_2$};
		\node [style=none] (11) at (0.875, 1.5) {};
	\end{pgfonlayer}
	\begin{pgfonlayer}{edgelayer}
		\draw (7.center) to (2);
		\draw (0) to (2);
		\draw (2) to (1);
		\draw (11.center) to (10);
		\draw (0) to (10);
		\draw (1) to (10);
		\draw [style=hadamard edge] (2) to (10);
	\end{pgfonlayer}
\end{tikzpicture}
} \ , \
		\scalebox{0.8}{\begin{tikzpicture}
	\begin{pgfonlayer}{nodelayer}
		\node [style=Z dot] (0) at (-0.75, -1.25) {};
		\node [style=X dot] (1) at (0.75, -1.25) {};
		\node [style=qutrit Z phase] (2) at (-1.5, 0.25) {$\alpha_1$ \nodepart{lower} $\beta_1$};
		\node [style=none] (7) at (-1.5, 1.5) {};
		\node [style=qutrit Z phase] (8) at (0, 1.25) {$\alpha_2$ \nodepart{lower} $\beta_2$};
		\node [style=none] (9) at (0, 2.5) {};
		\node [style=qutrit Z phase] (10) at (1.5, 0.25) {$\alpha_3$ \nodepart{lower} $\beta_3$};
		\node [style=none] (11) at (1.5, 1.5) {};
	\end{pgfonlayer}
	\begin{pgfonlayer}{edgelayer}
		\draw (7.center) to (2);
		\draw (9.center) to (8);
		\draw (0) to (2);
		\draw (2) to (1);
		\draw (0) to (8);
		\draw (8) to (1);
		\draw (11.center) to (10);
		\draw (0) to (10);
		\draw (1) to (10);
		\draw [style=hadamard edge] (2) to (8);
		\draw [style=hadamard edge] (8) to (10);
		\draw [style=hadamard edge] (2) to (10);
	\end{pgfonlayer}
\end{tikzpicture}
}
	~ \right\}
\end{equation}

If a !-box has no corner decoration, the unfolded spiders within are not connected by any edge, just as in the usual !-box notation.

\begin{theorem}\label{thm:eliminate_P_spiders}
	\eliminatePSpidersStatement
\end{theorem}
\begin{theorem}\label{thm:eliminate_N_spiders}
	\eliminateNSpidersStatement
\end{theorem}
\begin{theorem}\label{thm:eliminate_M_spiders}
	\eliminateMSpidersStatement
\end{theorem}

Proofs of these theorems are found in Appendix \ref{thm:eliminate_P_spiders_appendix}, \ref{thm:eliminate_N_spiders_appendix} and \ref{thm:eliminate_M_spiders_appendix} respectively. We can now combine them into an algorithm for efficiently simplifying a \emph{closed} graph-like ZX-diagram. First note that after applying any one of the three elimination theorems to such a diagram, and perhaps removing parallel $H$- or $H^\dagger$-edges via \eqref{eq:h_edges_are_mod_3}, we again have a graph-like diagram.

\begin{theorem}\label{thm:simplification_algorithm_works}
	Given any closed graph-like ZX-diagram, the following algorithm will always terminate after a finite number of steps, returning an equivalent graph-like ZX-diagram with no \Nspiders, \Pspiders, or adjacent pairs of \Mspiders. Repeat the steps below until no rule matches. After each step, apply \eqref{eq:h_edges_are_mod_3} as needed until the resulting diagram is graph-like:
	\begin{enumerate}
		\item Eliminate a \Pspider\ via Theorem~\ref{thm:eliminate_P_spiders}.
		\item Eliminate an \Nspider\ via Theorem~\ref{thm:eliminate_N_spiders}.
		\item Eliminate two adjacent \Mspiders\ via Theorem~\ref{thm:eliminate_M_spiders}.
	\end{enumerate}
	\begin{proof}
		At every step the total number of spiders decreases by at least one, so since we start with a finite diagram the algorithm terminates after a finite number of steps. By construction, when it does so we are left with an equivalent graph-like ZX-diagram with no \Nspiders, \Pspiders, or adjacent pairs of \Mspiders.
	\end{proof}
\end{theorem}
In particular, if we start with a stabilizer diagram, we can eliminate all but perhaps one spider, depending on whether the initial number of \Mspiders\ was odd or even. This is because no step introduces any non-stabilizer phases. The algorithm above could be extended to a \emph{non-closed} graph-like diagrams as in \cite[Theorem 5.4]{graph_theoretic_simplification} as part of a qutrit circuit optimisation algorithm, by supplementing with a method for circuit extraction method. We leave this for future work.
\section{Case Studies}

In this section we present two problems which can naturally be cast in tensor network form.
These problem families are interesting in that they show a transition in complexity from easy to hard when the dimension $d$ carried by the wires is greater than a specific value. We recast known results about evaluating the \emph{Jones polynomial}, and finally we briefly look at graph colouring.

\subsection{Jones Polynomial at Lattice Roots of Unity}


A \emph{knot} $K$ is a circle embedded in $\mathbb{R}^3$.
A set of knots tangled together make a \emph{link} $L$.
A link $L$ can be represented by a \emph{link diagram}
by projecting it on to the plane
but retaining the information of over- or under-crossings.
We say that $L\simeq L'$ iff the diagram of link $L$ can be deformed to that of link $L'$ without cutting or gluing strands, or passing strands through each other.
The Jones polynomial $V_L(t)$
is a Laurent polynomial in a variable $t\in\mathbb{C}$
and is a \emph{link invariant}.
This means that
$V_L(t)\neq V_{L'}(t)\Rightarrow L\not\simeq L'$.

In general, computing $V_L(t)$
is exponentially costly in the number of crossings $c$,
something made explicit when one uses the Kauffman bracket method \cite{Kauffman2001}.
Exactly evaluating the Jones polynomial at points $t\in\mathbb{C}$ is \#P-hard,
except at the \emph{lattice roots of unity} $\Lambda = \{ \pm 1, \pm i, \pm e^{i 2\pi/3}, \pm e^{i 4\pi/3} \}$,
where it can be evaluated at cost $O(poly(c))$ \cite{jaeger_vertigan_welsh_1990}.

Additively approximating the Jones polynomial at non-lattice roots of unity is a paradigmatic BQP-complete complete problem \cite{Aharonov_2008,kuperberg2014hard}.
Topological quantum computation \cite{Freedman2002,pachos_2012} is the most natural model for such knot theoretic questions. In this model, quantum states are defined in the fusion space of \emph{anyons}, emergent quasiparticles with nontrivial exchange statistics arising in two-dimensional exotic phases of matter.
Quantum computation is performed by creating anyons from the vacuum, then braiding them, and finally fusing them back to the vacuum, where braids play the role of unitary gates.
The world-lines of the anyons define a closed braid, i.e. a link.
Such a link encodes a quantum amplitude corresponding to its Jones polynomial evaluated at a root of unity depending on the anyon theory at hand \cite{Witten1989}. 
Specifically, in the case of $\text{SU}(2)_k$ anyons, the Jones polynomial is evaluated at $t(k) = e^{i 2\pi/(2+k)}$ \cite{Rowell_2018}.

Also, the evaluation of the Jones polynomial at certain points can be expressed, up to an efficiently computable scalar that depends on the link diagram, as the partition function $Z_{G_L}(d)$ of a $d$-state Potts model with suitable spin-spin interactions \cite{RevModPhys.64.1099}.
This Potts model is defined on a signed graph $G_L$, obtained as follows.
The link diagram is bicoloured checkerboard-style,
then every coloured area is mapped to a vertex and every crossing is mapped to a signed edge according to
its orientation relative to the surrounding colours, as in \eqref{eq:link-tn} below. 
The relation between the point $t(d)$ at which the Jones polynomial is evaluated and the dimension $d$ of the spins is $d = t + t^{-1} +2$,
which can be solved for $t(d)$.

Note the correspondence between the dimension $d$ in the Potts approach and the level $k$ in the anyon-braiding approach to the Jones polynomial: $\{t(d) | ~ d \in\{1,2,3,4\} \} = \{ t(k) | ~ k\in\{1,2,4,\infty\} \} \subseteq \Lambda$.
This is consistent with the fact that braiding $\text{SU}(2)_2$ anyons (Ising) or $\text{SU}(2)_4$ anyons is not universal (unless the $\text{SU}(2)_4$ anyons are augmented by fusion and measurements \cite{Levaillant_2015}).

The partition function $Z_{G_L}(d\in\mathbb{N})$ can be expressed as a closed tensor network in terms of phaseless (green) $d$-dimensional $Z$-spiders connected via wires that go through \emph{$\pm$-boxes} \cite{PhysRevE.100.033303}:
\begin{equation}
\includegraphics[scale=0.35, valign=c]{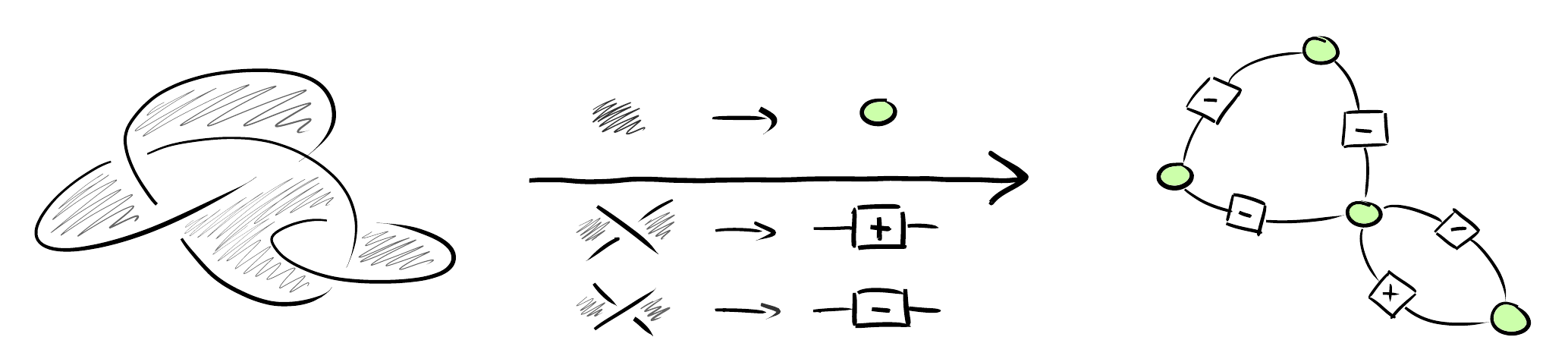}
\label{eq:link-tn}
\end{equation}
The $\pm$-boxes have the following concrete interpretation as $d \times d$ matrices:
\begin{equation}\label{eq:pm_tensor}
	\left\llbracket \ \begin{tikzpicture}
	\begin{pgfonlayer}{nodelayer}
		\node [style=box] (0) at (0, 0) {$\pm$};
		\node [style=none] (1) at (0, -1.25) {};
		\node [style=none] (2) at (0, 1.25) {};
	\end{pgfonlayer}
	\begin{pgfonlayer}{edgelayer}
		\draw (2.center) to (1.center);
	\end{pgfonlayer}
\end{tikzpicture}
 \ \right\rrbracket ~~ = ~~  
	\begin{pmatrix}
		-t(d)^{\mp 1} & 1 & \dots & 1 \\
		1 & -t(d)^{\mp 1} & \dots & 1 \\
		\vdots & \vdots & \ddots & \vdots\\
		1 & 1 & \dots & -t(d)^{\mp 1}
	\end{pmatrix}
\end{equation}
The $\pm$-matrices above for $d \in \{2, 4\}$
are equal up to a scalar to concrete interpretations of qubit stabilizer ZX-diagrams, and the $\pm$-matrices for $d=3$ of qutrit stabilizer ZX-diagrams (see Appendix \ref{prop:pm_maps_zx_appendix}):
	\begin{equation}
		\left\llbracket \ \begin{tikzpicture}
	\begin{pgfonlayer}{nodelayer}
		\node [style=box] (0) at (0, 0) {$\pm$};
		\node [style=none] (1) at (0, -1.25) {};
		\node [style=none] (2) at (0, 1.25) {};
	\end{pgfonlayer}
	\begin{pgfonlayer}{edgelayer}
		\draw (2.center) to (1.center);
	\end{pgfonlayer}
\end{tikzpicture}
 \ \right\rrbracket_{d=2} \hspace{-10pt}\simeq 
		\left\llbracket \ \begin{tikzpicture}
	\begin{pgfonlayer}{nodelayer}
		\node [style=none] (0) at (0, -1) {};
		\node [style=none] (1) at (0, 1) {};
		\node [style=X phase dot] (6) at (0, 0) {$\pm\frac{\pi}{2}$};
	\end{pgfonlayer}
	\begin{pgfonlayer}{edgelayer}
		\draw (0.center) to (1.center);
	\end{pgfonlayer}
\end{tikzpicture}
 \ \right\rrbracket ~, 
		\qquad
		\left\llbracket \ \begin{tikzpicture}
	\begin{pgfonlayer}{nodelayer}
		\node [style=box] (0) at (0, 0) {$\pm$};
		\node [style=none] (1) at (0, -1.25) {};
		\node [style=none] (2) at (0, 1.25) {};
	\end{pgfonlayer}
	\begin{pgfonlayer}{edgelayer}
		\draw (2.center) to (1.center);
	\end{pgfonlayer}
\end{tikzpicture}
 \ \right\rrbracket_{d=3} \hspace{-10pt}\simeq
		\left\llbracket \ \begin{tikzpicture}
	\begin{pgfonlayer}{nodelayer}
		\node [style=qutrit X phase] (0) at (0, 0) {$\pm 1$ \nodepart{lower} $\pm 1$};
		\node [style=none] (3) at (0, -1) {};
		\node [style=none] (4) at (0, 1) {};
	\end{pgfonlayer}
	\begin{pgfonlayer}{edgelayer}
		\draw (3.center) to (4.center);
	\end{pgfonlayer}
\end{tikzpicture}
 \ \right\rrbracket ~,
		\qquad
		\left\llbracket \ \begin{tikzpicture}
	\begin{pgfonlayer}{nodelayer}
		\node [style=box] (0) at (0, 0) {$\pm$};
		\node [style=none] (1) at (0, -1.25) {};
		\node [style=none] (2) at (0, 1.25) {};
	\end{pgfonlayer}
	\begin{pgfonlayer}{edgelayer}
		\draw (2.center) to (1.center);
	\end{pgfonlayer}
\end{tikzpicture}
 \ \right\rrbracket_{d=4} \hspace{-10pt}\simeq 
		\left\llbracket \ \begin{tikzpicture}
	\begin{pgfonlayer}{nodelayer}
		\node [style=X phase dot] (0) at (-1, 0) {$\pi$};
		\node [style=X phase dot] (1) at (1, 0) {$\pi$};
		\node [style=none] (2) at (-1, 1) {};
		\node [style=none] (3) at (1, 1) {};
		\node [style=none] (4) at (-1, -1) {};
		\node [style=none] (5) at (1, -1) {};
		\node [style=hadamard] (7) at (0, 0) {};
	\end{pgfonlayer}
	\begin{pgfonlayer}{edgelayer}
		\draw (4.center) to (2.center);
		\draw (5.center) to (3.center);
		\draw (0) to (1);
	\end{pgfonlayer}
\end{tikzpicture}
 \ \right\rrbracket
	\end{equation}
Since these generators decompose as stabilizer diagrams,
$Z_{G_L}(d\in\{2,3,4\})$ can be evaluated efficiently
via stabilizer ZX-diagram simplification.
Thus, we recover the known result that evaluating the Jones polynomial at $t\in\Lambda$ is in P.
On the other hand, computing $Z_{G_L}(d\geq 5)$ is \#P-hard.

\subsection{Graph Colouring}

Finally, let us briefly look at the \emph{graph colouring problem}. A \emph{$d$-colouring} of a graph $G$ is an assignment of colours $\{1, ..., d\}$ to the vertices of $G$ so that no neighbouring vertices have the same colour. Given a graph $G$ and an integer $d$, we wish to count the number of such $d$-colourings. Again, this problem can be interpreted as the zero-temperature
partition function of an antiferromagnetic $d$-state Potts model \cite{ChromaticPotts}.
Through the lens of our graphical exposition we see that counting problems and computing partition functions are essentially the same problem, since both can be straightforwardly encoded as closed tensor networks.

Given a graph $G$, the graph colouring problem can be encoded as a ZX-diagram as follows. Every vertex of the graph is mapped to a $d$-dimensional phaseless (green) $Z$-spider which copies spin states so that they can enter into the $X$-boxes, each of which in turn enforces that spin states entering it are not the same.
\begin{equation}\label{eq:G-tn}
	\includegraphics[scale=0.4, valign=c]{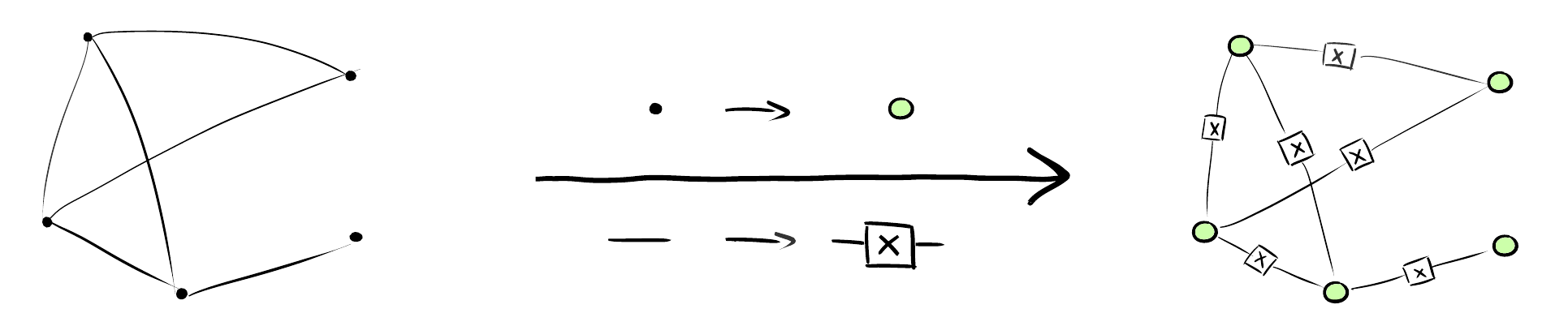}
\end{equation}
The $X$-boxes have the following concrete interpretation as $d \times d$ matrices,
and for $d=2$ they are just the Pauli X:
\begin{equation}\label{eq:X_tensor}
	\left\llbracket \ \begin{tikzpicture}
	\begin{pgfonlayer}{nodelayer}
		\node [style=box] (0) at (0, 0) {$X$};
		\node [style=none] (1) at (0, -1.25) {};
		\node [style=none] (2) at (0, 1.25) {};
	\end{pgfonlayer}
	\begin{pgfonlayer}{edgelayer}
		\draw (2.center) to (1.center);
	\end{pgfonlayer}
\end{tikzpicture}
 \ \right\rrbracket ~~ = ~~  
	\begin{pmatrix}
		0 & 1 & \dots & 1 \\
		1 & 0 & \dots & 1 \\
		\vdots & \vdots & \ddots & \vdots\\
		1 & 1 & \dots & 0
	\end{pmatrix} ~~,\qquad
			\left\llbracket \  \ \right\rrbracket_{d=2} \simeq \left\llbracket \ \begin{tikzpicture}
	\begin{pgfonlayer}{nodelayer}
		\node [style=none] (0) at (0, -1) {};
		\node [style=none] (1) at (0, 1) {};
		\node [style=X phase dot] (6) at (0, 0) {$\pm\pi$};
	\end{pgfonlayer}
	\begin{pgfonlayer}{edgelayer}
		\draw (0.center) to (1.center);
	\end{pgfonlayer}
\end{tikzpicture}
 \ \right\rrbracket
\end{equation}
Counting k-colourings for $k\geq 3$ is a canonical \#P-complete problem.
For $k=0,1,2$ the problem is in P \cite{jaeger_vertigan_welsh_1990}.
For $k=2$ this is witnessed by the fact that the problem reduces to simplifying a qubit stabilizer diagram,
which can be done efficiently.
However, for $d=3$ the $X$-matrix cannot be expressed as a stabilizer qutrit diagram; we prove this using our qutrit elimination theorems in Appendix \ref{prop:X_box_not_stab}.
This is consistent with the fact that counting 3-colourings is \#P-complete.

\section{Discussion and Outlook}

In this work, we leveraged local complementation and pivot operations to flesh out non-trivial simplifications strategies for qutrit ZX diagrams. We expect these results to have immediate uses in qutrit circuit optimisation. 

Somewhat tangentially, we have also used our new simplification strategies to provide complexity-theoretic insight into certain tensor network problems that demonstrate a `step-change' in complexity when the dimension $d$ carried by the wires is greater than a particular value. When such problems are translated into the ZX-calculus, this step-change corresponds to whether a diagram is inside or outside of the efficiently reducible stabilizer fragment of the calculus.



A main path for future work entails the generalisation of our stabilizer simplification rules to qudits of higher dimensions. Recent work on the completeness of the stabilizer fragment of the qudit ZX-calculus for odd prime $d$ should provide insight and motivation, as well as some very elegant simplifications to the calculus \cite{Booth_2022}. Among these simplifications are a restoration of the maxim `only topology matters' \cite{Carette_2021}, and a representation of any stabilizer phase as a tuple $(x, y) \in \mathbb{Z}_d \times \mathbb{Z}_d$.
It would be useful to rederive the qutrit results given here in the language of Ref. \cite{Carette_2021}.
The relevant notion therein is that of `flexsymmetry', which has also been employed in recent relevant work on qutrit ZX-calculus \cite{van_de_Wetering_2022}. 

On a related note, Ref. \cite{Booth_2022} concludes by asking about extending their stabilizer fragment completeness results to non-prime dimensions $d$. In the first author's MSc theis, we defined a parametrisation of stabilizer phases as tuples $(x, y)$ from some group isomorphic to $\mathbb{Z}_d \times \mathbb{Z}_d$ valid for \emph{all} dimensions $d$, not just odd prime ones \cite[Theorem 5.2]{TeagueMasters}. Although finding such a parametrisation is not the main obstacle to extending these completeness results to non-prime dimensions, we feel obliged to refer to this result in case it can contribute towards this endeavour, as to the extent of our knowledge it has not appeared yet in the literature.

The other primary direction of further applied research is in circuit extraction \cite{backens2020again} for qudit circuits, where one could hope to find graph-theoretic simplification \cite{graph_theoretic_simplification} strategies analogous to those for the case of qubits.


\section{Acknowledgements}
We wish to thank Niel de Beaudrap, Aleks Kissinger, Stefanos 
Kourtis and Quanlong Wang for inspiring and helpful discussions, as well as the QPL 2022 and QPL 2022 reviewers for valuable feedback.
KM acknowledges financial support from the Royal Commission for the Exhibition of 1851 through a postdoctoral research fellowship, while ATT thanks the Einstein Foundation (Einstein Research Unit on Quantum Devices) for funding.

\bibliographystyle{eptcs}

\appendix

\section{Qubit ZX-Calculus}\label{app:qubit_zx_calculus}

\begin{figure}[h]
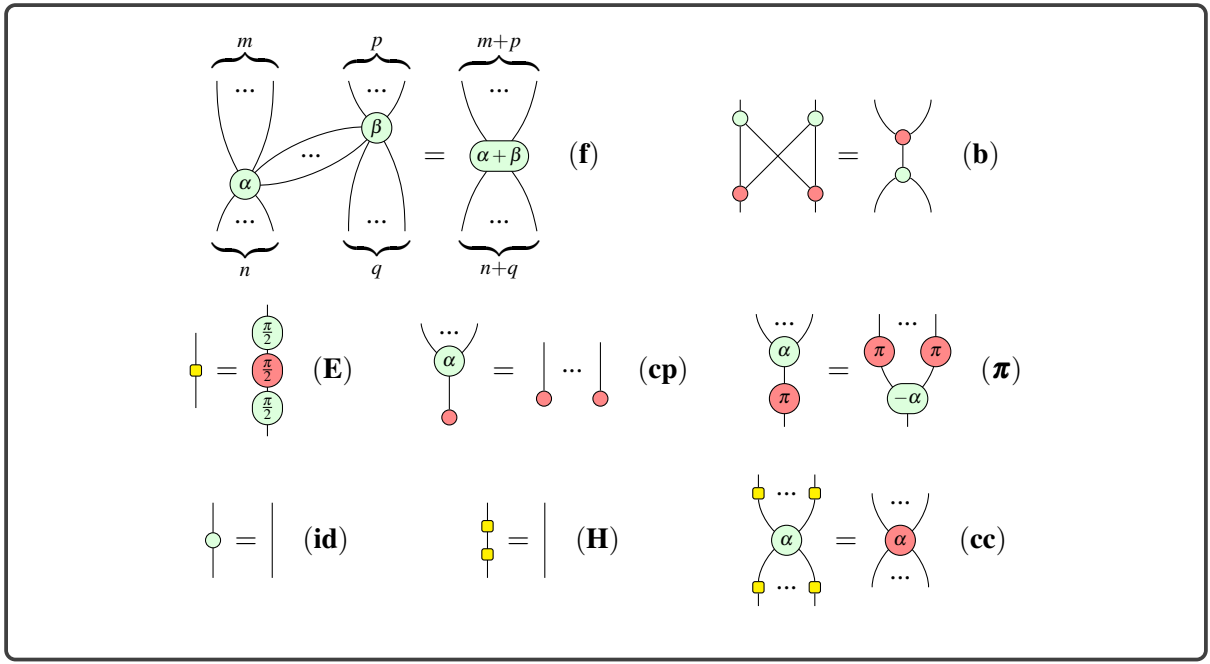

	\begin{tcolorbox}[colback=white]
		\begin{equation*}
		\vspace{-5pt}
			\input{qubit_rules/fusion/lhs.tikz} \ = \ 
			\input{qubit_rules/fusion/rhs.tikz} \quad \hypertarget{qubit_rule_fusion}{\mathbf{(f)}}
			\hspace{50pt}
			\input{qubit_rules/bialgebra/lhs.tikz} \ = \
			\input{qubit_rules/bialgebra/rhs.tikz} \quad \hypertarget{qubit_rule_bialgebra}{\mathbf{(b)}}
		\end{equation*}
		\vspace{5pt}
		\begin{equation*}
			\input{qubit_hadamard/yellow_box.tikz} \ = \ 
			\input{qubit_hadamard/decomposed.tikz} \quad \hypertarget{qubit_rule_euler}{\mathbf{(E)}}
			\hspace{25pt}
			\input{qubit_rules/copy/lhs.tikz} \ = \ \
			\input{qubit_rules/copy/rhs.tikz} \quad \hypertarget{qubit_rule_copy}{\mathbf{(cp)}}
			\hspace{25pt}
			\input{qubit_rules/pi/lhs.tikz} \ = \
			\input{qubit_rules/pi/rhs.tikz} \quad \hypertarget{qubit_rule_pi}{(\bm{\pi})}
		\end{equation*}
		\vspace{5pt}
		\begin{equation*}
			\input{qubit_rules/identity/lhs.tikz} \ = \
			\input{qubit_rules/identity/rhs.tikz} \quad \hypertarget{qubit_rule_id}{\mathbf{(id)}}
			\hspace{50pt}
			\input{qubit_rules/hadamard/lhs.tikz} \ = \
			\input{qubit_rules/hadamard/rhs.tikz} \quad \hypertarget{qubit_rule_hadamard}{\mathbf{(H)}}
			\hspace{50pt}
			\input{qubit_rules/colour_change/lhs.tikz} \ = \
			\input{qubit_rules/colour_change/rhs.tikz} \quad \hypertarget{qubit_rule_colour_change}{\mathbf{(cc)}}
		\end{equation*}
		\vspace{3pt}
	\end{tcolorbox}
	\vspace{5pt}
	\caption{Rewrite rules for the qubit ZX-calculus where spiders are interpreted as tensors over $\mathbb{C}$.}
	\label{fig:qubit_ZX_rules}
	\vspace{-1pt}
\end{figure}

The well-known rewrite rules of the qubit ZX-calculus are shown in Fig.\ref{fig:qubit_ZX_rules}. 
The \emph{stabilizer fragment} of the calculus consists of all diagrams in which all phases are $\alpha=\frac{\pi n}{2}$, $n\in\mathbb{Z}$.
In \cite[Theorem 5.4]{graph_theoretic_simplification} the authors give an efficient algorithm for simplifying any qubit ZX-diagram to an equivalent diagram with fewer spiders.
The algorithm consists of consecutive applications of spider-eliminating rewrites.

Every diagram is equivalent to a \emph{graph-like} diagram:
every spider is green,
every edge is a Hadamard edge,
there are no parallel edges or self-loops,
every input and output wire is connected to a spider
and every spider has at most one input or output wire.
The following derivable rules are key to this:
\begin{equation}\label{eq:maintain_graph_like}
	\begin{tikzpicture}
	\begin{pgfonlayer}{nodelayer}
		\node [style=none] (9) at (0, -2) {...};
		\node [style=none] (10) at (-1, -2.5) {};
		\node [style=Z phase dot] (11) at (0, -1) {$\beta$};
		\node [style=none] (12) at (1, -2.5) {};
		\node [style=none] (13) at (0, 2) {...};
		\node [style=none] (14) at (-1, 2.5) {};
		\node [style=none] (15) at (1, 2.5) {};
		\node [style=Z phase dot] (16) at (0, 1) {$\alpha$};
	\end{pgfonlayer}
	\begin{pgfonlayer}{edgelayer}
		\draw [bend left, looseness=0.75] (11) to (12.center);
		\draw [bend right, looseness=0.75] (11) to (10.center);
		\draw [bend right, looseness=0.75] (16) to (15.center);
		\draw [bend left, looseness=0.75] (16) to (14.center);
		\draw [style=hadamard edge, bend right=60, looseness=1.25] (16) to (11);
		\draw [style=hadamard edge, bend left=60, looseness=1.25] (16) to (11);
	\end{pgfonlayer}
\end{tikzpicture}
 \ = \  
	\begin{tikzpicture}
	\begin{pgfonlayer}{nodelayer}
		\node [style=none] (9) at (0, -2) {...};
		\node [style=none] (10) at (-1, -2.5) {};
		\node [style=Z phase dot] (11) at (0, -1) {$\beta$};
		\node [style=none] (12) at (1, -2.5) {};
		\node [style=none] (13) at (0, 2) {...};
		\node [style=none] (14) at (-1, 2.5) {};
		\node [style=none] (15) at (1, 2.5) {};
		\node [style=Z phase dot] (16) at (0, 1) {$\alpha$};
	\end{pgfonlayer}
	\begin{pgfonlayer}{edgelayer}
		\draw [bend left, looseness=0.75] (11) to (12.center);
		\draw [bend right, looseness=0.75] (11) to (10.center);
		\draw [bend right, looseness=0.75] (16) to (15.center);
		\draw [bend left, looseness=0.75] (16) to (14.center);
	\end{pgfonlayer}
\end{tikzpicture}
 \ ,
	\hspace{50pt}
	\begin{tikzpicture}
	\begin{pgfonlayer}{nodelayer}
		\node [style=none] (13) at (0, 1) {...};
		\node [style=none] (14) at (-1, 1.5) {};
		\node [style=none] (15) at (1, 1.5) {};
		\node [style=Z phase dot] (16) at (0, 0) {$\alpha$};
		\node [style=none] (17) at (0, -1.25) {};
	\end{pgfonlayer}
	\begin{pgfonlayer}{edgelayer}
		\draw [bend right, looseness=0.75] (16) to (15.center);
		\draw [bend left, looseness=0.75] (16) to (14.center);
		\draw [in=180, out=-150, looseness=1.50] (16) to (17.center);
		\draw [in=0, out=-30, looseness=1.50] (16) to (17.center);
	\end{pgfonlayer}
\end{tikzpicture}
 \ = \  
	\begin{tikzpicture}
	\begin{pgfonlayer}{nodelayer}
		\node [style=none] (13) at (0, 1) {...};
		\node [style=none] (14) at (-1, 1.5) {};
		\node [style=none] (15) at (1, 1.5) {};
		\node [style=Z phase dot] (16) at (0, 0) {$\alpha$};
	\end{pgfonlayer}
	\begin{pgfonlayer}{edgelayer}
		\draw [bend right, looseness=0.75] (16) to (15.center);
		\draw [bend left, looseness=0.75] (16) to (14.center);
	\end{pgfonlayer}
\end{tikzpicture}
 \ ,
	\hspace{50pt}
	\begin{tikzpicture}
	\begin{pgfonlayer}{nodelayer}
		\node [style=none] (13) at (0, 1) {...};
		\node [style=none] (14) at (-1, 1.5) {};
		\node [style=none] (15) at (1, 1.5) {};
		\node [style=Z phase dot] (16) at (0, 0) {$\alpha$};
		\node [style=none] (17) at (0, -1.25) {};
	\end{pgfonlayer}
	\begin{pgfonlayer}{edgelayer}
		\draw [bend right, looseness=0.75] (16) to (15.center);
		\draw [bend left, looseness=0.75] (16) to (14.center);
		\draw [style=hadamard edge, in=180, out=-150, looseness=1.50] (16) to (17.center);
		\draw [style=hadamard edge, in=0, out=-30, looseness=1.50] (16) to (17.center);
	\end{pgfonlayer}
\end{tikzpicture}
 \ = \  
	\begin{tikzpicture}
	\begin{pgfonlayer}{nodelayer}
		\node [style=none] (13) at (0, 1) {...};
		\node [style=none] (14) at (-1, 1.5) {};
		\node [style=none] (15) at (1, 1.5) {};
		\node [style=Z phase dot] (16) at (0, 0) {$\alpha + \pi$};
	\end{pgfonlayer}
	\begin{pgfonlayer}{edgelayer}
		\draw [bend right, looseness=0.75] (16) to (15.center);
		\draw [bend left, looseness=0.75] (16) to (14.center);
	\end{pgfonlayer}
\end{tikzpicture}

\end{equation}
Then the following two rewrite rules, derived via
\emph{local complementation} and \emph{pivoting},
can be used to eliminate spiders:
	\begin{equation}\label{eq:eliminate_clifford}
		\begin{tikzpicture}
	\begin{pgfonlayer}{nodelayer}
		\node [style=Z phase dot] (0) at (-5, 1) {$\ \pm\frac\pi2\ $};
		\node [style=Z phase dot] (2) at (-7.5, 0.5) {$\alpha_1$};
		\node [style=Z phase dot] (4) at (-2.25, 0.5) {$\alpha_n$};
		\node [style=none] (5) at (-8.5, -0.5) {};
		\node [style=none] (6) at (-7.5, -0.5) {};
		\node [style=none] (7) at (-2.25, -0.5) {};
		\node [style=none] (8) at (-1.25, -0.5) {};
		\node [style=none] (9) at (-5, -0.25) {...};
		\node [style=none] (10) at (-8, -0.5) {...};
		\node [style=none] (11) at (-1.75, -0.5) {...};
		\node [style=none] (16) at (0.25, -0.75) {$=$};
		\node [style=none] (17) at (10, -0.5) {};
		\node [style=none] (18) at (1.75, -0.5) {};
		\node [style=none] (19) at (9.5, -0.5) {...};
		\node [style=none] (20) at (9, -0.5) {};
		\node [style=Z phase dot] (24) at (2.75, 0.5) {$\ \alpha_1\!\mp\!\frac\pi2\ $};
		\node [style=none] (26) at (2.75, -0.5) {};
		\node [style=none] (27) at (2.25, -0.5) {...};
		\node [style=Z phase dot] (30) at (9, 0.5) {$\ \alpha_n \!\mp\! \frac\pi2\ $};
		\node [style=none] (31) at (-6.5, -2) {};
		\node [style=Z phase dot] (32) at (-6.5, -1) {$\alpha_2$};
		\node [style=none] (33) at (-7, -2) {...};
		\node [style=none] (34) at (-7.5, -2) {};
		\node [style=Z phase dot] (35) at (-3.25, -1) {$\ \alpha_{n\!-\!1}\ $};
		\node [style=none] (36) at (-3.25, -2) {};
		\node [style=none] (37) at (-2.25, -2) {};
		\node [style=none] (38) at (-2.75, -2) {...};
		\node [style=Z phase dot] (39) at (4, -1) {$\ \alpha_2\!\mp\!\frac\pi2\ $};
		\node [style=none] (40) at (3.5, -2) {...};
		\node [style=none] (41) at (3, -2) {};
		\node [style=none] (42) at (4, -2) {};
		\node [style=Z phase dot] (43) at (7.75, -1) {$\ \alpha_{n\!-\!1} \!\mp\! \frac\pi2\ $};
		\node [style=none] (44) at (8.75, -2) {};
		\node [style=none] (45) at (8.25, -2) {...};
		\node [style=none] (46) at (7.75, -2) {};
		\node [style=none] (47) at (6, 0) {...};
	\end{pgfonlayer}
	\begin{pgfonlayer}{edgelayer}
		\draw (5.center) to (2);
		\draw (6.center) to (2);
		\draw (7.center) to (4);
		\draw (8.center) to (4);
		\draw [style=hadamard edge] (2) to (0);
		\draw [style=hadamard edge] (4) to (0);
		\draw (18.center) to (24);
		\draw (26.center) to (24);
		\draw (20.center) to (30);
		\draw (17.center) to (30);
		\draw (34.center) to (32);
		\draw (31.center) to (32);
		\draw (36.center) to (35);
		\draw (37.center) to (35);
		\draw (41.center) to (39);
		\draw (42.center) to (39);
		\draw (46.center) to (43);
		\draw (44.center) to (43);
		\draw [style=hadamard edge] (24) to (30);
		\draw [style=hadamard edge] (30) to (43);
		\draw [style=hadamard edge] (43) to (39);
		\draw [style=hadamard edge] (39) to (24);
		\draw [style=hadamard edge] (24) to (43);
		\draw [style=hadamard edge] (39) to (30);
		\draw [style=hadamard edge] (0) to (32);
		\draw [style=hadamard edge] (0) to (35);
	\end{pgfonlayer}
\end{tikzpicture}

	\end{equation}
	\begin{equation}\label{eq:eliminate_pauli}
		\begin{tikzpicture}
	\begin{pgfonlayer}{nodelayer}
		\node [style=Z phase dot] (0) at (-11.5, 1.5) {$j\pi$};
		\node [style=Z phase dot] (2) at (-13.25, 1) {$\alpha_1$};
		\node [style=none] (16) at (-3.5, 0) {$=$};
		\node [style=Z phase dot] (32) at (-13.25, -0.75) {$\alpha_n$};
		\node [style=Z phase dot] (37) at (-9, -2.25) {$\beta_1$};
		\node [style=Z phase dot] (41) at (-9, -0.5) {$\beta_n$};
		\node [style=Z phase dot] (45) at (-5, 1) {$\gamma_1$};
		\node [style=Z phase dot] (49) at (-5, -0.75) {$\gamma_n$};
		\node [style=Z phase dot] (51) at (-6.5, 1.5) {$k\pi$};
		\node [style=none, rotate=90] (52) at (-13.25, 0.125) {...};
		\node [style=none, rotate=90] (53) at (-9, -1.375) {...};
		\node [style=none, rotate=90] (54) at (-5, 0.125) {...};
		\node [style=Z phase dot] (55) at (-1.25, -0.25) {$\ \alpha_n+k\pi\ $};
		\node [style=Z phase dot] (56) at (3.5, -1.5) {$\ \beta_n+(j+k+1)\pi\ $};
		\node [style=none, rotate=90] (57) at (3.5, -2.375) {...};
		\node [style=Z phase dot] (58) at (3.5, -3.25) {$\ \beta_1+(j+k+1)\pi\ $};
		\node [style=Z phase dot] (59) at (8.5, 1.5) {$\ \gamma_1+j\pi\ $};
		\node [style=Z phase dot] (60) at (-1.25, 1.5) {$\ \alpha_1+k\pi\ $};
		\node [style=none, rotate=90] (61) at (8.5, 0.625) {...};
		\node [style=none, rotate=90] (62) at (-1.25, 0.625) {...};
		\node [style=Z phase dot] (63) at (8.5, -0.25) {$\ \gamma_n+j\pi\ $};
		\node [style=none] (64) at (-12.75, 1.75) {};
		\node [style=none] (65) at (-13.75, 1.75) {};
		\node [style=none] (66) at (-13.25, 1.75) {...};
		\node [style=none] (67) at (-13.75, -1.5) {};
		\node [style=none] (68) at (-13.25, -1.5) {...};
		\node [style=none] (69) at (-12.75, -1.5) {};
		\node [style=none] (70) at (-9.5, -3) {};
		\node [style=none] (71) at (-9, -3) {...};
		\node [style=none] (72) at (-8.5, -3) {};
		\node [style=none] (73) at (-9.5, 0.25) {};
		\node [style=none] (74) at (-9, 0.25) {...};
		\node [style=none] (75) at (-8.5, 0.25) {};
		\node [style=none] (76) at (-5.5, 1.75) {};
		\node [style=none] (77) at (-5, 1.75) {...};
		\node [style=none] (78) at (-4.5, 1.75) {};
		\node [style=none] (79) at (-5, -1.5) {...};
		\node [style=none] (80) at (-5.5, -1.5) {};
		\node [style=none] (81) at (-4.5, -1.5) {};
		\node [style=none] (82) at (-1.75, 2.25) {};
		\node [style=none] (83) at (-1.25, 2.25) {...};
		\node [style=none] (84) at (-0.75, 2.25) {};
		\node [style=none] (85) at (-1.25, -1) {...};
		\node [style=none] (86) at (-1.75, -1) {};
		\node [style=none] (87) at (-0.75, -1) {};
		\node [style=none] (88) at (3, -0.75) {};
		\node [style=none] (89) at (3.5, -0.75) {...};
		\node [style=none] (90) at (4, -0.75) {};
		\node [style=none] (91) at (3.5, -4) {...};
		\node [style=none] (92) at (3, -4) {};
		\node [style=none] (93) at (4, -4) {};
		\node [style=none] (94) at (8, 2.25) {};
		\node [style=none] (95) at (8.5, 2.25) {...};
		\node [style=none] (96) at (9, 2.25) {};
		\node [style=none] (97) at (8.5, -1) {...};
		\node [style=none] (98) at (8, -1) {};
		\node [style=none] (99) at (9, -1) {};
	\end{pgfonlayer}
	\begin{pgfonlayer}{edgelayer}
		\draw [style=hadamard edge] (2) to (0);
		\draw [style=hadamard edge] (0) to (32);
		\draw [style=hadamard edge] (0) to (51);
		\draw [style=hadamard edge] (0) to (37);
		\draw [style=hadamard edge] (0) to (41);
		\draw [style=hadamard edge] (51) to (37);
		\draw [style=hadamard edge] (51) to (41);
		\draw [style=hadamard edge] (51) to (45);
		\draw [style=hadamard edge] (51) to (49);
		\draw [style=hadamard edge] (60) to (56);
		\draw [style=hadamard edge] (60) to (58);
		\draw [style=hadamard edge] (55) to (56);
		\draw [style=hadamard edge] (55) to (58);
		\draw [style=hadamard edge] (59) to (56);
		\draw [style=hadamard edge] (59) to (58);
		\draw [style=hadamard edge] (56) to (63);
		\draw [style=hadamard edge] (63) to (58);
		\draw [style=hadamard edge] (60) to (59);
		\draw [style=hadamard edge] (60) to (63);
		\draw [style=hadamard edge] (55) to (63);
		\draw [style=hadamard edge] (55) to (59);
		\draw (64.center) to (2);
		\draw (2) to (65.center);
		\draw (67.center) to (32);
		\draw (69.center) to (32);
		\draw (70.center) to (37);
		\draw (72.center) to (37);
		\draw (41) to (75.center);
		\draw (41) to (73.center);
		\draw (45) to (78.center);
		\draw (45) to (76.center);
		\draw (49) to (80.center);
		\draw (49) to (81.center);
		\draw (60) to (82.center);
		\draw (60) to (84.center);
		\draw (59) to (94.center);
		\draw (59) to (96.center);
		\draw (98.center) to (63);
		\draw (99.center) to (63);
		\draw (86.center) to (55);
		\draw (87.center) to (55);
		\draw (56) to (88.center);
		\draw (56) to (90.center);
		\draw (92.center) to (58);
		\draw (93.center) to (58);
	\end{pgfonlayer}
\end{tikzpicture}

	\end{equation}
For more details,
see \cite[Section 4]{graph_theoretic_simplification}.
Eq.~\eqref{eq:eliminate_clifford} says that we can remove any spider with phase $\pm\frac{\pi}{2}$ at the cost of performing a local complementation at said spider.
Furthermore,
Eq.~\eqref{eq:eliminate_pauli} says we can remove any pair of spiders with phases in $\{0, \pi\}$ connected by a Hadamard edge at the cost of performing a pivot along said edge.
After each application of \eqref{eq:eliminate_clifford} or \eqref{eq:eliminate_pauli}, we can use \eqref{eq:maintain_graph_like} to remove any parallel Hadamard edges and ensure the diagram remains graph-like.


\section{Qutrit ZX-Calculus}\label{app:qutrit_zx_calculus}

We now give a fuller definition of the qutrit ZX-calculus. Our presentation aims for clarity and accessibility; for a more rigourous description, see \cite{harny_completeness}.

\begin{definition}\label{def:qutrit_ZX_rules}
	The \textit{qutrit ZX-calculus} is a graphical calculus generated by the following diagrams, where $\alpha, \beta \in [0, 2 \pi)$:

	\begin{equation}
		\left\{ \quad
			\input{qutrit_generators/spiders/Z_a_b.tikz} \quad , \qquad 
			\input{qutrit_generators/spiders/X_a_b.tikz} \quad , \qquad
			\begin{tikzpicture}
	\begin{pgfonlayer}{nodelayer}
		\node [style=none] (3) at (0, 1) {};
		\node [style=none] (4) at (0, -1) {};
		\node [style=qutrit hadamard] (5) at (0, 0) {};
	\end{pgfonlayer}
	\begin{pgfonlayer}{edgelayer}
		\draw (3.center) to (4.center);
	\end{pgfonlayer}
\end{tikzpicture}
 \quad , \qquad
			\begin{tikzpicture}
	\begin{pgfonlayer}{nodelayer}
		\node [style=none] (0) at (-1, 1) {};
		\node [style=none] (1) at (1, -1) {};
		\node [style=none] (2) at (-1, -1) {};
		\node [style=none] (3) at (1, 1) {};
	\end{pgfonlayer}
	\begin{pgfonlayer}{edgelayer}
		\draw [in=-90, out=90, looseness=0.75] (2.center) to (3.center);
		\draw [in=90, out=-90, looseness=0.75] (0.center) to (1.center);
	\end{pgfonlayer}
\end{tikzpicture}
 \quad , \qquad
			\begin{tikzpicture}
	\begin{pgfonlayer}{nodelayer}
		\node [style=none] (3) at (0, 1) {};
		\node [style=none] (4) at (0, -1) {};
	\end{pgfonlayer}
	\begin{pgfonlayer}{edgelayer}
		\draw (3.center) to (4.center);
	\end{pgfonlayer}
\end{tikzpicture}

		\quad \right\}
	\end{equation}

	and their adjoints $(-)^\dagger$. Adjoints are found by swapping inputs and outputs and negating any decorations - recall negation is mod $2\pi$ for general spider phases, and mod $3$ for integer spider phases and Hadamard boxes. Thus the two rightmost generators are self-adjoint, whereas the first three satisfy: 

	\begin{equation}
		\left(\ \scalebox{0.9}{\input{qutrit_generators/spiders/Z_a_b_labelled.tikz}}\ \right)^\dagger = \ \scalebox{0.9}{\input{qutrit_generators/spiders/Z_a_b_adjoint_labelled.tikz}} \quad , 
		\hspace{15pt}
		\left(\ \scalebox{0.9}{\input{qutrit_generators/spiders/X_a_b_labelled.tikz}}\ \right)^\dagger = \ \scalebox{0.9}{\input{qutrit_generators/spiders/X_a_b_adjoint_labelled.tikz}} \quad , 
		\hspace{15pt}
		\left(~~~~\right)^\dagger = \ 
	\end{equation}

	These generators can be composed in parallel ($\otimes$) and sequentially ($\circ$), and the resulting diagrams are governed by the rewrite rules in Figure \ref{fig:qutrit_ZX_rules}, wherein addition is modulo $2\pi$. The fusion rule $\qutritRuleFusion$ applies to spiders of the same colour connected by at least one wire. Importantly, all the rules hold under taking adjoints, where for diagrams $D$ and $E$ we have:

	\begin{equation}
		(D \otimes E)^\dagger = D^\dagger \otimes E^\dagger\ , 
		\hspace{50pt}
		(D \circ E)^\dagger = D^\dagger \circ E^\dagger
	\end{equation}

	Furthermore, it can be derived that all but the commutation equations $\qutritRuleCommute$ and the colour change equations $\qutritRuleColourChange$ continue to hold when the roles of green and red (i.e. $Z$ and $X$) are interchanged. For these four exceptions, however, analogous equations can be derived from the existing ones; for example, the corresponding colour change equations will be relevant for us later.
\end{definition}
	
\begin{proposition}
	The following equations are derivable in the qutrit ZX-calculus:
	\begin{equation}\label{eq:derived_colour_change}
		\begin{tikzpicture}
	\begin{pgfonlayer}{nodelayer}
		\node [style=none] (0) at (0, 1) {...};
		\node [style=none] (1) at (-1, 1.75) {};
		\node [style=none] (2) at (1, 1.75) {};
		\node [style=none] (3) at (0, -1) {...};
		\node [style=none] (4) at (-1, -1.75) {};
		\node [style=qutrit X phase] (5) at (0, 0) {$\alpha$ \nodepart{lower} $\beta$};
		\node [style=none] (6) at (1, -1.75) {};
	\end{pgfonlayer}
	\begin{pgfonlayer}{edgelayer}
		\draw [in=-45, out=90, looseness=0.75] (6.center) to (5);
		\draw [in=-135, out=90, looseness=0.75] (4.center) to (5);
		\draw [in=45, out=-90, looseness=0.75] (2.center) to (5);
		\draw [in=135, out=-90, looseness=0.75] (1.center) to (5);
	\end{pgfonlayer}
\end{tikzpicture}
 \quad = \quad \begin{tikzpicture}
	\begin{pgfonlayer}{nodelayer}
		\node [style=none] (1) at (0, 1) {...};
		\node [style=none] (8) at (0, -1) {...};
		\node [style=qutrit Z phase] (10) at (0, 0) {$\alpha$ \nodepart{lower} $\beta$};
		\node [style=qutrit hadamard] (35) at (-1, 1.25) {};
		\node [style=qutrit hadamard] (36) at (1, 1.25) {};
		\node [style=none] (37) at (-1, 1.75) {};
		\node [style=none] (38) at (1, 1.75) {};
		\node [style=qutrit hadamard adjoint] (39) at (-1, -1.25) {};
		\node [style=qutrit hadamard adjoint] (40) at (1, -1.25) {};
		\node [style=none] (41) at (-1, -1.75) {};
		\node [style=none] (42) at (1, -1.75) {};
	\end{pgfonlayer}
	\begin{pgfonlayer}{edgelayer}
		\draw (37.center) to (35);
		\draw (38.center) to (36);
		\draw [bend right=15] (10) to (36);
		\draw [bend left=15] (10) to (35);
		\draw (41.center) to (39);
		\draw (42.center) to (40);
		\draw [bend left=15] (10) to (40);
		\draw [bend right=15] (10) to (39);
	\end{pgfonlayer}
\end{tikzpicture}
 \quad ,
		\hspace{50pt}
		\begin{tikzpicture}
	\begin{pgfonlayer}{nodelayer}
		\node [style=none] (0) at (0, 1) {...};
		\node [style=none] (1) at (-1, 1.75) {};
		\node [style=none] (2) at (1, 1.75) {};
		\node [style=none] (3) at (0, -1) {...};
		\node [style=none] (4) at (-1, -1.75) {};
		\node [style=qutrit X phase] (5) at (0, 0) {$\beta$ \nodepart{lower} $\alpha$};
		\node [style=none] (6) at (1, -1.75) {};
	\end{pgfonlayer}
	\begin{pgfonlayer}{edgelayer}
		\draw [in=-45, out=90, looseness=0.75] (6.center) to (5);
		\draw [in=-135, out=90, looseness=0.75] (4.center) to (5);
		\draw [in=45, out=-90, looseness=0.75] (2.center) to (5);
		\draw [in=135, out=-90, looseness=0.75] (1.center) to (5);
	\end{pgfonlayer}
\end{tikzpicture}
 \quad = \quad \begin{tikzpicture}
	\begin{pgfonlayer}{nodelayer}
		\node [style=none] (1) at (0, 1) {...};
		\node [style=none] (8) at (0, -1) {...};
		\node [style=qutrit Z phase] (10) at (0, 0) {$\alpha$ \nodepart{lower} $\beta$};
		\node [style=qutrit hadamard adjoint] (35) at (-1, 1.25) {};
		\node [style=qutrit hadamard adjoint] (36) at (1, 1.25) {};
		\node [style=none] (37) at (-1, 1.75) {};
		\node [style=none] (38) at (1, 1.75) {};
		\node [style=qutrit hadamard] (39) at (-1, -1.25) {};
		\node [style=qutrit hadamard] (40) at (1, -1.25) {};
		\node [style=none] (41) at (-1, -1.75) {};
		\node [style=none] (42) at (1, -1.75) {};
	\end{pgfonlayer}
	\begin{pgfonlayer}{edgelayer}
		\draw (37.center) to (35);
		\draw (38.center) to (36);
		\draw [bend right=15] (10) to (36);
		\draw [bend left=15] (10) to (35);
		\draw (41.center) to (39);
		\draw (42.center) to (40);
		\draw [bend left=15] (10) to (40);
		\draw [bend right=15] (10) to (39);
	\end{pgfonlayer}
\end{tikzpicture}

	\end{equation}
	\begin{proof}
		Add $H$- and $H^\dagger$-boxes to both sides of the original colour change equations in such a way that we can then cancel Hadamards on the legs of the red spiders via $\qutritRuleHadamard$.
	\end{proof}
\end{proposition}

\begin{lemma}\label{lem:h_edges_are_mod_3_appendix} 
	\HEdgesAreModThreeStatement
	\begin{proof}
		It is shown in \cite[Lemma 2.8]{qutrit_euler} that the qutrit ZX-calculus satisfies the following `Hopf law':
			\begin{equation}\label{eq:qutrit_hopf}
				\begin{tikzpicture}
	\begin{pgfonlayer}{nodelayer}
		\node [style=none] (17) at (-0.75, 2.25) {};
		\node [style=none] (18) at (0.75, 2.25) {};
		\node [style=none] (19) at (0, 2) {...};
		\node [style=qutrit Z phase] (20) at (0, 1) {$\alpha$ \nodepart{lower} $\beta$};
		\node [style=none] (21) at (-0.75, -2.25) {};
		\node [style=none] (22) at (0.75, -2.25) {};
		\node [style=none] (23) at (0, -2) {...};
		\node [style=qutrit X phase] (24) at (0, -1) {$\gamma$ \nodepart{lower} $\delta$};
	\end{pgfonlayer}
	\begin{pgfonlayer}{edgelayer}
		\draw [bend right=15] (17.center) to (20);
		\draw [bend right=15] (20) to (18.center);
		\draw [bend left=15] (21.center) to (24);
		\draw [bend left=15] (24) to (22.center);
		\draw [bend right=60] (20) to (24);
		\draw [bend left=60] (20) to (24);
		\draw (20) to (24);
	\end{pgfonlayer}
\end{tikzpicture}
 \quad = \quad 
				\begin{tikzpicture}
	\begin{pgfonlayer}{nodelayer}
		\node [style=none] (25) at (-0.75, 2.25) {};
		\node [style=none] (26) at (0.75, 2.25) {};
		\node [style=none] (27) at (0, 2) {...};
		\node [style=qutrit Z phase] (28) at (0, 1) {$\alpha$ \nodepart{lower} $\beta$};
		\node [style=none] (29) at (0, -2) {...};
		\node [style=qutrit X phase] (30) at (0, -1) {$\gamma$ \nodepart{lower} $\delta$};
		\node [style=none] (31) at (-0.75, -2.25) {};
		\node [style=none] (32) at (0.75, -2.25) {};
	\end{pgfonlayer}
	\begin{pgfonlayer}{edgelayer}
		\draw [bend right=15] (25.center) to (28);
		\draw [bend right=15] (28) to (26.center);
		\draw [bend left=15] (31.center) to (30);
		\draw [bend left=15] (30) to (32.center);
	\end{pgfonlayer}
\end{tikzpicture}

			\end{equation}
		Therefore, to prove the first equality of the first equation, we can argue as follows:
			\begin{equation}
				\input{hadamard_lemmas/3_h_edges_vanish/1.tikz} \quad \xeq{\qutritRuleHadamard} \quad
				\input{hadamard_lemmas/3_h_edges_vanish/2.tikz} \quad \xeq{\eqref{eq:derived_colour_change}} \quad
				\input{hadamard_lemmas/3_h_edges_vanish/3.tikz} \quad \xeq{\eqref{eq:qutrit_hopf}} \quad
				\input{hadamard_lemmas/3_h_edges_vanish/4.tikz} \quad \xeq{\qutritRuleColourChange} \quad
				\input{hadamard_lemmas/3_h_edges_vanish/disconnected.tikz}
			\end{equation}
		The second equality of the first equation is proved analogously, and the other two equations are proved in \cite[Lemma 3.4]{gong2017equivalence}.
		\end{proof}
\end{lemma}
\section{Pivoting in the Qutrit ZX-Calculus}

Next we prove the pivot equality from Theorem~\ref{thm:local_pivot_equality}. We require the following: since $\qutritRuleEuler$ holds under taking adjoints, and swapping the roles of red and green, we have:

\begin{equation}\label{eq:qutrit_hadamard_decompositions}
	\begin{tikzpicture}
	\begin{pgfonlayer}{nodelayer}
		\node [style=none] (3) at (0, 1) {};
		\node [style=none] (4) at (0, -1) {};
		\node [style=qutrit hadamard] (5) at (0, 0) {};
	\end{pgfonlayer}
	\begin{pgfonlayer}{edgelayer}
		\draw (3.center) to (4.center);
	\end{pgfonlayer}
\end{tikzpicture}
 \ = \ 
	\begin{tikzpicture}
	\begin{pgfonlayer}{nodelayer}
		\node [style=none] (3) at (0, 2) {};
		\node [style=none] (4) at (0, -2) {};
		\node [style=qutrit X phase] (5) at (0, 0) {$2$ \nodepart{lower} $2$};
		\node [style=qutrit Z phase] (6) at (0, 1.25) {$2$ \nodepart{lower} $2$};
		\node [style=qutrit Z phase] (7) at (0, -1.25) {$2$ \nodepart{lower} $2$};
	\end{pgfonlayer}
	\begin{pgfonlayer}{edgelayer}
		\draw (3.center) to (4.center);
	\end{pgfonlayer}
\end{tikzpicture}
 \ = \ 
	\begin{tikzpicture}
	\begin{pgfonlayer}{nodelayer}
		\node [style=none] (3) at (0, 2) {};
		\node [style=none] (4) at (0, -2) {};
		\node [style=qutrit Z phase] (5) at (0, 0) {$2$ \nodepart{lower} $2$};
		\node [style=qutrit X phase] (6) at (0, 1.25) {$2$ \nodepart{lower} $2$};
		\node [style=qutrit X phase] (7) at (0, -1.25) {$2$ \nodepart{lower} $2$};
	\end{pgfonlayer}
	\begin{pgfonlayer}{edgelayer}
		\draw (3.center) to (4.center);
	\end{pgfonlayer}
\end{tikzpicture}
 ~~ ,
	\hspace{50pt} 
	\begin{tikzpicture}
	\begin{pgfonlayer}{nodelayer}
		\node [style=qutrit hadamard adjoint] (0) at (0, 0) {};
		\node [style=none] (1) at (0, 1) {};
		\node [style=none] (2) at (0, -1) {};
	\end{pgfonlayer}
	\begin{pgfonlayer}{edgelayer}
		\draw (2.center) to (1.center);
	\end{pgfonlayer}
\end{tikzpicture}
 \ = \
	\begin{tikzpicture}
	\begin{pgfonlayer}{nodelayer}
		\node [style=none] (3) at (0, 2) {};
		\node [style=none] (4) at (0, -2) {};
		\node [style=qutrit Z phase] (5) at (0, 0) {$1$ \nodepart{lower} $1$};
		\node [style=qutrit X phase] (6) at (0, 1.25) {$1$ \nodepart{lower} $1$};
		\node [style=qutrit X phase] (7) at (0, -1.25) {$1$ \nodepart{lower} $1$};
	\end{pgfonlayer}
	\begin{pgfonlayer}{edgelayer}
		\draw (3.center) to (4.center);
	\end{pgfonlayer}
\end{tikzpicture}
 \ = \ 
	\begin{tikzpicture}
	\begin{pgfonlayer}{nodelayer}
		\node [style=none] (3) at (0, 2) {};
		\node [style=none] (4) at (0, -2) {};
		\node [style=qutrit X phase] (5) at (0, 0) {$1$ \nodepart{lower} $1$};
		\node [style=qutrit Z phase] (6) at (0, 1.25) {$1$ \nodepart{lower} $1$};
		\node [style=qutrit Z phase] (7) at (0, -1.25) {$1$ \nodepart{lower} $1$};
	\end{pgfonlayer}
	\begin{pgfonlayer}{edgelayer}
		\draw (3.center) to (4.center);
	\end{pgfonlayer}
\end{tikzpicture}

\end{equation}

\begin{theorem}\label{thm:local_pivot_equality_appendix} \textbf{/\ Theorem~\ref{thm:local_pivot_equality}.} 
	\qutritPivotEqualityStatement
	\begin{proof}
		There are four cases ($a, w_{i,j} \in \{1,2\}$), which split into two pairs of symmetric cases: $a = w_{i,j}$ and $a \neq w_{i,j}$. We show just one case - the $1$-pivot along $ij$ of weight $1$ - the remaining cases being analogous. We again employ the !-notation \eqref{eq:bang_box}. An asterisk \textcolor{red}{$*_a$} next to a spider means that at the next step an $a$-local complementation will be performed at this spider.
		\begingroup
			\allowdisplaybreaks
			\setlength{\jot}{20pt}
			\begin{alignat*}{2}
				&\quad &&\input{proper_local_pivot/a_1/w_1/bang/1.tikz} \\
				&\xeq{\ref{thm:local_comp_equality}} \quad
				&&\input{proper_local_pivot/a_1/w_1/bang/2.tikz} \\
				&\xeq{\ref{thm:local_comp_equality}} \quad
				&&\input{proper_local_pivot/a_1/w_1/bang/3.tikz} \\
				&\xeq{\ref{thm:local_comp_equality}} \quad
				&&\input{proper_local_pivot/a_1/w_1/bang/4.tikz} \\
				&\xeqq{\eqref{eq:qutrit_hadamard_decompositions}}{\qutritRuleFusion} \quad
				&&\input{proper_local_pivot/a_1/w_1/bang/5.tikz}
			\end{alignat*}
		\endgroup

	\end{proof}
\end{theorem}
\section{Spider Elimination Rules for Qutrits}
Now we prove the three elimination theorems for $\mathcal{P}$-, $\mathcal{N}$- and \Mspiders. All three require the following two lemmas.

\begin{lemma}\label{lem:leg_flip}
	The following `leg flip' equation holds in the qutrit ZX-calculus:
	\begin{equation}
		\begin{tikzpicture}
	\begin{pgfonlayer}{nodelayer}
		\node [style=none] (16) at (-0.25, 1) {...};
		\node [style=none] (17) at (-1.25, 1.5) {};
		\node [style=none] (18) at (0.5, 1.5) {};
		\node [style=none] (19) at (0, -1) {...};
		\node [style=none] (20) at (-1.25, -1.5) {};
		\node [style=none] (21) at (1.25, 1.5) {};
		\node [style=qutrit X phase] (22) at (1.25, 1.5) {$\alpha$ \nodepart{lower} $\beta$};
		\node [style=none] (23) at (1.25, -1.5) {};
		\node [style=Z dot] (24) at (0, 0) {};
	\end{pgfonlayer}
	\begin{pgfonlayer}{edgelayer}
		\draw [bend left, looseness=0.75] (24) to (23.center);
		\draw [bend right, looseness=0.75] (24) to (21.center);
		\draw [bend right, looseness=0.75] (24) to (20.center);
		\draw [bend right=15, looseness=0.75] (24) to (18.center);
		\draw [bend left, looseness=0.75] (24) to (17.center);
	\end{pgfonlayer}
\end{tikzpicture}
 = \begin{tikzpicture}
	\begin{pgfonlayer}{nodelayer}
		\node [style=none] (3) at (0, 1) {...};
		\node [style=none] (4) at (-1.25, 1.5) {};
		\node [style=none] (6) at (-0.25, -1) {...};
		\node [style=none] (7) at (-1.25, -1.5) {};
		\node [style=none] (9) at (0.5, -1.5) {};
		\node [style=none] (10) at (1.25, 1.5) {};
		\node [style=Z dot] (11) at (0, 0) {};
		\node [style=qutrit X phase] (14) at (1.25, -1.5) {$\beta$ \nodepart{lower} $\alpha$};
	\end{pgfonlayer}
	\begin{pgfonlayer}{edgelayer}
		\draw [bend right, looseness=0.75] (11) to (10.center);
		\draw [bend left=15, looseness=0.75] (11) to (9.center);
		\draw [bend right, looseness=0.75] (11) to (7.center);
		\draw [bend left, looseness=0.75] (11) to (4.center);
		\draw [bend left, looseness=0.75] (11) to (14);
	\end{pgfonlayer}
\end{tikzpicture}

	\end{equation}
	\begin{proof}
		\begin{equation}
			 \ \xeq{\qutritRuleFusion} \ 
			\begin{tikzpicture}
	\begin{pgfonlayer}{nodelayer}
		\node [style=none] (3) at (0, 1) {...};
		\node [style=none] (4) at (-1.25, 1.5) {};
		\node [style=none] (6) at (-0.25, -1) {...};
		\node [style=none] (7) at (-1.25, -1.5) {};
		\node [style=none] (9) at (0.5, -1.5) {};
		\node [style=none] (10) at (1.25, 1.5) {};
		\node [style=Z dot] (11) at (0, 0) {};
		\node [style=Z dot] (12) at (1.75, -1.5) {};
		\node [style=X dot] (13) at (3.5, 0) {};
		\node [style=X dot] (14) at (5.275, -1.5) {};
		\node [style=qutrit X phase] (15) at (5.275, -1.5) {$\alpha$ \nodepart{lower} $\beta$};
	\end{pgfonlayer}
	\begin{pgfonlayer}{edgelayer}
		\draw [in=180, out=-15] (11) to (12);
		\draw [bend right, looseness=0.75] (11) to (10.center);
		\draw [bend left=15, looseness=0.75] (11) to (9.center);
		\draw [bend right, looseness=0.75] (11) to (7.center);
		\draw [bend left, looseness=0.75] (11) to (4.center);
		\draw [in=-180, out=0] (12) to (13);
		\draw [in=0, out=180] (14) to (13);
	\end{pgfonlayer}
\end{tikzpicture}
 \ \xeq{\qutritRuleSnake} \ 
			\begin{tikzpicture}
	\begin{pgfonlayer}{nodelayer}
		\node [style=none] (3) at (0, 1) {...};
		\node [style=none] (4) at (-1.25, 1.5) {};
		\node [style=none] (6) at (-0.25, -1) {...};
		\node [style=none] (7) at (-1.25, -1.5) {};
		\node [style=none] (9) at (0.5, -1.5) {};
		\node [style=none] (10) at (1.25, 1.5) {};
		\node [style=qutrit X phase] (14) at (1.25, -3.25) {$\alpha$ \nodepart{lower} $\beta$};
		\node [style=qutrit hadamard] (15) at (1.25, -2.25) {};
		\node [style=qutrit hadamard] (16) at (1.25, -1.5) {};
		\node [style=none] (17) at (1.25, -1.5) {};
		\node [style=Z dot] (18) at (0, 0) {};
	\end{pgfonlayer}
	\begin{pgfonlayer}{edgelayer}
		\draw [bend left, looseness=0.75] (18) to (17.center);
		\draw [bend right, looseness=0.75] (18) to (10.center);
		\draw [bend left=15, looseness=0.75] (18) to (9.center);
		\draw [bend right, looseness=0.75] (18) to (7.center);
		\draw [bend left, looseness=0.75] (18) to (4.center);
		\draw (14) to (17.center);
	\end{pgfonlayer}
\end{tikzpicture}
 \ \xeq{\qutritRuleColourChange} \  
			\begin{tikzpicture}
	\begin{pgfonlayer}{nodelayer}
		\node [style=none] (3) at (0, 1) {...};
		\node [style=none] (4) at (-1.25, 1.5) {};
		\node [style=none] (6) at (-0.25, -1) {...};
		\node [style=none] (7) at (-1.25, -1.5) {};
		\node [style=none] (9) at (0.5, -1.5) {};
		\node [style=none] (10) at (1.25, 1.5) {};
		\node [style=qutrit hadamard] (16) at (1.25, -1.5) {};
		\node [style=none] (17) at (1.25, -1.5) {};
		\node [style=Z dot] (18) at (0, 0) {};
		\node [style=qutrit Z phase] (19) at (1.25, -2.5) {$\beta$ \nodepart{lower} $\alpha$};
	\end{pgfonlayer}
	\begin{pgfonlayer}{edgelayer}
		\draw [bend left, looseness=0.75] (18) to (17.center);
		\draw [bend right, looseness=0.75] (18) to (10.center);
		\draw [bend left=15, looseness=0.75] (18) to (9.center);
		\draw [bend right, looseness=0.75] (18) to (7.center);
		\draw [bend left, looseness=0.75] (18) to (4.center);
		\draw (19) to (17.center);
	\end{pgfonlayer}
\end{tikzpicture}
 \ \xeq{\qutritRuleColourChange} \  
			
		\end{equation}
	\end{proof}
\end{lemma}

\begin{lemma}\label{lem:substantial_m_copy}
	The following more substantial `$\mathcal{M}$-copy' rule holds in the qutrit ZX-calculus, for any \Mspider\ state (i.e. $m \in \{0, 1, 2\}$ below):
	\begin{equation}
		\begin{tikzpicture}
	\begin{pgfonlayer}{nodelayer}
		\node [style=qutrit X phase] (0) at (0, -1.75) {$m$ \nodepart{lower} $-m$};
		\node [style=none] (1) at (-1.25, 1.5) {};
		\node [style=none] (3) at (0, 1.25) {...};
		\node [style=qutrit Z phase] (4) at (0, 0) {$\alpha$ \nodepart{lower} $\beta$};
		\node [style=none] (5) at (1.25, 1.5) {};
	\end{pgfonlayer}
	\begin{pgfonlayer}{edgelayer}
		\draw [bend right, looseness=0.75] (4) to (5.center);
		\draw [bend left, looseness=0.75] (4) to (1.center);
		\draw (0) to (4);
	\end{pgfonlayer}
\end{tikzpicture}
 = \begin{tikzpicture}
	\begin{pgfonlayer}{nodelayer}
		\node [style=qutrit X phase] (0) at (-1, -1) {$m$ \nodepart{lower} $-m$};
		\node [style=qutrit X phase] (1) at (1, -1) {$m$ \nodepart{lower} $-m$};
		\node [style=none] (3) at (-1, 1.25) {};
		\node [style=none] (4) at (1, 1.25) {};
		\node [style=none] (5) at (0, 0.5) {...};
	\end{pgfonlayer}
	\begin{pgfonlayer}{edgelayer}
		\draw (0) to (3.center);
		\draw (1) to (4.center);
	\end{pgfonlayer}
\end{tikzpicture}

	\end{equation}
	\begin{proof}
		First we prove that an \Mspider\ with non-trivial phase (i.e. $m \in \{1, 2\}$) satisfies a copy rule exactly like the rule $\qutritRuleZeroCopy$:
		\begin{equation}\label{eq:non_trivial_m_copy}
			\begin{tikzpicture}
	\begin{pgfonlayer}{nodelayer}
		\node [style=qutrit X phase] (2) at (0, -0.5) {$m$ \nodepart{lower} $-m$};
		\node [style=none] (4) at (-0.75, 2) {};
		\node [style=Z dot] (5) at (0, 1) {};
		\node [style=none] (6) at (0.75, 2) {};
	\end{pgfonlayer}
	\begin{pgfonlayer}{edgelayer}
		\draw [bend right, looseness=0.75] (5) to (6.center);
		\draw [bend left, looseness=0.75] (5) to (4.center);
		\draw (2) to (5);
	\end{pgfonlayer}
\end{tikzpicture}
 \ \xeq{\qutritRuleFusion} \ 
			\begin{tikzpicture}
	\begin{pgfonlayer}{nodelayer}
		\node [style=qutrit X phase] (2) at (0, -0.5) {$m$ \nodepart{lower} $-m$};
		\node [style=none] (4) at (-0.75, 2) {};
		\node [style=Z dot] (5) at (0, 1) {};
		\node [style=none] (6) at (0.75, 2) {};
		\node [style=X dot] (7) at (0, -2) {};
	\end{pgfonlayer}
	\begin{pgfonlayer}{edgelayer}
		\draw [bend right, looseness=0.75] (5) to (6.center);
		\draw [bend left, looseness=0.75] (5) to (4.center);
		\draw (2) to (5);
		\draw (7) to (2);
	\end{pgfonlayer}
\end{tikzpicture}
 \ \xeq{\qutritRuleMCopy} \ 
			\begin{tikzpicture}
	\begin{pgfonlayer}{nodelayer}
		\node [style=Z dot] (5) at (0, -1.25) {};
		\node [style=X dot] (7) at (0, -2.25) {};
		\node [style=qutrit X phase] (8) at (-0.75, 0) {$m$ \nodepart{lower} $-m$};
		\node [style=qutrit X phase] (9) at (0.75, 0) {$m$ \nodepart{lower} $-m$};
		\node [style=none] (10) at (-0.75, 1.25) {};
		\node [style=none] (11) at (0.75, 1.25) {};
	\end{pgfonlayer}
	\begin{pgfonlayer}{edgelayer}
		\draw (7) to (5);
		\draw [bend right=15] (5) to (9);
		\draw [bend left=15] (5) to (8);
		\draw (8) to (10.center);
		\draw (9) to (11.center);
	\end{pgfonlayer}
\end{tikzpicture}
 \ \xeq{\qutritRuleZeroCopy} \ 
			\begin{tikzpicture}
	\begin{pgfonlayer}{nodelayer}
		\node [style=X dot] (7) at (-0.75, -1.25) {};
		\node [style=qutrit X phase] (8) at (-0.75, 0) {$m$ \nodepart{lower} $-m$};
		\node [style=qutrit X phase] (9) at (0.75, 0) {$m$ \nodepart{lower} $-m$};
		\node [style=none] (10) at (-0.75, 1.25) {};
		\node [style=none] (11) at (0.75, 1.25) {};
		\node [style=X dot] (12) at (0.75, -1.25) {};
	\end{pgfonlayer}
	\begin{pgfonlayer}{edgelayer}
		\draw (8) to (10.center);
		\draw (9) to (11.center);
		\draw (7) to (8);
		\draw (12) to (9);
	\end{pgfonlayer}
\end{tikzpicture}
 \ \xeq{\qutritRuleFusion} \ 
			\begin{tikzpicture}
	\begin{pgfonlayer}{nodelayer}
		\node [style=qutrit X phase] (8) at (-0.75, 0) {$m$ \nodepart{lower} $-m$};
		\node [style=qutrit X phase] (9) at (0.75, 0) {$m$ \nodepart{lower} $-m$};
		\node [style=none] (10) at (-0.75, 1.25) {};
		\node [style=none] (11) at (0.75, 1.25) {};
	\end{pgfonlayer}
	\begin{pgfonlayer}{edgelayer}
		\draw (8) to (10.center);
		\draw (9) to (11.center);
	\end{pgfonlayer}
\end{tikzpicture}

		\end{equation}
		Then we prove the case $\alpha = \beta = 0$ by induction:
		\begin{equation}\label{eq:m_copy_through_trivial}
			\begin{tikzpicture}
	\begin{pgfonlayer}{nodelayer}
		\node [style=qutrit X phase] (2) at (0, -1) {$m$ \nodepart{lower} $-m$};
		\node [style=none] (4) at (-1, 1.75) {};
		\node [style=none] (6) at (0.5, 1.75) {};
		\node [style=none] (7) at (-0.25, 1.5) {...};
		\node [style=Z dot] (8) at (0, 0.5) {};
		\node [style=none] (9) at (1, 1.75) {};
	\end{pgfonlayer}
	\begin{pgfonlayer}{edgelayer}
		\draw [bend right, looseness=0.75] (8) to (9.center);
		\draw [bend right=15, looseness=0.75] (8) to (6.center);
		\draw [bend left, looseness=0.75] (8) to (4.center);
		\draw (2) to (8);
	\end{pgfonlayer}
\end{tikzpicture}
 \ \xeq{\qutritRuleFusion} \ 
			\begin{tikzpicture}
	\begin{pgfonlayer}{nodelayer}
		\node [style=qutrit X phase] (0) at (0, -1) {$m$ \nodepart{lower} $-m$};
		\node [style=none] (1) at (-1.75, 2.25) {};
		\node [style=none] (3) at (-0.75, 2) {...};
		\node [style=none] (5) at (1.5, 2.25) {};
		\node [style=Z dot] (6) at (0, 0.25) {};
		\node [style=none] (7) at (0.25, 2.25) {};
		\node [style=Z dot] (8) at (-0.75, 1) {};
	\end{pgfonlayer}
	\begin{pgfonlayer}{edgelayer}
		\draw [in=-90, out=15, looseness=0.75] (6) to (5.center);
		\draw (0) to (6);
		\draw [bend right, looseness=0.75] (8) to (7.center);
		\draw [bend left, looseness=0.75] (8) to (1.center);
		\draw [bend left] (6) to (8);
	\end{pgfonlayer}
\end{tikzpicture}
 \ \xeq{\eqref{eq:non_trivial_m_copy}} \ 
			\begin{tikzpicture}
	\begin{pgfonlayer}{nodelayer}
		\node [style=none] (1) at (-1.75, 1.5) {};
		\node [style=none] (3) at (-0.75, 1.25) {...};
		\node [style=none] (5) at (1.25, 1.5) {};
		\node [style=none] (7) at (0.25, 1.5) {};
		\node [style=Z dot] (8) at (-0.75, 0.25) {};
		\node [style=qutrit X phase] (9) at (-0.75, -1) {$m$ \nodepart{lower} $-m$};
		\node [style=qutrit X phase] (10) at (1.25, -1) {$m$ \nodepart{lower} $-m$};
	\end{pgfonlayer}
	\begin{pgfonlayer}{edgelayer}
		\draw [bend right, looseness=0.75] (8) to (7.center);
		\draw [bend left, looseness=0.75] (8) to (1.center);
		\draw (9) to (8);
		\draw (10) to (5.center);
	\end{pgfonlayer}
\end{tikzpicture}
 \ \xeq{\text{ind.}} \ 
			\begin{tikzpicture}
	\begin{pgfonlayer}{nodelayer}
		\node [style=none] (3) at (-0.75, 0.25) {...};
		\node [style=none] (5) at (-1.75, 1) {};
		\node [style=qutrit X phase] (10) at (-1.75, -1) {$m$ \nodepart{lower} $-m$};
		\node [style=none] (11) at (1.75, 1) {};
		\node [style=qutrit X phase] (12) at (1.75, -1) {$m$ \nodepart{lower} $-m$};
		\node [style=none] (13) at (0.25, 1) {};
		\node [style=qutrit X phase] (14) at (0.25, -1) {$m$ \nodepart{lower} $-m$};
	\end{pgfonlayer}
	\begin{pgfonlayer}{edgelayer}
		\draw (10) to (5.center);
		\draw (12) to (11.center);
		\draw (14) to (13.center);
	\end{pgfonlayer}
\end{tikzpicture}

		\end{equation}
		Which finally allows us to prove the full statement, where the last equality is just dropping the scalar term:
		\begin{equation}
			 \ \xeq{\qutritRuleFusion} \ 
			\begin{tikzpicture}
	\begin{pgfonlayer}{nodelayer}
		\node [style=qutrit X phase] (0) at (0, -1.75) {$m$ \nodepart{lower} $-m$};
		\node [style=none] (1) at (-1.25, 1.5) {};
		\node [style=none] (2) at (0.5, 1.5) {};
		\node [style=none] (3) at (-0.25, 1.25) {...};
		\node [style=Z dot] (4) at (0, 0) {};
		\node [style=none] (5) at (1.25, 1.5) {};
		\node [style=qutrit Z phase] (6) at (1.25, 1.5) {$\alpha$ \nodepart{lower} $\beta$};
	\end{pgfonlayer}
	\begin{pgfonlayer}{edgelayer}
		\draw [bend right, looseness=0.75] (4) to (5.center);
		\draw [bend right=15, looseness=0.75] (4) to (2.center);
		\draw [bend left, looseness=0.75] (4) to (1.center);
		\draw (0) to (4);
	\end{pgfonlayer}
\end{tikzpicture}
 \ \xeq{\eqref{eq:m_copy_through_trivial}} \ 
			\begin{tikzpicture}
	\begin{pgfonlayer}{nodelayer}
		\node [style=qutrit X phase] (0) at (-1.75, -1) {$m$ \nodepart{lower} $-m$};
		\node [style=qutrit X phase] (1) at (0.25, -1) {$m$ \nodepart{lower} $-m$};
		\node [style=qutrit X phase] (2) at (1.75, -1) {$m$ \nodepart{lower} $-m$};
		\node [style=none] (3) at (-1.75, 1.25) {};
		\node [style=none] (4) at (0.25, 1.25) {};
		\node [style=none] (5) at (-0.75, 0.5) {...};
		\node [style=none] (6) at (1.75, 0.75) {};
		\node [style=qutrit Z phase] (7) at (1.75, 0.75) {$\alpha$ \nodepart{lower} $\beta$};
	\end{pgfonlayer}
	\begin{pgfonlayer}{edgelayer}
		\draw (0) to (3.center);
		\draw (1) to (4.center);
		\draw (2) to (7);
	\end{pgfonlayer}
\end{tikzpicture}
 \ = \
			
		\end{equation}
	\end{proof}
\end{lemma}

\subsection{\Pspider\ Elimination}

The \Pspider\ elimination theorem additionally requires a lemma allowing us to turn a \Pspider\ state of one colour into a \Pspider\ state of the other. We will use this lemma its adjoint form in the proof of the main theorem.

\begin{lemma}\label{lem:P_state_colour_change}
	The following rule holds in the qutrit ZX-calculus, for $p \in \{1, 2\}$:
	\begin{equation}
		\begin{tikzpicture}
	\begin{pgfonlayer}{nodelayer}
		\node [style=qutrit Z phase] (0) at (0, -0.5) {$p$ \nodepart{lower} $p$};
		\node [style=none] (1) at (0, 1) {};
	\end{pgfonlayer}
	\begin{pgfonlayer}{edgelayer}
		\draw (0) to (1.center);
	\end{pgfonlayer}
\end{tikzpicture}
 \ = \ \begin{tikzpicture}
	\begin{pgfonlayer}{nodelayer}
		\node [style=none] (7) at (0, 1) {};
		\node [style=qutrit X phase] (9) at (0, -0.5) {$-p$ \nodepart{lower} $-p$};
	\end{pgfonlayer}
	\begin{pgfonlayer}{edgelayer}
		\draw (9) to (7.center);
	\end{pgfonlayer}
\end{tikzpicture}

	\end{equation}
	\begin{proof}
		\begin{equation}
			 \ \ \xeq{\qutritRuleColourChange} \ \ 
			\begin{tikzpicture}
	\begin{pgfonlayer}{nodelayer}
		\node [style=qutrit X phase] (0) at (0, -0.5) {$p$ \nodepart{lower} $p$};
		\node [style=none] (1) at (0, 1) {};
		\node [style=qutrit hadamard unknown] (5) at (0, 0.5) {$p$};
	\end{pgfonlayer}
	\begin{pgfonlayer}{edgelayer}
		\draw (0) to (1.center);
	\end{pgfonlayer}
\end{tikzpicture}
 \ \ \xeq{\eqref{eq:qutrit_hadamard_decompositions}} \ \ 
			\begin{tikzpicture}
	\begin{pgfonlayer}{nodelayer}
		\node [style=qutrit X phase] (6) at (0, -2.425) {$p$ \nodepart{lower} $p$};
		\node [style=none] (7) at (0, 2.75) {};
		\node [style=qutrit Z phase] (8) at (0, 0.5) {$-p$ \nodepart{lower} $-p$};
		\node [style=qutrit X phase] (9) at (0, 2) {$-p$ \nodepart{lower} $-p$};
		\node [style=qutrit X phase] (10) at (0, -1) {$-p$ \nodepart{lower} $-p$};
	\end{pgfonlayer}
	\begin{pgfonlayer}{edgelayer}
		\draw (6) to (7.center);
	\end{pgfonlayer}
\end{tikzpicture}
 \ \ \xeq{\qutritRuleFusion} \ \ 
			\begin{tikzpicture}
	\begin{pgfonlayer}{nodelayer}
		\node [style=none] (7) at (0, 1.75) {};
		\node [style=qutrit Z phase] (8) at (0, -0.5) {$-p$ \nodepart{lower} $-p$};
		\node [style=qutrit X phase] (9) at (0, 1) {$-p$ \nodepart{lower} $-p$};
		\node [style=X dot] (10) at (0, -1.75) {};
	\end{pgfonlayer}
	\begin{pgfonlayer}{edgelayer}
		\draw (10) to (7.center);
	\end{pgfonlayer}
\end{tikzpicture}
 \ \ \xeq{\ref{lem:substantial_m_copy}} \ \ 
			\begin{tikzpicture}
	\begin{pgfonlayer}{nodelayer}
		\node [style=none] (7) at (0, 1) {};
		\node [style=qutrit X phase] (9) at (0, -0.5) {$-p$ \nodepart{lower} $-p$};
		\node [style=X dot] (10) at (0, -1.75) {};
	\end{pgfonlayer}
	\begin{pgfonlayer}{edgelayer}
		\draw (10) to (7.center);
	\end{pgfonlayer}
\end{tikzpicture}
 \ \ \xeq{\qutritRuleFusion} \ \ 
			
		\end{equation}
	\end{proof}
\end{lemma}

\begin{theorem}\label{thm:eliminate_P_spiders_appendix} \textbf{/\ Theorem~\ref{thm:eliminate_P_spiders}.} 
	\eliminatePSpidersStatement
	\begin{proof}
		For clarity of presentation we only show the case $\qutritZphase{p}{p} = \qutritZphase{1}{1}$, the other case being near-identical.
		\begingroup
			\allowdisplaybreaks
			\setlength{\jot}{20pt}
				\begin{align*}
					&\ &&\input{eliminate/P_spiders/bang/p_1/1.tikz} 
					&&&\xeq{\qutritRuleFusion} 
					&&&&\input{eliminate/P_spiders/bang/p_1/2.tikz}
					&&&&&\xeq{\ref{thm:local_comp_equality}} 
					&&&&&&\input{eliminate/P_spiders/bang/p_1/3.tikz} \\
					&\xeqq{\ref{lem:P_state_colour_change}}{\qutritRuleFusion} 
					&&\input{eliminate/P_spiders/bang/p_1/4.tikz}
					&&&\xeq{\ref{lem:leg_flip}} 
					&&&&\input{eliminate/P_spiders/bang/p_1/5.tikz} 
					&&&&&\xeq{\ref{lem:substantial_m_copy}} 
					&&&&&&\input{eliminate/P_spiders/bang/p_1/6.tikz} \\
					&\xeq{\eqref{eq:qutrit_dashed_lines}}
					&&\input{eliminate/P_spiders/bang/p_1/7.tikz} 
					&&&\xeq{\qutritRuleColourChange} 
					&&&&\input{eliminate/P_spiders/bang/p_1/8.tikz}
					&&&&&\xeq{\qutritRuleFusion} 
					&&&&&&\input{eliminate/P_spiders/bang/p_1/9.tikz} \\
				\end{align*}
		\endgroup
	\end{proof}
\end{theorem}

\subsection{\Nspider\ Elimination}

Proving the corresponding \Nspider\ elimination theorem again requires a lemma allowing us to turn an \Nspider\ state of one colour into an \Nspider\ state of the other.

\begin{lemma}\label{lem:N_state_colour_change}
	The following rules hold in the qutrit ZX-calculus:
	\begin{equation}
		\begin{tikzpicture}
	\begin{pgfonlayer}{nodelayer}
		\node [style=qutrit Z phase] (0) at (0, 0) {$0$ \nodepart{lower} $1$};
		\node [style=none] (1) at (0, 1) {};
	\end{pgfonlayer}
	\begin{pgfonlayer}{edgelayer}
		\draw (0) to (1.center);
	\end{pgfonlayer}
\end{tikzpicture}
 \ = \ \begin{tikzpicture}
	\begin{pgfonlayer}{nodelayer}
		\node [style=qutrit X phase] (0) at (0, 0) {$2$ \nodepart{lower} $0$};
		\node [style=none] (6) at (0, 1) {};
	\end{pgfonlayer}
	\begin{pgfonlayer}{edgelayer}
		\draw (0) to (6.center);
	\end{pgfonlayer}
\end{tikzpicture}
 \ , \qquad
		\begin{tikzpicture}
	\begin{pgfonlayer}{nodelayer}
		\node [style=qutrit Z phase] (0) at (0, 0) {$0$ \nodepart{lower} $2$};
		\node [style=none] (1) at (0, 1) {};
	\end{pgfonlayer}
	\begin{pgfonlayer}{edgelayer}
		\draw (0) to (1.center);
	\end{pgfonlayer}
\end{tikzpicture}
 \ = \ \begin{tikzpicture}
	\begin{pgfonlayer}{nodelayer}
		\node [style=qutrit X phase] (0) at (0, 0) {$0$ \nodepart{lower} $1$};
		\node [style=none] (6) at (0, 1) {};
	\end{pgfonlayer}
	\begin{pgfonlayer}{edgelayer}
		\draw (0) to (6.center);
	\end{pgfonlayer}
\end{tikzpicture}
 \ , \qquad
		\begin{tikzpicture}
	\begin{pgfonlayer}{nodelayer}
		\node [style=qutrit Z phase] (0) at (0, 0) {$1$ \nodepart{lower} $0$};
		\node [style=none] (1) at (0, 1) {};
	\end{pgfonlayer}
	\begin{pgfonlayer}{edgelayer}
		\draw (0) to (1.center);
	\end{pgfonlayer}
\end{tikzpicture}
 \ = \ \begin{tikzpicture}
	\begin{pgfonlayer}{nodelayer}
		\node [style=qutrit X phase] (0) at (0, 0) {$0$ \nodepart{lower} $2$};
		\node [style=none] (6) at (0, 1) {};
	\end{pgfonlayer}
	\begin{pgfonlayer}{edgelayer}
		\draw (0) to (6.center);
	\end{pgfonlayer}
\end{tikzpicture}
 \ , \qquad
		\begin{tikzpicture}
	\begin{pgfonlayer}{nodelayer}
		\node [style=qutrit Z phase] (0) at (0, 0) {$2$ \nodepart{lower} $0$};
		\node [style=none] (1) at (0, 1) {};
	\end{pgfonlayer}
	\begin{pgfonlayer}{edgelayer}
		\draw (0) to (1.center);
	\end{pgfonlayer}
\end{tikzpicture}
 \ = \ \begin{tikzpicture}
	\begin{pgfonlayer}{nodelayer}
		\node [style=qutrit X phase] (0) at (0, 0) {$1$ \nodepart{lower} $0$};
		\node [style=none] (6) at (0, 1) {};
	\end{pgfonlayer}
	\begin{pgfonlayer}{edgelayer}
		\draw (0) to (6.center);
	\end{pgfonlayer}
\end{tikzpicture}

	\end{equation}
	\begin{proof}
		For any green \Nspider\ state with phase $\qutritZphase{n}{n'}$, we have a choice of two colour change rules which we could use to turn it into a red \Nspider\ state with a $H$- or $H^\dagger$-box on top:
		
		\begin{equation}
			\begin{tikzpicture}
	\begin{pgfonlayer}{nodelayer}
		\node [style=qutrit X phase] (0) at (0, 0) {$n$ \nodepart{lower} $n'$};
		\node [style=none] (1) at (0, 1.5) {};
		\node [style=qutrit hadamard adjoint] (5) at (0, 1) {};
	\end{pgfonlayer}
	\begin{pgfonlayer}{edgelayer}
		\draw (0) to (1.center);
	\end{pgfonlayer}
\end{tikzpicture}
 \ \xeq{\qutritRuleColourChange} \ 
			\begin{tikzpicture}
	\begin{pgfonlayer}{nodelayer}
		\node [style=qutrit Z phase] (0) at (0, 0) {$n$ \nodepart{lower} $n'$};
		\node [style=none] (1) at (0, 1.5) {};
	\end{pgfonlayer}
	\begin{pgfonlayer}{edgelayer}
		\draw (0) to (1.center);
	\end{pgfonlayer}
\end{tikzpicture}
 \ \xeq{\qutritRuleColourChange} \ 
			\begin{tikzpicture}
	\begin{pgfonlayer}{nodelayer}
		\node [style=qutrit X phase] (0) at (0, 0) {$n'$ \nodepart{lower} $n$};
		\node [style=none] (1) at (0, 1.5) {};
		\node [style=qutrit hadamard] (2) at (0, 1) {};
	\end{pgfonlayer}
	\begin{pgfonlayer}{edgelayer}
		\draw (0) to (1.center);
	\end{pgfonlayer}
\end{tikzpicture}

		\end{equation}

		Of these two choices, exactly one has a decomposition of of the $H$-/$H^\dagger$-box as in \eqref{eq:qutrit_hadamard_decompositions} that allows the bottom two red spiders to fuse into an \Mspider, which we can then move past the green spider above it via \ref{lem:substantial_m_copy}. For brevity we only show the case $\qutritZphase{n}{n'} = \qutritZphase{0}{1}$:

		\begin{equation}
			\begin{tikzpicture}
	\begin{pgfonlayer}{nodelayer}
		\node [style=qutrit Z phase] (0) at (0, -0.5) {$0$ \nodepart{lower} $1$};
		\node [style=none] (1) at (0, 1) {};
	\end{pgfonlayer}
	\begin{pgfonlayer}{edgelayer}
		\draw (0) to (1.center);
	\end{pgfonlayer}
\end{tikzpicture}
 \ \ \xeq{\qutritRuleColourChange} \ \ 
			\begin{tikzpicture}
	\begin{pgfonlayer}{nodelayer}
		\node [style=qutrit X phase] (0) at (0, -0.5) {$0$ \nodepart{lower} $1$};
		\node [style=none] (1) at (0, 1) {};
		\node [style=qutrit hadamard adjoint] (5) at (0, 0.5) {};
	\end{pgfonlayer}
	\begin{pgfonlayer}{edgelayer}
		\draw (0) to (1.center);
	\end{pgfonlayer}
\end{tikzpicture}
 \ \ \xeq{\eqref{eq:qutrit_hadamard_decompositions}} \ \ 
			\begin{tikzpicture}
	\begin{pgfonlayer}{nodelayer}
		\node [style=qutrit X phase] (0) at (0, -2) {$0$ \nodepart{lower} $1$};
		\node [style=none] (6) at (0, 2.5) {};
		\node [style=qutrit Z phase] (8) at (0, 0.5) {$1$ \nodepart{lower} $1$};
		\node [style=qutrit X phase] (9) at (0, 1.75) {$1$ \nodepart{lower} $1$};
		\node [style=qutrit X phase] (10) at (0, -0.75) {$1$ \nodepart{lower} $1$};
	\end{pgfonlayer}
	\begin{pgfonlayer}{edgelayer}
		\draw (0) to (6.center);
	\end{pgfonlayer}
\end{tikzpicture}
 \ \ \xeq{\qutritRuleFusion} \ \ 
			\begin{tikzpicture}
	\begin{pgfonlayer}{nodelayer}
		\node [style=qutrit X phase] (0) at (0, -1.25) {$1$ \nodepart{lower} $2$};
		\node [style=none] (6) at (0, 2) {};
		\node [style=qutrit Z phase] (8) at (0, 0) {$1$ \nodepart{lower} $1$};
		\node [style=qutrit X phase] (9) at (0, 1.25) {$1$ \nodepart{lower} $1$};
	\end{pgfonlayer}
	\begin{pgfonlayer}{edgelayer}
		\draw (0) to (6.center);
	\end{pgfonlayer}
\end{tikzpicture}
 \ \ \xeq{\ref{lem:substantial_m_copy}} \ \ 
			\begin{tikzpicture}
	\begin{pgfonlayer}{nodelayer}
		\node [style=qutrit X phase] (0) at (0, -0.75) {$1$ \nodepart{lower} $2$};
		\node [style=none] (6) at (0, 1.25) {};
		\node [style=qutrit X phase] (9) at (0, 0.5) {$1$ \nodepart{lower} $1$};
	\end{pgfonlayer}
	\begin{pgfonlayer}{edgelayer}
		\draw (0) to (6.center);
	\end{pgfonlayer}
\end{tikzpicture}
 \ \ \xeq{\qutritRuleFusion} \ \ 
			\begin{tikzpicture}
	\begin{pgfonlayer}{nodelayer}
		\node [style=qutrit X phase] (0) at (0, -0.5) {$2$ \nodepart{lower} $0$};
		\node [style=none] (6) at (0, 1) {};
	\end{pgfonlayer}
	\begin{pgfonlayer}{edgelayer}
		\draw (0) to (6.center);
	\end{pgfonlayer}
\end{tikzpicture}

		\end{equation}
	\end{proof}
\end{lemma}

\begin{corollary}\label{cor:N_effect}
	The following equations hold in the qutrit ZX-calculus, for $n \in \{1, 2\}$:
	\begin{equation}
		\begin{tikzpicture}
	\begin{pgfonlayer}{nodelayer}
		\node [style=qutrit Z phase] (0) at (0, 1) {$0$ \nodepart{lower} $n$};
		\node [style=qutrit X phase] (1) at (0, -0.5) {$-n$ \nodepart{lower} $-n$};
		\node [style=none] (2) at (0, -1.5) {};
	\end{pgfonlayer}
	\begin{pgfonlayer}{edgelayer}
		\draw (0) to (2.center);
	\end{pgfonlayer}
\end{tikzpicture}
 \ = \ \begin{tikzpicture}
	\begin{pgfonlayer}{nodelayer}
		\node [style=qutrit X phase] (1) at (0, 0) {$2$ \nodepart{lower} $1$};
		\node [style=none] (2) at (0, -1.25) {};
	\end{pgfonlayer}
	\begin{pgfonlayer}{edgelayer}
		\draw (2.center) to (1);
	\end{pgfonlayer}
\end{tikzpicture}
 \ , \qquad
		\begin{tikzpicture}
	\begin{pgfonlayer}{nodelayer}
		\node [style=qutrit Z phase] (0) at (0, 1) {$n$ \nodepart{lower} $0$};
		\node [style=qutrit X phase] (1) at (0, -0.5) {$-n$ \nodepart{lower} $-n$};
		\node [style=none] (2) at (0, -1.5) {};
	\end{pgfonlayer}
	\begin{pgfonlayer}{edgelayer}
		\draw (0) to (2.center);
	\end{pgfonlayer}
\end{tikzpicture}
 \ = \ \begin{tikzpicture}
	\begin{pgfonlayer}{nodelayer}
		\node [style=qutrit X phase] (1) at (0, 0) {$1$ \nodepart{lower} $2$};
		\node [style=none] (2) at (0, -1.25) {};
	\end{pgfonlayer}
	\begin{pgfonlayer}{edgelayer}
		\draw (2.center) to (1);
	\end{pgfonlayer}
\end{tikzpicture}

	\end{equation}
	\begin{proof}
		Again we only prove one case, the other three being analogous. Each case uses \ref{lem:N_state_colour_change} in its adjoint form - recall that the adjoint of a spider is found by swapping inputs and outputs and negating angles.
		\begin{equation}
			\begin{tikzpicture}
	\begin{pgfonlayer}{nodelayer}
		\node [style=qutrit Z phase] (0) at (0, 1) {$0$ \nodepart{lower} $1$};
		\node [style=qutrit X phase] (1) at (0, -0.5) {$-1$ \nodepart{lower} $-1$};
		\node [style=none] (2) at (0, -1.5) {};
	\end{pgfonlayer}
	\begin{pgfonlayer}{edgelayer}
		\draw (0) to (2.center);
	\end{pgfonlayer}
\end{tikzpicture}
 \ \ = \ \ 
			\begin{tikzpicture}
	\begin{pgfonlayer}{nodelayer}
		\node [style=qutrit Z phase] (0) at (0, 1) {$0$ \nodepart{lower} $-2$};
		\node [style=qutrit X phase] (1) at (0, -0.5) {$-1$ \nodepart{lower} $-1$};
		\node [style=none] (2) at (0, -1.5) {};
	\end{pgfonlayer}
	\begin{pgfonlayer}{edgelayer}
		\draw (0) to (2.center);
	\end{pgfonlayer}
\end{tikzpicture}
 \ \ \xeq{\ref{lem:N_state_colour_change}} \ \ 
			\begin{tikzpicture}
	\begin{pgfonlayer}{nodelayer}
		\node [style=qutrit X phase] (0) at (0, 1) {$0$ \nodepart{lower} $-1$};
		\node [style=qutrit X phase] (1) at (0, -0.5) {$-1$ \nodepart{lower} $-1$};
		\node [style=none] (2) at (0, -1.5) {};
	\end{pgfonlayer}
	\begin{pgfonlayer}{edgelayer}
		\draw (0) to (2.center);
	\end{pgfonlayer}
\end{tikzpicture}
 \ \ \xeq{\qutritRuleFusion} \ \ 
			\begin{tikzpicture}
	\begin{pgfonlayer}{nodelayer}
		\node [style=qutrit X phase] (0) at (0, 0) {$-1$ \nodepart{lower} $-2$};
		\node [style=none] (2) at (0, -1.25) {};
	\end{pgfonlayer}
	\begin{pgfonlayer}{edgelayer}
		\draw (0) to (2.center);
	\end{pgfonlayer}
\end{tikzpicture}
 \ \ = \ \ 
			\begin{tikzpicture}
	\begin{pgfonlayer}{nodelayer}
		\node [style=qutrit X phase] (0) at (0, 0) {$2$ \nodepart{lower} $1$};
		\node [style=none] (2) at (0, -1.25) {};
	\end{pgfonlayer}
	\begin{pgfonlayer}{edgelayer}
		\draw (0) to (2.center);
	\end{pgfonlayer}
\end{tikzpicture}

		\end{equation}
	\end{proof}
\end{corollary}

\begin{theorem}\label{thm:eliminate_N_spiders_appendix} \textbf{/\ Theorem~\ref{thm:eliminate_N_spiders}.}
	\eliminateNSpidersStatement
	\begin{proof}
		We prove the case where $x$ has phase \qutritZphase{0}{1}, the other cases being near-identical.
		\begingroup
			\allowdisplaybreaks
			\setlength{\jot}{20pt}
				\begin{align*}
					&\ &&\input{eliminate/N_spiders/0_n/bang/0_1/1.tikz} 
					&&&\xeq{\qutritRuleFusion} 
					&&&&\input{eliminate/N_spiders/0_n/bang/0_1/2.tikz}
					&&&&&\xeq{\ref{thm:local_comp_equality}} 
					&&&&&&\input{eliminate/N_spiders/0_n/bang/0_1/3.tikz} \\
					&\xeqq{\ref{cor:N_effect}}{\qutritRuleFusion} 
					&&\input{eliminate/N_spiders/0_n/bang/0_1/4.tikz}
					&&&\xeq{\ref{lem:leg_flip}} 
					&&&&\input{eliminate/N_spiders/0_n/bang/0_1/5.tikz} 
					&&&&&\xeq{\ref{lem:substantial_m_copy}} 
					&&&&&&\input{eliminate/N_spiders/0_n/bang/0_1/6.tikz} \\
					&\xeq{\eqref{eq:qutrit_dashed_lines}}
					&&\input{eliminate/N_spiders/0_n/bang/0_1/7.tikz} 
					&&&\xeq{\qutritRuleColourChange} 
					&&&&\input{eliminate/N_spiders/0_n/bang/0_1/8.tikz}
					&&&&&\xeq{\qutritRuleFusion} 
					&&&&&&\input{eliminate/N_spiders/0_n/bang/0_1/9.tikz} \\
				\end{align*}
		\endgroup
	\end{proof}
\end{theorem}
\subsection{\Mspider\ Elimination}

\begin{theorem}\label{thm:eliminate_M_spiders_appendix} \textbf{/\ Theorem~\ref{thm:eliminate_M_spiders}.}
	\eliminateMSpidersStatement
	\begin{proof}
		We show the case where $w_{ij} \eqdef w = 1$, with the case $w = 2$ being completely analogous. We can choose either a proper $1$-pivot or a proper $2$-pivot; both give the same result. Here we only show the former:
		\begingroup
			\allowdisplaybreaks
			\setlength{\jot}{20pt}
			\begin{align*}
				&\ &&\input{eliminate/M_spiders/bang/w_1/1.tikz} \\
				&\xeq{\qutritRuleFusion} 
				&&\input{eliminate/M_spiders/bang/w_1/2.tikz} \\
				&\xeq{\ref{thm:local_pivot_equality_appendix}} 
				&&\input{eliminate/M_spiders/bang/w_1/3.tikz} \\
				&\xeq{\qutritRuleColourChange}
				&&\input{eliminate/M_spiders/bang/w_1/4.tikz} \\
				&\xeq{\ref{lem:leg_flip}}
				&&\input{eliminate/M_spiders/bang/w_1/5.tikz} \\
				&\xeq{\ref{lem:substantial_m_copy}}
				&&\input{eliminate/M_spiders/bang/w_1/6.tikz} \\
				&= 
				&&\input{eliminate/M_spiders/bang/w_1/7.tikz} \\
				&\xeq{\qutritRuleColourChange}
				&&\input{eliminate/M_spiders/bang/w_1/8.tikz} \\
				&\xeq{\qutritRuleFusion}
				&&\input{eliminate/M_spiders/bang/w_1/9.tikz} \\
			\end{align*}
		\endgroup
	\end{proof}
\end{theorem}
\section{$\bf{\pm}$-Boxes in the ZX-Calculus}

We close by proving that the $\pm$-boxes in \eqref{eq:pm_tensor} correspond to stabilizer ZX-diagrams, while the qutrit $X$-box in \eqref{eq:X_tensor} does not.

\begin{proposition}\label{prop:pm_maps_zx_appendix} 
	The following equalities hold up to a scalar under the standard interpretation:
	\begin{equation}
		\left\llbracket \ \begin{tikzpicture}
	\begin{pgfonlayer}{nodelayer}
		\node [style=box] (0) at (0, 0) {$\pm$};
		\node [style=none] (1) at (0, -1.25) {};
		\node [style=none] (2) at (0, 1.25) {};
	\end{pgfonlayer}
	\begin{pgfonlayer}{edgelayer}
		\draw (2.center) to (1.center);
	\end{pgfonlayer}
\end{tikzpicture}
 \ \right\rrbracket_{d=2} \simeq 
		\left\llbracket \ \begin{tikzpicture}
	\begin{pgfonlayer}{nodelayer}
		\node [style=none] (0) at (0, -1) {};
		\node [style=none] (1) at (0, 1) {};
		\node [style=X phase dot] (6) at (0, 0) {$\pm\frac{\pi}{2}$};
	\end{pgfonlayer}
	\begin{pgfonlayer}{edgelayer}
		\draw (0.center) to (1.center);
	\end{pgfonlayer}
\end{tikzpicture}
 \ \right\rrbracket ~, 
		\hspace{30pt}
		\left\llbracket \ \begin{tikzpicture}
	\begin{pgfonlayer}{nodelayer}
		\node [style=box] (0) at (0, 0) {$\pm$};
		\node [style=none] (1) at (0, -1.25) {};
		\node [style=none] (2) at (0, 1.25) {};
	\end{pgfonlayer}
	\begin{pgfonlayer}{edgelayer}
		\draw (2.center) to (1.center);
	\end{pgfonlayer}
\end{tikzpicture}
 \ \right\rrbracket_{d=3} \simeq
		\left\llbracket \ \begin{tikzpicture}
	\begin{pgfonlayer}{nodelayer}
		\node [style=qutrit X phase] (0) at (0, 0) {$\pm 1$ \nodepart{lower} $\pm 1$};
		\node [style=none] (3) at (0, -1) {};
		\node [style=none] (4) at (0, 1) {};
	\end{pgfonlayer}
	\begin{pgfonlayer}{edgelayer}
		\draw (3.center) to (4.center);
	\end{pgfonlayer}
\end{tikzpicture}
 \ \right\rrbracket ~,
		\hspace{30pt}
		\left\llbracket \ \begin{tikzpicture}
	\begin{pgfonlayer}{nodelayer}
		\node [style=box] (0) at (0, 0) {$\pm$};
		\node [style=none] (1) at (0, -1.25) {};
		\node [style=none] (2) at (0, 1.25) {};
	\end{pgfonlayer}
	\begin{pgfonlayer}{edgelayer}
		\draw (2.center) to (1.center);
	\end{pgfonlayer}
\end{tikzpicture}
 \ \right\rrbracket_{d=4} \simeq 
		\left\llbracket \ \begin{tikzpicture}
	\begin{pgfonlayer}{nodelayer}
		\node [style=X phase dot] (0) at (-1, 0) {$\pi$};
		\node [style=X phase dot] (1) at (1, 0) {$\pi$};
		\node [style=none] (2) at (-1, 1) {};
		\node [style=none] (3) at (1, 1) {};
		\node [style=none] (4) at (-1, -1) {};
		\node [style=none] (5) at (1, -1) {};
		\node [style=hadamard] (7) at (0, 0) {};
	\end{pgfonlayer}
	\begin{pgfonlayer}{edgelayer}
		\draw (4.center) to (2.center);
		\draw (5.center) to (3.center);
		\draw (0) to (1);
	\end{pgfonlayer}
\end{tikzpicture}
 \ \right\rrbracket
	\end{equation}

	\begin{proof}
		Recalling $\omega = e^{i\frac{2\pi}{3}}$, the standard interpretations of phase gates in matrix form are:
		
		\begin{equation*}
			\left\llbracket \ \begin{tikzpicture}
	\begin{pgfonlayer}{nodelayer}
		\node [style=Z phase dot] (0) at (0, 0) {$\alpha$};
		\node [style=none] (1) at (0, 1) {};
		\node [style=none] (2) at (0, -1) {};
	\end{pgfonlayer}
	\begin{pgfonlayer}{edgelayer}
		\draw (1.center) to (2.center);
	\end{pgfonlayer}
\end{tikzpicture}
 \ \right\rrbracket = 
			\begin{pmatrix}
				1 & 0 \\
				0 & e^{i\alpha}
			\end{pmatrix} \ , \quad
			\left\llbracket \ \begin{tikzpicture}
	\begin{pgfonlayer}{nodelayer}
		\node [style=X phase dot] (0) at (0, 0) {$\alpha$};
		\node [style=none] (1) at (0, 1) {};
		\node [style=none] (2) at (0, -1) {};
	\end{pgfonlayer}
	\begin{pgfonlayer}{edgelayer}
		\draw (1.center) to (2.center);
	\end{pgfonlayer}
\end{tikzpicture}
 \ \right\rrbracket = 
			\frac{1}{2} \begin{pmatrix}
				1 + e^{i\alpha} & 1 - e^{i\alpha} \\
				1 - e^{i\alpha} & 1 + e^{i\alpha}
			\end{pmatrix} \ , \quad
			\left\llbracket \ \begin{tikzpicture}
	\begin{pgfonlayer}{nodelayer}
		\node [style=qutrit Z phase] (0) at (0, 0) {$\alpha$ \nodepart{lower} $\beta$};
		\node [style=none] (1) at (0, 1) {};
		\node [style=none] (2) at (0, -1) {};
	\end{pgfonlayer}
	\begin{pgfonlayer}{edgelayer}
		\draw (1.center) to (2.center);
	\end{pgfonlayer}
\end{tikzpicture}
 \ \right\rrbracket = 
			\begin{pmatrix}
				1 & 0 & 0\\
				0 & e^{i\alpha} & 0 \\
				0 & 0 & e^{i\beta}
			\end{pmatrix}
		\end{equation*}
		\begin{equation*}
			\left\llbracket \ \begin{tikzpicture}
	\begin{pgfonlayer}{nodelayer}
		\node [style=qutrit X phase] (0) at (0, 0) {$\alpha$ \nodepart{lower} $\beta$};
		\node [style=none] (1) at (0, 1) {};
		\node [style=none] (2) at (0, -1) {};
	\end{pgfonlayer}
	\begin{pgfonlayer}{edgelayer}
		\draw (1.center) to (2.center);
	\end{pgfonlayer}
\end{tikzpicture}
 \ \right\rrbracket = 
			\frac{1}{3} \begin{pmatrix}
				1 + e^{i\alpha} + e^{i\beta} & 1 + \bar{\omega}e^{i\alpha} + {\omega}e^{i\beta} & 1 + {\omega}e^{i\alpha} + \bar{\omega}e^{i\beta} \\
				1 + {\omega}e^{i\alpha} + \bar{\omega}e^{i\beta} & 1 + e^{i\alpha} + e^{i\beta} & 1 + \bar{\omega}e^{i\alpha} + {\omega}e^{i\beta} \\
				1 + \bar{\omega}e^{i\alpha} + {\omega}e^{i\beta} & 1 + {\omega}e^{i\alpha} + \bar{\omega}e^{i\beta} & 1 + e^{i\alpha} + e^{i\beta} \\
			\end{pmatrix}
		\end{equation*}

		So in the simplest case $d=2$ it is fairly straightforward to see that:

		\begin{equation}
			\left\llbracket \ \begin{tikzpicture}
	\begin{pgfonlayer}{nodelayer}
		\node [style=none] (0) at (0, -1) {};
		\node [style=none] (1) at (0, 1) {};
		\node [style=X phase dot] (6) at (0, 0) {$\pm\frac{\pi}{2}$};
	\end{pgfonlayer}
	\begin{pgfonlayer}{edgelayer}
		\draw (0.center) to (1.center);
	\end{pgfonlayer}
\end{tikzpicture}
 \ \right\rrbracket \ = \ 
			\frac{1}{2} \begin{pmatrix}
				1 \pm i & 1 \mp i \\
				1 \mp i & 1 \pm i \\
			\end{pmatrix} \ = \ 
			\frac{\sqrt{2}}{2} e^{\mp i \frac{\pi}{4}} \begin{pmatrix}
				\pm i & 1 \\
				1 & \pm i \\
			\end{pmatrix} \ = \ 
			\frac{\sqrt{2}}{2} e^{\mp i \frac{\pi}{4}} \left\llbracket \ \begin{tikzpicture}
	\begin{pgfonlayer}{nodelayer}
		\node [style=box] (0) at (0, 0) {$\pm$};
		\node [style=none] (1) at (0, -1.25) {};
		\node [style=none] (2) at (0, 1.25) {};
	\end{pgfonlayer}
	\begin{pgfonlayer}{edgelayer}
		\draw (2.center) to (1.center);
	\end{pgfonlayer}
\end{tikzpicture}
 \ \right\rrbracket_{d=2}
		\end{equation}

		The next case $d=3$ is proved similarly:

		\begin{equation}
		\begin{aligned}
				\left\llbracket \ \begin{tikzpicture}
	\begin{pgfonlayer}{nodelayer}
		\node [style=qutrit X phase] (0) at (0, 0) {$\pm 1$ \nodepart{lower} $\pm 1$};
		\node [style=none] (3) at (0, -1) {};
		\node [style=none] (4) at (0, 1) {};
	\end{pgfonlayer}
	\begin{pgfonlayer}{edgelayer}
		\draw (3.center) to (4.center);
	\end{pgfonlayer}
\end{tikzpicture}
 \ \right\rrbracket
				&= \frac{1}{3} \begin{pmatrix}
					1 + e^{\pm i\frac{2\pi}{3}} + e^{\pm i\frac{2\pi}{3}} & 1 + \bar{\omega}e^{\pm i\frac{2\pi}{3}} + {\omega}e^{\pm i\frac{2\pi}{3}} & 1 + {\omega}e^{\pm i\frac{2\pi}{3}} + \bar{\omega}e^{\pm i\frac{2\pi}{3}} \\
					1 + {\omega}e^{\pm i\frac{2\pi}{3}} + \bar{\omega}e^{\pm i\frac{2\pi}{3}} & 1 + e^{\pm i\frac{2\pi}{3}} + e^{\pm i\frac{2\pi}{3}} & 1 + \bar{\omega}e^{\pm i\frac{2\pi}{3}} + {\omega}e^{\pm i\frac{2\pi}{3}} \\
					1 + \bar{\omega}e^{\pm i\frac{2\pi}{3}} + {\omega}e^{\pm i\frac{2\pi}{3}} & 1 + {\omega}e^{\pm i\frac{2\pi}{3}} + \bar{\omega}e^{\pm i\frac{2\pi}{3}} & 1 + e^{\pm i\frac{2\pi}{3}} + e^{\pm i\frac{2\pi}{3}} \\
				\end{pmatrix} \\
				&= \frac{1}{3} \begin{pmatrix}
					\sqrt{3}e^{\pm i\frac{\pi}{2}} & \sqrt{3}e^{\mp i\frac{\pi}{6}} & \sqrt{3}e^{\mp i\frac{\pi}{6}} \\
					\sqrt{3}e^{\mp i\frac{\pi}{6}} & \sqrt{3}e^{\pm i\frac{\pi}{2}} & \sqrt{3}e^{\mp i\frac{\pi}{6}} \\
					\sqrt{3}e^{\mp i\frac{\pi}{6}} & \sqrt{3}e^{\mp i\frac{\pi}{6}} & \sqrt{3}e^{\pm i\frac{\pi}{2}} \\
				\end{pmatrix} \\
				&= \frac{\sqrt{3}}{3} e^{\mp i\frac{\pi}{6}}\begin{pmatrix}
					e^{\pm i\frac{2\pi}{3}} & 1 & 1 \\
					1 & e^{\pm i\frac{2\pi}{3}} & 1 \\
					1 & 1 & e^{\pm i\frac{2\pi}{3}} \\
				\end{pmatrix} \\
				&= \frac{\sqrt{3}}{3} e^{\mp i\frac{\pi}{6}} \left\llbracket \ \begin{tikzpicture}
	\begin{pgfonlayer}{nodelayer}
		\node [style=box] (0) at (0, 0) {$\pm$};
		\node [style=none] (1) at (0, -1.25) {};
		\node [style=none] (2) at (0, 1.25) {};
	\end{pgfonlayer}
	\begin{pgfonlayer}{edgelayer}
		\draw (2.center) to (1.center);
	\end{pgfonlayer}
\end{tikzpicture}
 \ \right\rrbracket_{d=3}
			\end{aligned}
		\end{equation}

		For the other qubit case $d=4$ we first note:

		\begin{equation}
			\left\llbracket \ \begin{tikzpicture}
	\begin{pgfonlayer}{nodelayer}
		\node [style=none] (0) at (0, -1) {};
		\node [style=none] (1) at (0, 1) {};
		\node [style=hadamard] (6) at (0, 0) {};
	\end{pgfonlayer}
	\begin{pgfonlayer}{edgelayer}
		\draw (0.center) to (1.center);
	\end{pgfonlayer}
\end{tikzpicture}
 \ \right\rrbracket = 
			\left\llbracket \  \ \right\rrbracket =
			\left\llbracket \ \qubitZphase{\frac{\pi}{2}} \ \right\rrbracket
			\left\llbracket \ \qubitXphase{\frac{\pi}{2}} \ \right\rrbracket
			\left\llbracket \ \qubitZphase{\frac{\pi}{2}} \ \right\rrbracket = 
			\frac{1}{2\sqrt{2}} \begin{pmatrix}
				1 & 1 \\
				1 & -1 \\
			\end{pmatrix}
		\end{equation}

		Then using the standard interpretation for spiders as in \eqref{eq:qubit_standard_interpretation}, we decompose the diagram in such a way that we can apply the standard interpretation:

		\begingroup
			\allowdisplaybreaks
				\begin{align*}
					\left\llbracket \ \begin{tikzpicture}
	\begin{pgfonlayer}{nodelayer}
		\node [style=X phase dot] (0) at (-1, 0) {$\pi$};
		\node [style=X phase dot] (1) at (1, 0) {$\pi$};
		\node [style=none] (2) at (-1, 1) {};
		\node [style=none] (3) at (1, 1) {};
		\node [style=none] (4) at (-1, -1) {};
		\node [style=none] (5) at (1, -1) {};
		\node [style=hadamard] (7) at (0, 0) {};
	\end{pgfonlayer}
	\begin{pgfonlayer}{edgelayer}
		\draw (4.center) to (2.center);
		\draw (5.center) to (3.center);
		\draw (0) to (1);
	\end{pgfonlayer}
\end{tikzpicture}
 \ \right\rrbracket 
					&= \left\llbracket \ \begin{tikzpicture}
	\begin{pgfonlayer}{nodelayer}
		\node [style=X phase dot] (0) at (-0.75, -1.25) {$\pi$};
		\node [style=none] (1) at (-1.75, 2) {};
		\node [style=none] (2) at (0.75, 2) {};
		\node [style=none] (3) at (-0.75, -2) {};
		\node [style=none] (5) at (1.75, -2) {};
		\node [style=X phase dot] (6) at (0.75, 1.25) {$\pi$};
		\node [style=hadamard] (7) at (0, 0) {};
		\node [style=none] (8) at (-2.25, -0.5) {};
		\node [style=none] (9) at (-2.25, -2) {};
		\node [style=none] (10) at (2.25, -2) {};
		\node [style=none] (11) at (2.25, -0.5) {};
		\node [style=none] (12) at (-2.25, 2) {};
		\node [style=none] (13) at (-2.25, 0.5) {};
		\node [style=none] (14) at (2.25, 0.5) {};
		\node [style=none] (15) at (2.25, 2) {};
		\node [style=none] (16) at (-0.75, -0.5) {};
		\node [style=none] (17) at (-0.75, 0.5) {};
		\node [style=none] (18) at (0.75, 0.5) {};
		\node [style=none] (19) at (0.75, -0.5) {};
		\node [style=none] (20) at (-0.75, 2) {};
		\node [style=none] (21) at (0.75, -2) {};
	\end{pgfonlayer}
	\begin{pgfonlayer}{edgelayer}
		\draw (3.center) to (0);
		\draw [in=-90, out=165, looseness=0.75] (0) to (1.center);
		\draw [in=90, out=-15, looseness=0.75] (6) to (5.center);
		\draw [in=-165, out=15] (0) to (6);
		\draw (2.center) to (6);
		\draw [style=empty] (19.center) to (21.center);
		\draw [style=empty] (21.center) to (10.center);
		\draw [style=empty] (10.center) to (11.center);
		\draw [style=empty] (19.center) to (11.center);
		\draw [style=empty] (11.center) to (14.center);
		\draw [style=empty] (19.center) to (18.center);
		\draw [style=empty] (18.center) to (14.center);
		\draw [style=empty] (14.center) to (15.center);
		\draw [style=empty] (15.center) to (20.center);
		\draw [style=empty] (20.center) to (17.center);
		\draw [style=empty] (17.center) to (18.center);
		\draw [style=empty] (17.center) to (16.center);
		\draw [style=empty] (16.center) to (19.center);
		\draw [style=empty] (16.center) to (8.center);
		\draw [style=empty] (8.center) to (9.center);
		\draw [style=empty] (9.center) to (21.center);
		\draw [style=empty] (8.center) to (13.center);
		\draw [style=empty] (13.center) to (17.center);
		\draw [style=empty] (13.center) to (12.center);
		\draw [style=empty] (12.center) to (20.center);
	\end{pgfonlayer}
\end{tikzpicture}
 \ \right\rrbracket \\
					&= \left(
						\left\llbracket \ \begin{tikzpicture}
	\begin{pgfonlayer}{nodelayer}
		\node [style=none] (0) at (0, -0.5) {};
		\node [style=none] (1) at (0, 0.5) {};
	\end{pgfonlayer}
	\begin{pgfonlayer}{edgelayer}
		\draw (1.center) to (0.center);
	\end{pgfonlayer}
\end{tikzpicture}
 \ \right\rrbracket \otimes 
						\left\llbracket \ \begin{tikzpicture}
	\begin{pgfonlayer}{nodelayer}
		\node [style=none] (7) at (-0.75, -1) {};
		\node [style=none] (8) at (0, 0.75) {};
		\node [style=X phase dot] (10) at (0, 0) {$\pi$};
		\node [style=none] (11) at (0.75, -1) {};
	\end{pgfonlayer}
	\begin{pgfonlayer}{edgelayer}
		\draw [bend left] (10) to (11.center);
		\draw (8.center) to (10);
		\draw [bend right] (10) to (7.center);
	\end{pgfonlayer}
\end{tikzpicture}
 \ \right\rrbracket
					\right)
					\left(
						\left\llbracket \ \begin{tikzpicture}
	\begin{pgfonlayer}{nodelayer}
		\node [style=none] (0) at (0, -0.5) {};
		\node [style=none] (1) at (0, 0.5) {};
	\end{pgfonlayer}
	\begin{pgfonlayer}{edgelayer}
		\draw (1.center) to (0.center);
	\end{pgfonlayer}
\end{tikzpicture}
 \ \right\rrbracket \otimes 
						\left\llbracket \ \begin{tikzpicture}
	\begin{pgfonlayer}{nodelayer}
		\node [style=none] (0) at (0, -0.5) {};
		\node [style=none] (1) at (0, 0.5) {};
		\node [style=hadamard] (2) at (0, 0) {};
	\end{pgfonlayer}
	\begin{pgfonlayer}{edgelayer}
		\draw (1.center) to (0.center);
	\end{pgfonlayer}
\end{tikzpicture}
 \ \right\rrbracket \otimes 
						\left\llbracket \ \begin{tikzpicture}
	\begin{pgfonlayer}{nodelayer}
		\node [style=none] (0) at (0, -0.5) {};
		\node [style=none] (1) at (0, 0.5) {};
	\end{pgfonlayer}
	\begin{pgfonlayer}{edgelayer}
		\draw (1.center) to (0.center);
	\end{pgfonlayer}
\end{tikzpicture}
 \ \right\rrbracket 
					\right)
					\left(
						\left\llbracket \ \begin{tikzpicture}
	\begin{pgfonlayer}{nodelayer}
		\node [style=none] (7) at (-0.75, 0.75) {};
		\node [style=none] (8) at (0, -1) {};
		\node [style=X phase dot] (10) at (0, -0.25) {$\pi$};
		\node [style=none] (11) at (0.75, 0.75) {};
	\end{pgfonlayer}
	\begin{pgfonlayer}{edgelayer}
		\draw [bend right] (10) to (11.center);
		\draw (8.center) to (10);
		\draw [bend left] (10) to (7.center);
	\end{pgfonlayer}
\end{tikzpicture}
 \ \right\rrbracket \otimes 
						\left\llbracket \ \begin{tikzpicture}
	\begin{pgfonlayer}{nodelayer}
		\node [style=none] (0) at (0, -0.5) {};
		\node [style=none] (1) at (0, 0.5) {};
	\end{pgfonlayer}
	\begin{pgfonlayer}{edgelayer}
		\draw (1.center) to (0.center);
	\end{pgfonlayer}
\end{tikzpicture}
 \ \right\rrbracket
					\right) \\
					&= \frac{\sqrt{2}}{8} \begin{pmatrix}
						-1 & 1 & 1 & 1 \\
						1 & -1 & 1 & 1 \\
						1 & 1 & -1 & 1 \\
						1 & 1 & 1 & -1 \\
					\end{pmatrix} \\
					&= \frac{\sqrt{2}}{8} \left\llbracket \ \begin{tikzpicture}
	\begin{pgfonlayer}{nodelayer}
		\node [style=box] (0) at (0, 0) {$\pm$};
		\node [style=none] (1) at (0, -1.25) {};
		\node [style=none] (2) at (0, 1.25) {};
	\end{pgfonlayer}
	\begin{pgfonlayer}{edgelayer}
		\draw (2.center) to (1.center);
	\end{pgfonlayer}
\end{tikzpicture}
 \ \right\rrbracket_{d=4}
				\end{align*}
		\endgroup
	\end{proof}
\end{proposition}

Now we prove the qutrit $X$-box is not a stabilizer diagram. The following lemma is required:
\begin{lemma}
	Every stabilizer state is equivalent to one of the following 12 diagrams:
	\begin{gather*}
		\qutritXstate{0}{0} ~~,\qquad 
		\qutritXstate{1}{2} ~~,\qquad 
		\qutritXstate{1}{2} ~~,\qquad 
		\qutritZstate{0}{0} ~~,\qquad 
		\qutritZstate{1}{2} ~~,\qquad 
		\qutritZstate{1}{2}\\[10pt]
		\qutritZstate{1}{1} ~~=~~ \qutritXstate{2}{2} ~~,\qquad 
		\qutritZstate{2}{2} ~~=~~ \qutritXstate{1}{1}\\[10pt]
		\qutritZstate{0}{1} ~~=~~ \qutritXstate{2}{0} ~~,\qquad 
		\qutritZstate{0}{2} ~~=~~ \qutritXstate{0}{1} ~~,\qquad 
		\qutritZstate{1}{0} ~~=~~ \qutritXstate{0}{2} ~~,\qquad 
		\qutritZstate{2}{0} ~~=~~ \qutritXstate{1}{0}
	\end{gather*}
	\begin{proof}
		We can use our elimination rules to prove this. A stabilizer state is a diagram with no inputs and one output, in which every phase component is a multiple of $\frac{2\pi}{3}$. After putting it in graph-like form (via Proposition \ref{prop:every_diagram_is_graph_like_qutrit}), all but one spider will be interior. We'll call the non-interior one the \emph{boundary} spider. Our elimination rules say we can remove all interior $\mathcal{P}$- and $\mathcal{N}$-spiders, plus all pairs of connected interior $\mathcal{M}$-spiders. So applying these rules until we can do so no more, we have two cases. The easiest case is where end up with just the single boundary spider, which must have no inputs and one output. In the other case we get a single $\mathcal{M}$-spider connected to a boundary spider. The \Mspider\ can have no other legs, and the boundary spider must have exactly one other leg, which is the output of the overall diagram:
		\begin{equation}
			\begin{tikzpicture}
	\begin{pgfonlayer}{nodelayer}
		\node [style=qutrit Z phase] (0) at (0, -0.5) {$a$ \nodepart{lower} $b$};
		\node [style=none] (1) at (0, 0.75) {};
	\end{pgfonlayer}
	\begin{pgfonlayer}{edgelayer}
		\draw (1.center) to (0);
	\end{pgfonlayer}
\end{tikzpicture}
 \quad \text{ or } \quad \begin{tikzpicture}
	\begin{pgfonlayer}{nodelayer}
		\node [style=qutrit Z phase] (0) at (0, 0.75) {$a$ \nodepart{lower} $b$};
		\node [style=none] (1) at (0, 2) {};
		\node [style=qutrit Z phase] (2) at (0, -1.75) {$m$ \nodepart{lower} $-m$};
		\node [style=qutrit hadamard unknown] (3) at (0, -0.5) {$\pm1$};
	\end{pgfonlayer}
	\begin{pgfonlayer}{edgelayer}
		\draw (1.center) to (0);
		\draw (2) to (0);
	\end{pgfonlayer}
\end{tikzpicture}

		\end{equation}
		In the first case, we're done. In the second case, we need only note:
		\begin{equation*}
			 \quad \xeq{\qutritRuleColourChange} \quad
			\begin{tikzpicture}
	\begin{pgfonlayer}{nodelayer}
		\node [style=qutrit Z phase] (0) at (0, 0.75) {$a$ \nodepart{lower} $b$};
		\node [style=none] (1) at (0, 2) {};
		\node [style=qutrit X phase] (2) at (0, -0.75) {$\pm m$ \nodepart{lower} $\mp m$};
	\end{pgfonlayer}
	\begin{pgfonlayer}{edgelayer}
		\draw (1.center) to (0);
		\draw (2) to (0);
	\end{pgfonlayer}
\end{tikzpicture}
 \quad \xeq{\ref{lem:substantial_m_copy}} \quad
			\begin{tikzpicture}
	\begin{pgfonlayer}{nodelayer}
		\node [style=none] (1) at (0, 0.75) {};
		\node [style=qutrit X phase] (2) at (0, -0.5) {$\pm m$ \nodepart{lower} $\mp m$};
	\end{pgfonlayer}
	\begin{pgfonlayer}{edgelayer}
		\draw (1.center) to (2);
	\end{pgfonlayer}
\end{tikzpicture}

		\end{equation*}







	\end{proof}
\end{lemma}	

\begin{proposition}\label{prop:X_box_not_stab}
	The qutrit $X$-box defined below is not a stabilizer diagram:
	\begin{equation*}
		\left\llbracket \  \ \right\rrbracket ~~ = ~~  
		\sum_{i,j=0}^{d-1} (1 -  \delta_{ij} ) \ket{i}\bra{j} ~~ = ~~  
		\begin{pmatrix}
			0 & 1 & 1 \\
			1 & 0 & 1 \\
			1 & 1 & 0
		\end{pmatrix}
	\end{equation*}
	\begin{proof}
		Stabilizer diagrams are closed under composition (they form a group). Hence if the $X$-box is stabilizer, it must send the phaseless red spider state to a stabilizer state. This has matrix:
		\begin{equation*}
			\left\llbracket~~ \begin{tikzpicture}
	\begin{pgfonlayer}{nodelayer}
		\node [style=X dot] (0) at (0, -1) {};
		\node [style=none] (1) at (0, 1.25) {};
		\node [style=box] (2) at (0, 0.25) {$X$};
	\end{pgfonlayer}
	\begin{pgfonlayer}{edgelayer}
		\draw (1.center) to (0);
	\end{pgfonlayer}
\end{tikzpicture}
 ~~\right\rrbracket ~~=~~
			\left\llbracket~~ \begin{tikzpicture}
	\begin{pgfonlayer}{nodelayer}
		\node [style=none] (0) at (0, -1) {};
		\node [style=none] (1) at (0, 1) {};
		\node [style=box] (2) at (0, 0) {$X$};
	\end{pgfonlayer}
	\begin{pgfonlayer}{edgelayer}
		\draw (1.center) to (0.center);
	\end{pgfonlayer}
\end{tikzpicture}
 ~~\right\rrbracket \left\llbracket~~ \begin{tikzpicture}
	\begin{pgfonlayer}{nodelayer}
		\node [style=X dot] (0) at (0, -0.5) {};
		\node [style=none] (1) at (0, 0.5) {};
	\end{pgfonlayer}
	\begin{pgfonlayer}{edgelayer}
		\draw (1.center) to (0);
	\end{pgfonlayer}
\end{tikzpicture}
 ~~\right\rrbracket ~~=~~
			\begin{pmatrix}
				0 & 1 & 1 \\
				1 & 0 & 1 \\
				1 & 1 & 0
			\end{pmatrix}
			\begin{pmatrix}
				1 \\
				0 \\
				0 
			\end{pmatrix} ~~=~~
			\begin{pmatrix}
				0 \\
				1 \\
				1 
			\end{pmatrix}
		\end{equation*}
		But, up to a scalar, this is not equal to any of the 12 stabilizer states above:
		\begin{equation*}
			\left\llbracket~~ \begin{tikzpicture}
	\begin{pgfonlayer}{nodelayer}
		\node [style=none] (1) at (0, 0.5) {};
		\node [style=X dot] (2) at (0, -0.5) {};
	\end{pgfonlayer}
	\begin{pgfonlayer}{edgelayer}
		\draw (1.center) to (2);
	\end{pgfonlayer}
\end{tikzpicture}
 ~~\right\rrbracket = \begin{pmatrix}1 \\ 0 \\ 0\end{pmatrix}, \quad
			\left\llbracket~~ \begin{tikzpicture}
	\begin{pgfonlayer}{nodelayer}
		\node [style=none] (1) at (0, 0.75) {};
		\node [style=qutrit X phase] (2) at (0, -0.5) {$1$ \nodepart{lower} $2$};
	\end{pgfonlayer}
	\begin{pgfonlayer}{edgelayer}
		\draw (1.center) to (2);
	\end{pgfonlayer}
\end{tikzpicture}
 ~~\right\rrbracket = \begin{pmatrix}0 \\ 1 \\ 0\end{pmatrix}, \quad
			\left\llbracket~~ \begin{tikzpicture}
	\begin{pgfonlayer}{nodelayer}
		\node [style=none] (1) at (0, 0.75) {};
		\node [style=qutrit X phase] (2) at (0, -0.5) {$2$ \nodepart{lower} $1$};
	\end{pgfonlayer}
	\begin{pgfonlayer}{edgelayer}
		\draw (1.center) to (2);
	\end{pgfonlayer}
\end{tikzpicture}
 ~~\right\rrbracket = \begin{pmatrix}0 \\ 0 \\ 1\end{pmatrix}, \quad
			\left\llbracket~~ \begin{tikzpicture}
	\begin{pgfonlayer}{nodelayer}
		\node [style=none] (1) at (0, 0.75) {};
		\node [style=qutrit Z phase] (2) at (0, -0.5) {$a$ \nodepart{lower} $b$};
	\end{pgfonlayer}
	\begin{pgfonlayer}{edgelayer}
		\draw (1.center) to (2);
	\end{pgfonlayer}
\end{tikzpicture}
 ~~\right\rrbracket = \begin{pmatrix}1 \\ \omega^a \\ \omega^b\end{pmatrix}
		\end{equation*}
	\end{proof}
\end{proposition}

\end{document}